\documentclass[final,5p,times,twocolumn,authoryear]{elsarticle}

\usepackage[authoryear]{natbib}
\usepackage{graphicx}
\usepackage{subcaption}
\usepackage{amssymb,amsmath}
\usepackage{pdflscape}
\usepackage{adjustbox}

\usepackage{booktabs}
\usepackage{lipsum}
\usepackage{epsfig}
\usepackage{txfonts}
\usepackage{xcolor}
\usepackage{tabularx}
\usepackage{footnote}
\usepackage{soul}
\usepackage{multirow}
\usepackage{multicol}
\usepackage{mathpazo}
\usepackage{lineno,hyperref}
\usepackage{textcomp}
\usepackage[utf8]{inputenc}
\usepackage{booktabs}

\usepackage{amsthm}

\journal{High Energy Astrophysics}
\UseRawInputEncoding

\begin{document}

\begin{frontmatter}

%% Title, authors and addresses

%% use the tnoteref command within \title for footnotes;
%% use the tnotetext command for theassociated footnote;
%% use the fnref command within \author or \affiliation for footnotes;
%% use the fntext command for theassociated footnote;
%% use the corref command within \author for corresponding author footnotes;
%% use the cortext command for theassociated footnote;
%% use the ead command for the email address,
%% and the form \ead[url] for the home page:
%% \title{Title\tnoteref{label1}}
%% \tnotetext[label1]{}
%% \author{Name\corref{cor1}\fnref{label2}}
%% \ead{email address}
%% \ead[url]{home page}
%% \fntext[label2]{}
%% \cortext[cor1]{Zahir Shah}
%% \affiliation{organization={},
%%            addressline={}, 
%%            city={},
%%            postcode={}, 
%%            state={},
%%            country={}}
%% \fntext[label3]{}

\title{Probing Spectral Evolution and  Intrinsic Variability of Mkn\,421: A Multi-Epoch AstroSat Study of X-ray Spectra}

%% use optional labels to link authors explicitly to addresses:
%% \author[label1,label2]{}
%% \affiliation[label1]{organization={},
%%             addressline={},
%%             city={},
%%             postcode={},
%%             state={},
%%             country={}}
%%
%% \affiliation[label2]{organization={},
%%             addressline={},
%%             city={},
%%             postcode={},
%%             state={},
%%             country={}}

%

\author{Sikandar Akbar\corref{cor1}\fnref{label1}}
\ead{darprince46@gmail.com}
\affiliation[label1]{Department of Physics, University of Kashmir, Srinagar 190006, India}
\cortext[cor1]{Corresponding author: Sikandar Akbar}

\author{Zahir Shah\corref{cor1}\fnref{label2}}
\ead{shahzahir4@gmail.com}
\affiliation[label2]{Department of Physics, Central University of Kashmir, Ganderbal 191201, India.}

\cortext[cor1]{Corresponding author: Zahir Shah}

\author{Ranjeev Misra\fnref{label3}}
\affiliation[label3]{Inter-University Centre for Astronomy and Astrophysics,  Post Bag  4, Ganeshkhind, Pune-411007, India}

\author{Naseer Iqbal\fnref{label1}}

\author{Javaid Tantry\fnref{label1}}

\begin{abstract}
Our study presents a time-resolved X-ray spectral analysis of Mkn\,421, using AstroSat observations
taken during different epochs between 2016 and 2019. The variability of the source in X-rays is
utilised to investigate the evolution of its spectral properties. Each observation period was divided
into segments of about 10 ks, and we employed three forms of particle distributions like broken-power law (BPL), log-parabola (LP), and power-law with maximum electron energy ( $\xi-max$ model) undergoing synchrotron losses to fit the broad X-ray spectrum in each segment. We observed that all of these models provided good fit to the spectra. In case of broken-power law model, we investigated the relationship between normalized particle density at an energy less than the break energy and the index before the break.
The results revealed an inverse correlation between the index and particle density with no time delay. Additionally, correlations between spectral parameters were used to determine the pivot energy. We observed that the pivot energy remained same across the observations. For $\xi-{max}$ and LP model, we define analogous pivot energies and show that they also do not vary, indicating the
model independent nature of the result. The constant pivot energy suggests that the source’s variability arises from
index variations and not due to changes in the normalization. Consequently, parameters such as magnetic field strength, doppler factor etc  do not contribute to the source’s variability.  Instead, variations are primarily associated with the acceleration or escape timescales of emitted particles within the source.

% This constancy is consistent within the $\xi-{max}$ and LP model, indicating the model's independence of the pivot energy. The constant pivot energy suggests that the source’s variability arises from index variations and is unaffected by normalization. Consequently, factors such as magnetic fields, doppler factor etc  do not contribute to the source’s variability.  Instead, variations are primarily associated with the acceleration or escape timescales of emitted particles within the source.
\end{abstract}
%%Graphical abstract
%\begin{graphicalabstract}
%\includegraphics{grabs}
%\end{graphicalabstract}

%%Research highlights
%\begin{highlights}
%\item Research highlight 1
%\item Research highlight 2
%\end{highlights}

\begin{keyword}
galaxies: active - BL Lacertae objects: general - BL Lacertae objects: individual: Mkn\,421 - galaxies: jets –
X-rays: galaxies.
%% keywords here, in the form: keyword \sep keyword, up to a maximum of 6 keywords
%% PACS codes here, in the form: \PACS code \sep code

%% MSC codes here, in the form: \MSC code \sep code
%% or \MSC[2008] code \sep code (2000 is the default)
\end{keyword}

\end{frontmatter}

%\tableofcontents

%% \linenumbers

%% main text

\label{introduction}

Blazars are subclass of active galactic nuclei (AGN) for which non-thermal emission from a relativistic jet is dominant.  This relativistic jet being directed proximal to the line of sight of an observer leads to the  Doppler boosting of the emission from blazars \citep{1995PASP..107..803U, 2003ApJ...596..847B}.
 Blazars are known for several distinctive features, including strong polarization in both optical and radio wavelengths, superluminal motion observed in radio observations, and rapid variability across the electromagnetic spectrum. The  variability in blazars ranges  on timescales from minutes to years, as reported in various studies \citep{1997ARA&A..35..445U, bhatta2020nature, bhatta2016multifrequency}\\
 
Blazars exhibit a distinctive spectral energy distribution (SED)  spanning from radio to $\gamma$-rays \citep{ghisellini1997optical}. The SED consist of two peaks: the first peak is observed  in optical/UV/X-ray wavelengths, while the second peak is observed in $\gamma$-rays. The first peak is usually attributed to the synchrotron radiation from relativistic particles in the jet (\cite{a0c40398058740ba8a38abc601fcfaf8}), while second peak is still being investigated and is currently understood by both leptonic  process via Inverse Compton \citep[IC;][]{bottcher2013leptonic, 10.1093/mnras/stx1194}  and hadronic emission processes \citep{mannheim1993proton, mucke2001proton}. In blazars,  the IC process can occur through synchrotron self-Compton  \citep[SSC;][]{1974ApJ...188..353J, maraschi1992jet, 1993ApJ...407...65G} and External Compton  \citep[EC;][]{1992A&A...256L..27D, sikora1994comptonization, 2000ApJ...545..107B, 10.1093/mnras/stx1194}. 
On the basis of synchrotron peak in the SED and optical spectral properties,  blazars can be broadly classified into  Flat-Spectrum Radio Quasars (FSRQ) and  BL Lac objects (BL Lacs). FSRQs are more luminous and show strong emission line features in their optical spectrum \citep{1995PASP..107..803U}. The synchrotron emission from  FSRQs  peaks at $\sim$  $10^{13-14}$ Hz. 
On the other hand, BL\,Lac objects are less luminous and show weak or no emission lines in their optical spectrum . The synchrotron emission from BL\,Lacs  peaks at $\sim$  $10^{14-16}$ Hz \citep{abdo2010spectral}. 
In recent classification,  BL lacs are categorised  into  low synchrotron-peaked BL\,Lac (LBL: $\nu_{p,syn}$$<$ $10^{14}$ Hz), 
intermediate synchrotron-peaked  BL\,Lac (IBL: $10^{14}$$<$ $\nu_{p,syn}$ $<$ $10^{15}$ Hz) and    high synchrotron-peaked BL\,Lac ( HBL: $\nu_{p,syn}$ $>$ $10^{15}$ Hz)  \citep{abdo2010spectral}.

Mkn\,421 is an HBL source located at a redshift $z\sim 0.031$. It is among the brightest nearby very-high-energy (VHE) sources in the extragalactic universe.  Mkn\,421 was first detected in $\gamma$-rays by 
the Energetic Gamma Ray Experiment Telescope \citep[EGRET;][]{lin1992detection}.  It was also the first confirmed TeV blazar observed by Whipple telescope \citep{punch1992detection, petry1996detection}.
 It has been extensively studied using different observatories across the electromagnetic spectrum \citep{fossati2008multiwavelength, shukla2012multiwavelength, sinha2016long, krawczynski2001simultaneous}.
It exhibits flux variability across all energy bands, particularly in X-ray/$\gamma$-ray bands, on timescales ranging from minutes to days, as reported in several studies \citep[see e.g.,][]{doi:10.1146/annurev.aa.33.090195.001115, falomo2014optical, goyal2020blazar}. The high variability indicate that the electrons responsible for the emission are highly energetic and have a short cooling time \citep{2007ApJ...664L..71A, ackermann2016minute}.
While number of  correlation studies have indicated a zero time lag between X-ray and TeV emissions from Mkn\,421 in different flux states \citep[see e.g.,][]{amenomori2003multi, 2006ApJ...641..740R, aleksic2015unprecedented, 2016ApJ...819..156B,  2021A&A...647A..88A},  it is noteworthy that a time lag of $2.1 \pm 0.7$ ks was observed between X-ray and TeV emissions during the 2001 flaring of Mkn\,421 \citep{fossati2008multiwavelength}.
 Additionally, investigations focusing on the correlation between optical and X-ray/TeV emissions, such as \citep{macomb1995multiwavelength, cao2013particle}, have reported no significant correlation. These findings lend support to hadronic models and two-zone SSC models.

Observations of Mkn\,421 in both flaring and quiescent states have revealed a curvature in its X-ray spectrum, with the log parabola model providing the best fit \citep{10.1093/mnras/sty2003, 2004A&A...413..489M, tramacere2007signatures}.
In addition to the log-parabola model, different physical models have also been used to reproduce the curvature in the X-ray spectrum in various studies \citep{10.1093/mnras/staa3022, 10.1093/mnras/stac1964, 2021MNRAS.508.5921H}. These studies constrain the physical models by studying  the correlation among the model parameters. 
For example, \citet{2021MNRAS.508.5921H} demonstrated that it is possible to differentiate between spectrally degenerate models  by correlating the spectral parameters. Additionally, with the help of correlation studies one can gain insight into the acceleration mechanism taking place in these systems. Following the same approach, \cite{10.1093/mnras/stac1964} extended the correlation study among model parameters to long term light curve of Mkn\,421,  and discovered that their results are consistent with the short-term evolution of the source.
Apart from these studies,  \citet{2022MNRAS.513.1662M} conducted a detailed spectral study of Mkn\,421 during January 2019 flaring period. Their findings demonstrated that the broad X-ray spectrum could be effectively modeled using a broken power-law model, both in time-averaged and time-resolved scenarios.

Blazar variability has been extensively studied using various theoretical models. Among these, two-zone models have gained significant attention for their ability to explain both steady-state and flaring emissions. \citet{1998A&A...333..452K} introduced a framework where particle acceleration occurs near the shock front, while radiation is emitted in a downstream region. This separation highlights the interplay between acceleration and radiative processes, providing critical insights into the multi-wavelength behavior of blazars like Mkn 421. \citet{2021MNRAS.505.2712D} analyzed the February 2010 flare of Mrk 421 using a two-zone model, consisting of a steady-state emitting zone responsible for quiescent emission and a transient acceleration zone for flaring activity. Their model  predicts the harder-when-brighter trend in soft X-rays as a result of particle acceleration processes, highlighting the dynamic interplay between acceleration and cooling mechanisms. Time-dependent models incorporating particle transport and acceleration mechanisms provide deeper insights into the formation of X-ray spectra and time lags in blazars. \citet{Lewis_2016} developed a physical framework using transport equations that account for stochastic acceleration, synchrotron losses, and shock acceleration. Applying this model to Mrk 421, they reproduced both the observed X-ray spectra and time lags, providing important constraints on the physical processes operating in the jet. \citet{2018ApJ...854...66K} analyzed the spectral and flux variability of Mrk 421 during selected epochs, identifying correlations consistent with first- and second-order Fermi acceleration. Their findings emphasize the influence of turbulence and magnetic field variations in shaping the observed X-ray behavior, offering valuable insights into the underlying particle acceleration mechanisms.

\cite{2021MNRAS.504.5485S} conducted a detailed time resolved X-ray spectral study of HBL\,1ES\,1959+650. The authors showed that  synchrotron emission of a broken power-law electron distribution provided a better fit to the X-ray spectrum than the synchrotron emission of log-parabola model. Further, the authors reported an anticorrelation between the the normalized particle density  with the index  below the break energy. However, when introducing a delay of approximately 60 ks, a strong positive correlation between the particle density and index was observed. The modelling of such  time delays is used to constrain the physical parameters responsible for emission. 
The investigation of short-term variability properties in jet of Mkn\,421 can offer insight into their particle acceleration and emission mechanism. Our study presents an X-ray analysis of Mkn\,421 using AstroSat observations taken at various epochs from 2016 to 2019.  Following a methodology similar  to \citet{2021MNRAS.504.5485S}, we perform a time-resolved spectral analysis to establish correlations between spectral parameters and gain insight into the system's nature. The paper is organized as follows: Section 1 covers data reduction, Section 2 presents the results of time variability, time-resolved spectral analysis, and a detailed correlation study. Section 3 offers a summary of our work, followed by an in-depth discussion.

\section{DATA  REDUCTION }

AstroSat telescope, as India's first multiwavelength observatory, has the unique capability of observing from optical/UV to soft-hard X-ray ranges simultaneously  \citep{2017JApA...38...27A}. It comprises of five payloads, including the Ultra-Violet Imaging Telescope (UVIT) which observes in the 130-300 nm range \citep{2017JApA...38...28T,2017AJ....154..128T}, the Soft X-ray focusing Telescope (SXT) which observes in the 0.3-8.0 keV range \citep{2016SPIE.9905E..1ES,2017JApA...38...29S}, the Large Area X-ray Proportional Counter (LAXPC) which observes in the 3-80 keV range \citep{2016SPIE.9905E..1DY}, and the Cadmium Zinc Telluride Imager (CZTI) which observes in the 10-100 keV range \citep{2017CSci..113..595R}. During the period 2016-2019, Mkn\,421 was observed by AstroSat at different epochs, the details of the observation time and total exposure time of these observations  are  provided in Table \ref{table_ast-obs}. 
The data were obtained from the AstroSat data archive ASTROBROWSE. Further details on the data reduction for the SXT and LAXPC observations are described below.

\subsection{Soft X-ray Telescope (SXT)}
The SXT is an X-ray imaging telescope with a focal length of 2 meters, operating in the 0.3-8.0 keV energy range (\cite{2017JApA...38...29S}. It uses a e2V CCD-22 as its primary detector, which is located at the common focus of all the mirrors. The telescope has an angular resolution of 2 arcmin and a field of view with a diameter of $\sim$ 40 arcmin.

AstroSat observed Mkn\,421 in photon counting (PC) mode during different epochs between 2016-2019. The Level-1 data of the source was processed using SXT pipeline version 1.4b (AS1SXTLevel2-1.4b; release date  2019-01-03). 
The SXTEVTMERGER tool was utilized to merge the level-2 event files from different orbits. Science products such as light curves, images, and spectra were extracted using XSELECT tool version V2.4m distributed with HEASOFT v. 6.29. A circle with a radius of 16 arcmin centered on the source, which included more than 95 percent of the photon counts, was used to extract light curves and spectra. An off-axis auxiliary response file (ARF) appropriate for the specific source region generated by using the sxtARFModule tool, along with a background spectrum 'SkyBkg comb EL3p5 Cl Rd16p0 v01.pha', provided by the SXT POC team, and the response matrix file (RMF) 'sxt pc mat g0to12.rmf' were used for spectral analysis.  In order to ensure optimal counts per bin, "ftgrouppha" command was used.

 \subsection{Large Area Proportional Counters
(LAXPC)}

LAXPC is one of the major payloads onboard AstroSat. It is non-focusing instrument with a large total effective area  of ($\sim$ 6000  $cm^{2}$. It  consist of three X-ray proportional counter units namely LAXPC10, LAXPC20, and LAXPC30.
These counters operate in the energy range of 3-80 keV and have a high time resolution of 10 $\mu$s \citep{2016SPIE.9905E..1DY, 2017ApJS..231...10A, 2017JApA...38...30A, 2017ApJ...835..195M}. In our work, we utilized data exclusively from LAXPC20, as LAXPC30 was turned off due to abnormal gain changes in March 2018 \citep{2017ApJS..231...10A}, while the application of standard response files to LAXPC10 was not feasible due to high voltage adjustments that occurred in the spring of 2018.\\
We processed and analyzed the data using the LAXPCSOFT software (version: LAXPCsoftware22Aug15). First, we created a level-2 event file using the command laxpc$\_$make$\_$event, and then generated a good time interval (gti) file using laxpc$\_$make$\_$stdgti to remove intervals of South Atlantic Anomaly (SAA) and Earth occultation. We used the gti file to create the source light curve and spectra. Since Mkn\,421 is faint source for LAXPC, we utilized the scheme of faint source background algorithm, in order to extract the background spectra and light curves for LAXPC 20. The faint source background algorithm  is implemented as a part of LAXPCSOFT \citep{misra2021alternative}.

We performed background subtraction using the standard FTool lcmath command. In this work,  we limited the spectral analysis to 3-19 keV since the background dominated the spectra at higher energies.

\begin{table}
    \centering
    \caption{Observations of Mkn\,421 during the flaring events between 2016 and 2019 using the LAXPC and SXT instruments.}
    \scalebox{0.7}{ % Scale the table to 80% of its original size
    \begin{tabular}{lllll} \hline
    Observation &  Observation ID & Instrument & Energy & Exposure  \\ 
      &&&(keV)&(ks)\\
      \hline
        
     S1 & G05\_201T01\_9000000478 & SXT & 0.7-7.0  & 11.00 \\
     & (2 June, 2016) & LAXPC20 & 3.0-30.0 & 27.60 \\
     S2 & A02\_005T01\_9000000948& SXT & 0.7-7.0  & 98.72  \\
     & (3-8 Jan, 2017) & LAXPC20 & 3.0-30.0  & 171.10 \\
     
     S3 & T01\_218T01\_9000001852 & SXT & 0.7-7.0  & 25.00\\
     & (19-20 Jan, 2018)& LAXPC20 & 3.0-30.0 & 31.00 \\
     
     S4 & A05\_015T01\_9000002650 & SXT & 0.7-7.0  & 111.94 \\
     & (10-19 Jan, 2019) & LAXPC20 & 3.0-30.0  & 150.40 \\
    S5 & A05\_204T01\_9000002856 & SXT & 0.7-7.0  & 104.90 \\
     & (23-28 April, 2019) & LAXPC20 & 3.0-30.0  & 146.00 \\
     \hline
     
    \end{tabular}
    }
    \label{table_ast-obs}
\end{table}

\begin{table}
    \centering
    \caption{The fractional root-mean-square variability of the source from  the 100 second binned light curves observed by SXT and LAXPC20 during various time periods.}
    \scalebox{0.75}{ % Scale the table to 80% of its original size
    \begin{tabular}{llll}
    \hline
       Observation &Observation ID  &Instrument& $F_{rms}$  \\ \hline
       S1.& G05\_201T01\_9000000478 & SXT& $0.0644 \pm 0.0051$\\
       & (2 June, 2016) & LAXPC20& $0.0764 \pm 0.0037$\\
       
       S2 & A02\_005T01\_9000000948&SXT & $0.2103 \pm 0.0051$\\
       & (3-8 Jan, 2017)&LAXPC20 & $0.3159 \pm 0.0056$\\
       S3 & T01\_218T01\_9000001852 & SXT& $0.1645 \pm 0.0077$  \\
       & (19-20 Jan, 2018) &LAXPC20& $0.2609 \pm 0.0110$\\
       S4 & A05\_015T01\_9000002650 & SXT &  $0.1206 \pm 0.0028$\\
       & (10-19 Jan, 2019)&LAXPC20 & $0.1409 \pm 0.0028$\\
       S5 & A05\_204T01\_9000002856 &SXT& $0.1193  \pm 0.0028$\\
       & (23-28 April, 2019)&LAXPC20&$0.1614  \pm 0.0031$\\

        \hline

    \end{tabular}
    }
    \label{tab:fvar}
\end{table}

\begin{table}
    \centering
    \caption{The break energy of the broken power-law model obtained  by fitting the time-averaged X-ray spectrum of different observations.}
    \scalebox{0.95}{ % Scale the table to 80% of its original size
    
    \begin{tabular}{llc} \hline
          Observation & Observation ID& $\xi_{brk} (\sqrt{keV})$ \\ \hline
           S1 &  G05\_201T01\_9000000478 & 1.7\\
           S2 & A02\_005T01\_9000000948& 1.9\\
           S3 & T01\_218T01\_9000001852 &1.8\\
           S4 & A05\_015T01\_9000002650& 2.1 \\
        
           S5 & A05\_204T01\_9000002856& 2.5 \\
          
        \hline
        
    \end{tabular}
    }
    
    \label{tab:brk}
\end{table}

\section{ANALYSIS}

\subsection{Time variability}

\begin{figure*}
    \centering
    \begin{subfigure}{0.45\linewidth}
        \includegraphics[width=0.7\linewidth, angle=270]{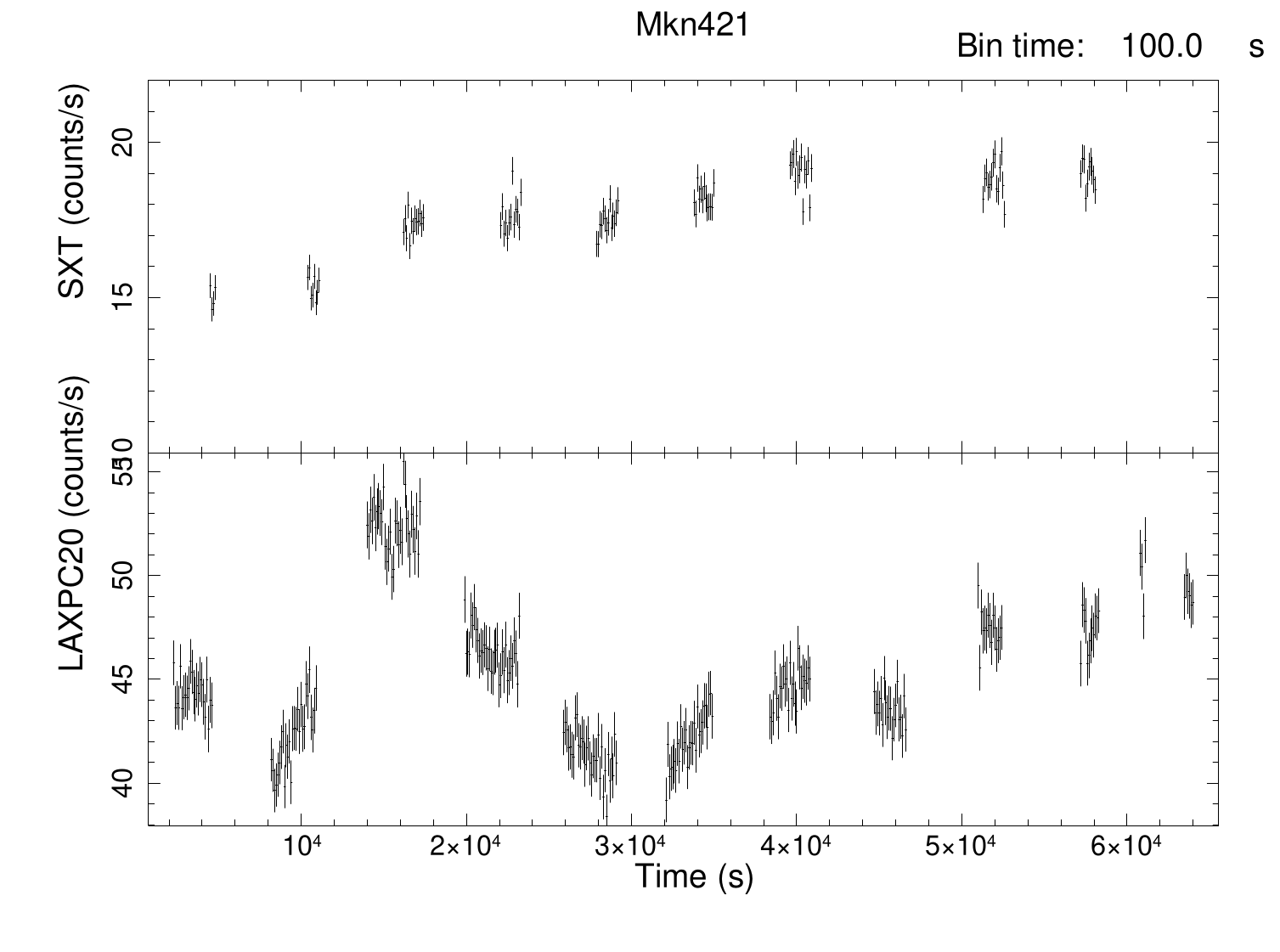}
    \end{subfigure}
    \hfill
    \begin{subfigure}{0.45\linewidth}
        \includegraphics[width=0.7\linewidth, angle=270]{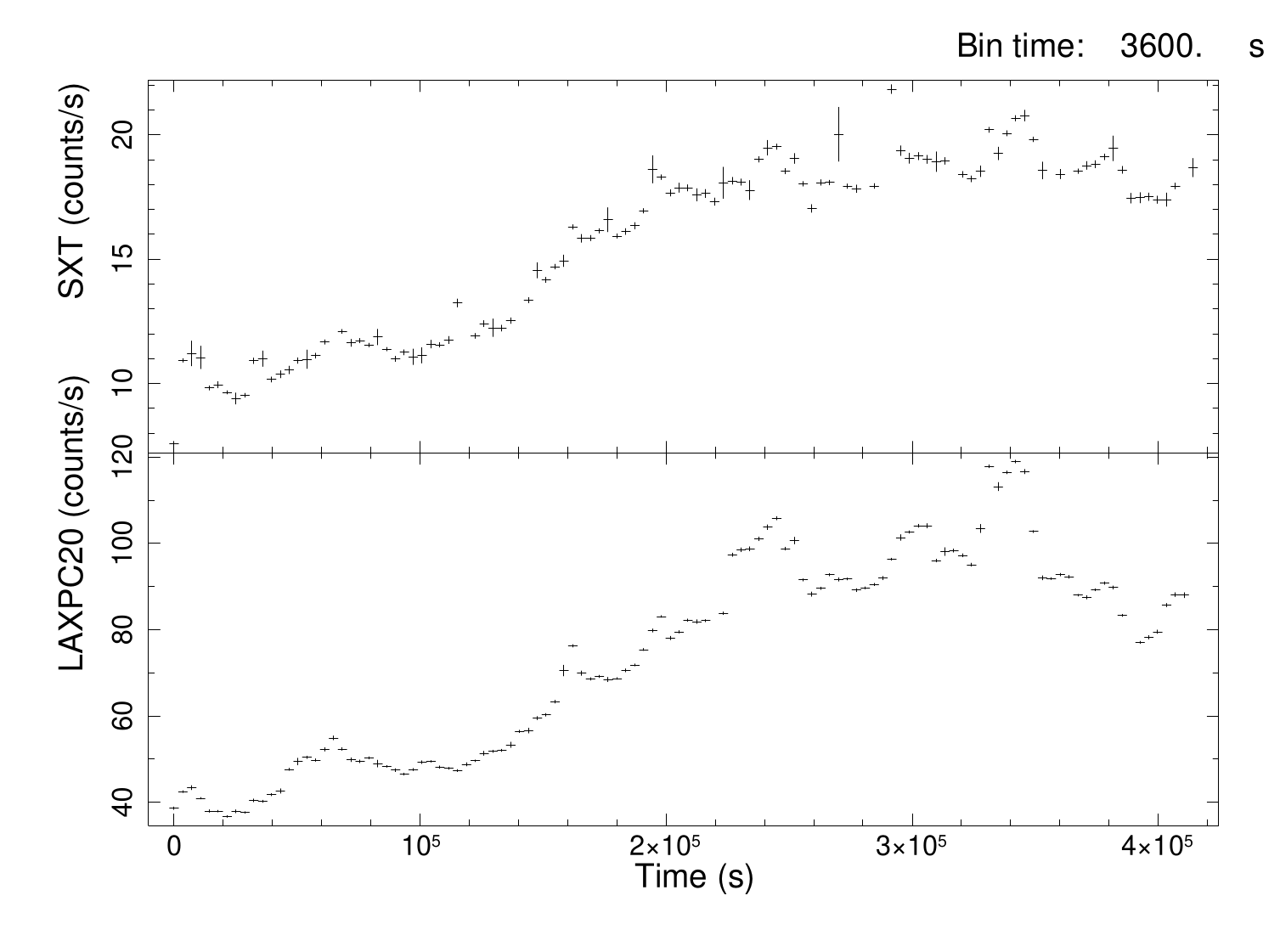}
    \end{subfigure}

    \vspace{0.5cm}

    \begin{subfigure}{0.45\linewidth}
        \includegraphics[width=0.7\linewidth, angle=270]{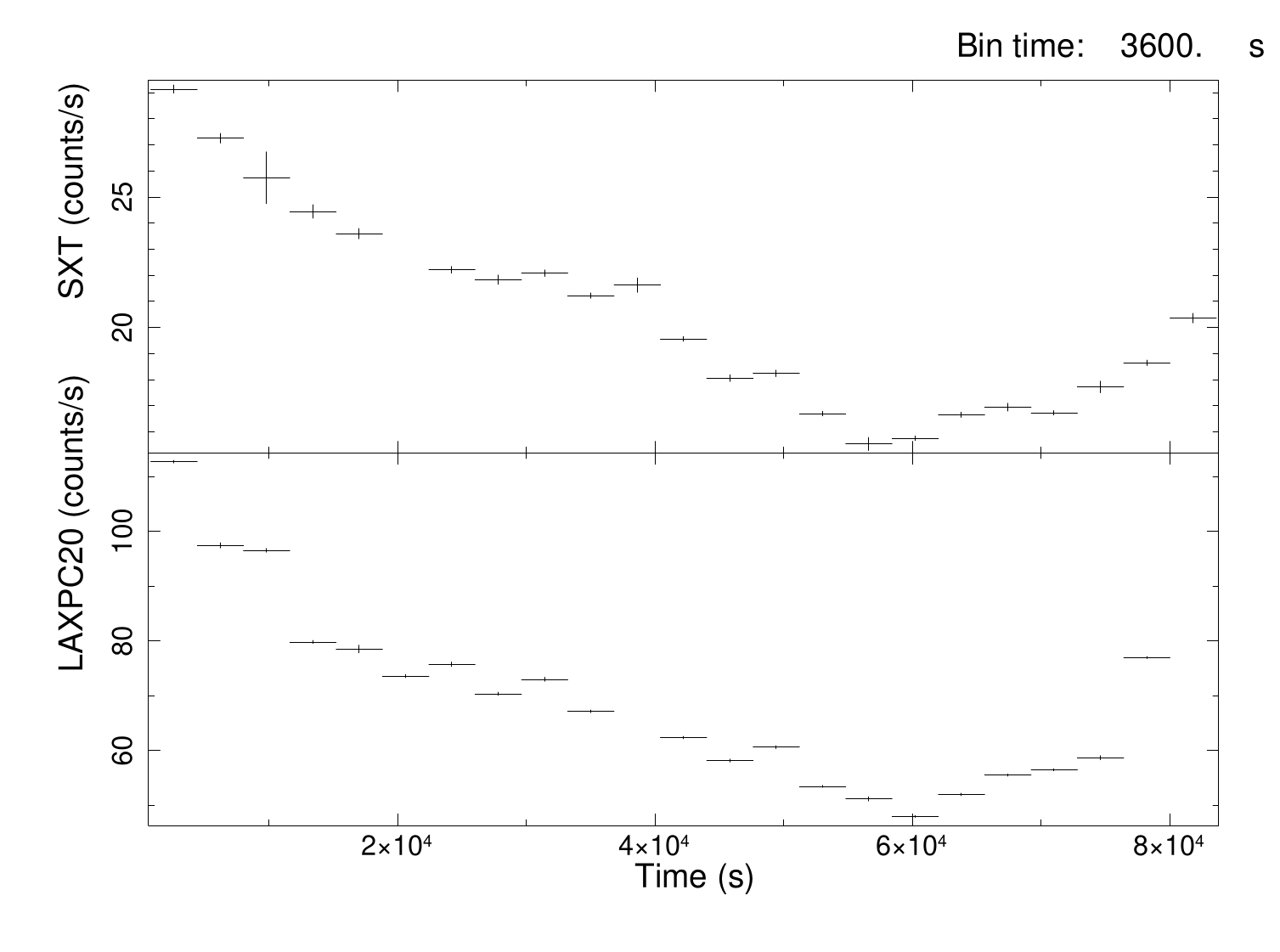}
    \end{subfigure}
    \hfill
    \begin{subfigure}{0.45\linewidth}
        \includegraphics[width=0.7\linewidth, angle=270]{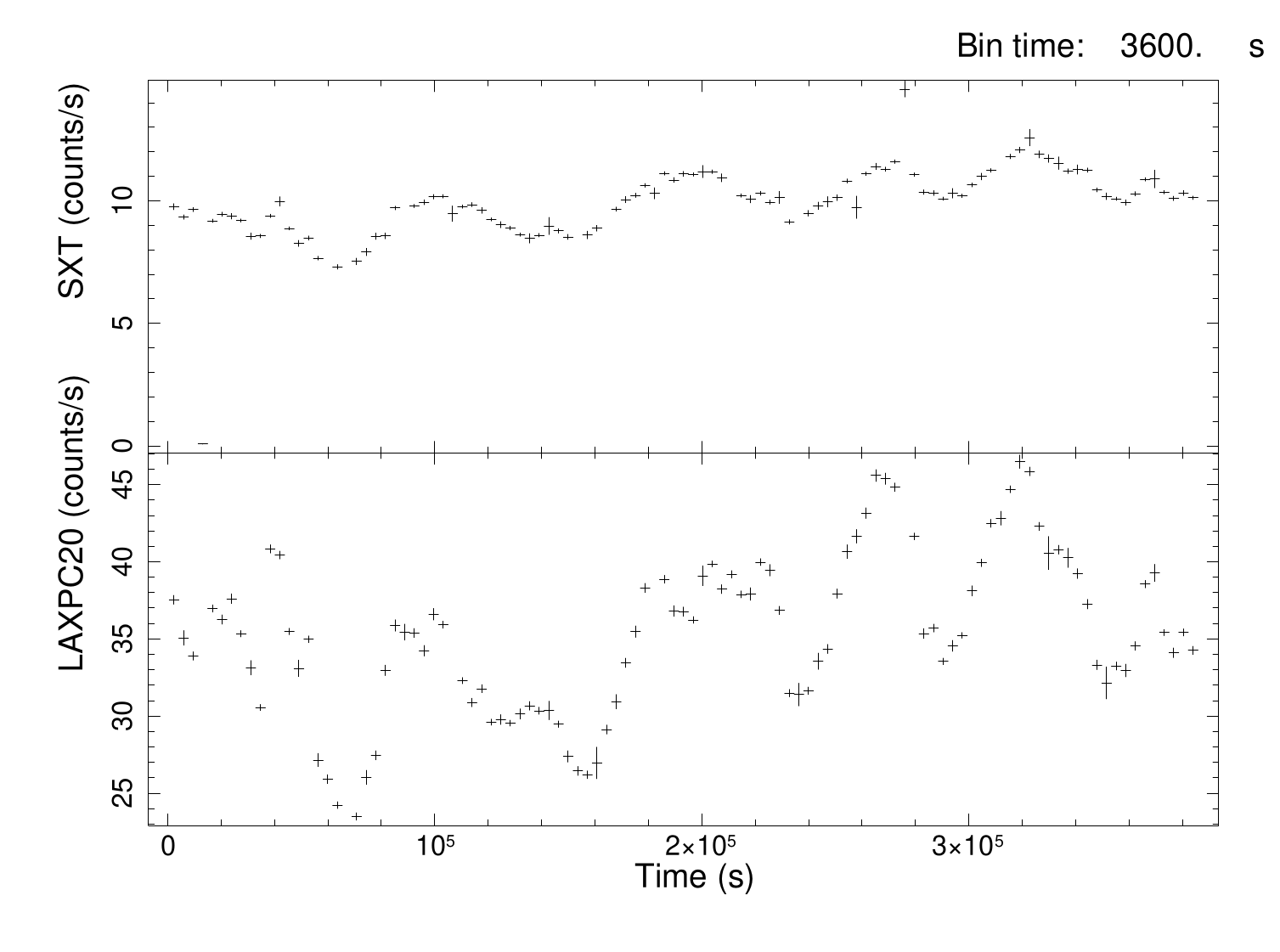}
    \end{subfigure}

    \vspace{0.5cm}

    \begin{subfigure}{0.45\linewidth}
        \includegraphics[width=0.7\linewidth, angle=270]{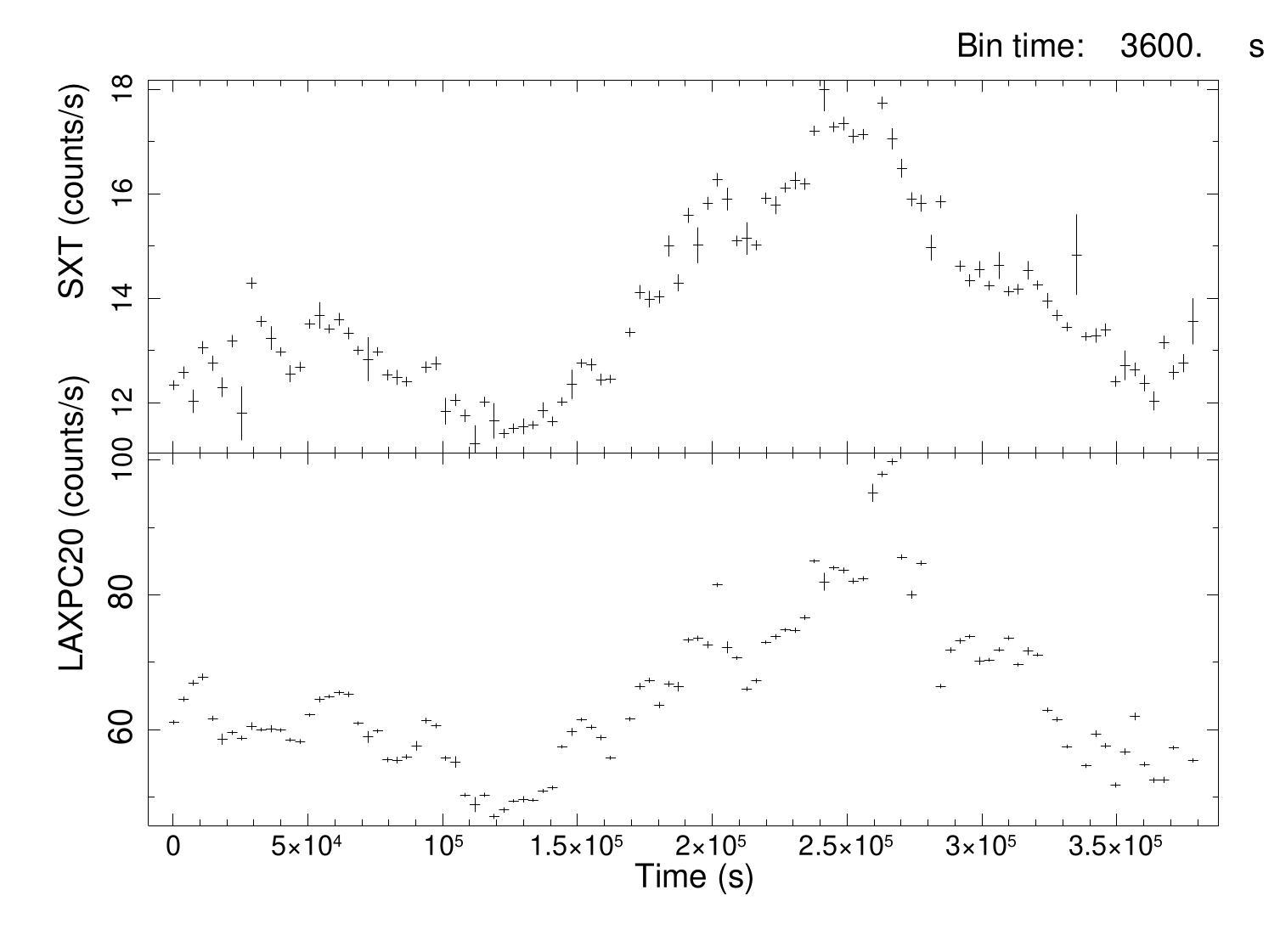}
    \end{subfigure}

    \caption{One-hour binned SXT and LAXPC light curves of Mkn\,421 obtained in the energy ranges 0.7-7 keV and 3-30 keV, respectively. Each panel corresponds to an individual observation. The upper left and right panels correspond to the S1 and S2 observations, respectively. The middle left and right panels correspond to the S3 and S4 observations, respectively, and the bottom panel corresponds to the S5 observation.}
    \label{fig:lc}
\end{figure*}

AstroSat observed Mkn\,421 at different times during the time period  2016 to 2019 (MJD: 57398-58601). 
We generated the X-ray light curves of Mkn\,421 in the energy range 0.7-7 keV using the SXT instrument, ignoring spectra below 0.7 keV due to poor quantum efficiency \citep{2017JApA...38...29S}. The LAXPC20 instrument was also used to extract light curves in the energy range 3.0-30 keV,  ignoring spectra above 30 keV due to background noise.
The aim is to conduct a  detailed  X-ray spectral investigation of Mrk\,421 by performing the correlation study among the spectral parameters. Therefore, we limited our analysis to AstroSat observations with a sufficient number of data points.
 The specific details of the X-ray observation like observation periods, the total time of exposure etc  are given in Table \ref{table_ast-obs}, different observations in the table are identified as S1, S2, S3, S4 and S5. The  SXT and LAXPC light curves corresponding to these observations are shown in Figure \ref{fig:lc}. We used the FTOOL LCSTATS to check the variability of the X-ray light curves. Using this tool, the fractional root-mean-square ($F_{rms}$) variability were calculated from the 100 sec binned SXT and LAXPC20 light curves.  The values  of $F_{rms}$ are mentioned in Table \ref{tab:fvar}. 
With $F_{rms}$ values ranging from 0.06 to 0.21 in SXT and 0.08 to 0.32 in LAXPC, it is evident that the source exhibits significant variability in X-rays.

\subsection{X-ray spectral analysis}

We used XSPEC version: 12.12.0 to analyze the X-ray spectra of the source in the energy band   $\sim 0.7-19$ keV. The SXT spectral data were included from 0.7-7.0 keV, while LAXPC data were included from 3.0-19.0 keV. Background spectrum ‘$\rm SkyBkgcomb_-EL3p5_-Cl_-Rd16p0_-v01.pha$' was used to
estimate background count rate, while the LAXPC background spectra were generated using the code for the faint source background provided by the LAXPC team. We applied a systematic error of 3\% throughout the fitting procedure. We used ftgrouppha command which adjusts the "GROUPING" column within a pha file  to achieve the  optimal binning of the SXT Spectrum. We grouped the LAXPC spectra at a 5 per cent level to obtain three energy bins per resolution. 

 The X-ray spectrum of Mkn\,421 is attributed to synchrotron emission from non-thermal relativistic electrons.  Therefore, we  model the X-ray spectrum by assuming   that the emission originates from a spherical region with a radius of R. This emission region is filled with a  magnetic field, B, and an isotropic distribution of relativistic electrons, $n(\gamma)$, which experience synchrotron loss.
The pitch angle averaged power per unit frequency emitted by single electron can be obtained  as \citep{1986rpa..book.....R}
\begin{equation}
    P(\gamma,\nu) = \frac{\sqrt{3}\pi q^{3} B} {4 mc^{2}} F\left(\frac{\nu}{\nu_{c}}\right)
\end{equation}
where, 

$ \nu_{c}$=$\frac{3\gamma^{2}qB}{16mc}$ and $F\left(\frac{\nu}{\nu_{c}}\right)$ is synchrotron power function defined by \citep{1986rpa..book.....R}

\begin{equation}
    F(x) = x \int_{x}^{\infty} K_{5/3}(\psi) \,d\psi\, 
\end{equation} 
here $K_{5/3}(\psi)$ is the
modified Bessel function of order $5/3$. \\ 
The synchrotron emissivity resulting from a relativistic electron distribution n($\gamma$) can be obtained by utilizing the equation, 

\begin{equation}
    J_{syn}(\omega,\alpha)= \frac{1}{4\pi} \int_{\gamma_{min}}^{\gamma_{max}}P(\gamma,\omega,\alpha)n(\gamma)d\gamma
\end{equation}
If we substitute $\xi$= $\gamma$ $\sqrt{C}$ , with $C$ = $1.36 \times {10^{-11}}$$\frac{\delta B}{1+z}$, where z represents the redshift of the source and $\delta$ represents the jet Doppler factor, the amount of synchrotron flux that the observer receives at energy $\epsilon$ can be obtained by using the expression
\citep{begelman1984theory}
\begin{equation}\label{eq:syn_obs}
    F_{syn}(\epsilon)= \frac{\delta^{3}{(1+z)}}{d_L^2} V\mathbb{A} \int_{\xi_{min}}^{\xi_{max}}F\left(\frac{\epsilon}{\xi^{2}}\right)n(\xi) d\xi
\end{equation}
Here, $\mathbb{A}= \frac{\sqrt{3}\pi e^{3} B} {16 m_{e} c^{2} \sqrt{C}} $,
$V$ refers to the volume of the emission region and $d_L$ represents the luminosity distance.

Our approach involved numerical solving of Equation \ref{eq:syn_obs}  and incorporating it into XSPEC as a local convolution model, synconv $\otimes$ n($\xi $).
This enables modeling of the photon spectrum for any particle energy distribution n($\xi$).
 More precisely, in the convolved XSPEC model, the variable ``Energy" is defined  as $\xi$ = $\sqrt{C}\gamma$. 
 We considered three models for $n(\xi)$, namely the broken power-law (BPL), log-parabola (LP), and power-law with maximum electron energy ($\xi-{max}$) model for the analysis. The joint spectra from SXT and LAXPC 20 instruments were fitted with the  \textit{constant$\times$TBabs$\times$Synconv $\otimes$ n$(\xi)$}. 
The value for the neutral hydrogen column density ($N_{H}$), determined from the LAB survey (\cite{2005A&A...440..775K}), remained fixed at $1.33\times10^{20}$ $cm^{-2}$ throughout the fitting procedure.   
   The  Galactic absorption is taken into account via the XSPEC TBabs routine. To account for potential variations in the effective areas of the SXT and LAXPC units, a scaling factor was introduced to the SXT spectra. In all spectral fits, the 'gain' command was employed, with the gain slope fixed at 1, and the offset allowed to vary. All these models provide a satisfactory fit to the combined SXT and LAXPC spectra. Given the source's high variability in both soft and hard X-ray bands, we investigated the temporal evolution of its X-ray spectral properties. To achieve this, we employed time-resolved spectral analysis by segmenting the total observation period into 10 ks intervals, creating SXT and LAXPC 20 spectra for each segment. In this analysis, we exclusively utilized SXT and LAXPC 20 spectra due to the greater stability of the LAXPC 20 background compared to LAXPC 10. The subsequent section provides a detailed description of these models and their corresponding results.
\subsubsection{Broken power-law model}
We consider the Broken power-law model (BPL) distribution of electrons and fitted the combined SXT and LAXPC  spectra (0.7-19 keV) with \textit{constant$\times$TBabs$\times$Synconv $\otimes$ n$(\xi)$},  here n($\xi$)  is the BPL distribution, which is given by
\begin{equation}
%\[
    n(\xi)= K \times
\begin{cases}
   \left( \frac{\xi}{\xi_{ref}}\right)^{-\Gamma_1}, \hspace{0.1cm}  \xi_{min}<\xi< \xi_{brk}\\
    \left(\frac{{\xi}_{brk}}{\xi_{ref}}\right)^{\Gamma_2-\Gamma_1} \hspace{0.1cm}\left(\frac{\xi}{\xi_{ref}}\right)^{-\Gamma_2},  \hspace{0.1cm} \xi_{brk}< \xi<\xi_{max},
\end{cases}
%\]
\end{equation}
where $\Gamma_{1}$ and $\Gamma_{2}$  represent the indices before and after break energy $\xi_{brk}$, respectively,  K is the normalization of the particle density at the reference energy $\xi_{ref}$.
In the time resolved spectra analysis, we fixed the break energy ($\xi_{break}$) at the values given in Table \ref{tab:brk}, which were determined during the fitting of the time-averaged spectrum.
We allowed the three parameters - namely, $\Gamma_{1}$, $\Gamma_{2}$, and the normalization factor to vary freely.
The normalization of the broken power-law model is defined in such a way that it equals the particle density at the break energy. However, this choice is flexible, and the model can be redefined to have the normalization equal to the density at any selected particle energy $\xi$. In the case of longest duration observation (S4), we calculated the normalization and its associated error for various $\xi$ values of time averaged spectrum. We observed that for $\xi$ $=$ $1.25 \sqrt{keV}$, the error on the normalization  %at $\xi$ $=$ $1.25 \sqrt{keV}$, 
was minimum. Therefore, we adopted this value as the $\xi_{ref}$ for all the observations. It's worth noting that this energy is lower than the break energy for all observations (see Table \ref{tab:brk}).  Therefore, we investigated the correlation 
between  the variation in the lower energy index  ($\Delta\Gamma_{1}$;  deviation of $\Gamma_{1}$ from their means)
and 
the variation in the normalized particle density $\frac{\Delta n}{<n>}$ (where $<n>$ is the mean value of n)  at the reference energy $(\xi_{ref} )$. 
The resulting spectral parameters, along with their respective reduced-$\chi^{2}$ values, are presented in Table \ref{tab:t1},\ref{tab:t2},\ref{tab:t3},\ref{tab:t4} and \ref{tab:t5} in the appendix section.

Spearman rank correlation between  $\frac{\Delta n}{<n>}$ and  $\Delta$$\Gamma_{1}$ for all observations are given in Table \ref{tab:bkn-src}.
In all observations except S1, a negative correlation between normalized particle density variation and index variation is evident. However, for the S1 observation, we cannot reject the null hypothesis that particle density variation and index variation are uncorrelated (refer to the $P_{s}$ value in Table \ref{tab:bkn-src}). As a result, we have chosen not to include the S1 observation data in our subsequent analysis involving the BPL model.

Due to the fact that flux is defined as a function of the normalization and index, it can be inferred that anticorrelation is observed between flux and index (\cite{2021MNRAS.504.5485S}). This pattern of hardening when brightening is a phenomenon that has been observed  in blazars before.
We noted  that there is no time delay  between the normalized  particle density variation and the index variation in all of the observations, which is  shown in  the discrete correlation function (DCF) plots (see Figure \ref{fig:dcf} in appendix section).
The  plots  in the Figure \ref{fig:slope-bkn} emphasize  that the  magnitude of the particle density variation, $\mid$$\frac{\Delta n}{<n>}$$\mid$ and variation in the index $\mid$$\Delta$$\Gamma_{1}$$\mid$ for individual observations as well as for the combined observation are comparable. %\textbf{In \citet{2021MNRAS.504.5485S}, the variability in the spectral index ($\Gamma$) and particle density is modeled as arising from variations in the acceleration timescale ($\tau_{\text{acc}}$). A sinusoidal perturbation in $\tau_{\text{acc}}$ leads to correlated variations in $\Delta n / n$ and $\Delta \Gamma$, but the relationship between their amplitudes and the presence of a time lag depends  on the relative timescales of the system.  When the variation timescale is comparable to $\tau_{\text{acc}}$, the amplitude of $\Delta \Gamma$ can significantly exceed that of $\Delta n / n$. In this scenario, a notable time lag between the variations in $\Delta n / n$ and $\Delta \Gamma$ is expected, as described by Equation (14) in \citet{2021MNRAS.504.5485S}, which is not observed in our results. However, for slow variations, the phase lag term $\left(\phi_{\text{lag}} / \sin(\phi_{\text{lag}})\right)$ in Equation (14) approaches unity. This leads to the steady-state solution represented by Equation (8) in \citet{2021MNRAS.504.5485S}, where the amplitudes of $\Delta n / n$ is greater than $\Delta \Gamma$.} 
The amplitude of $\Delta \Gamma$ can  exceed that of $\Delta n / n$, when the variation timescale is comparable to the acceleration timescale ($\tau_{\text{acc}}$), but in that case the presence of a time lag between the variations in $\Delta n / n$ and $\Delta \Gamma$ is expected \citep{2021MNRAS.504.5485S}.

BPL  at some reference energy, ${\xi}_{ref}$ and pivot energy,  ${\xi}_{piv}$ before the break energy can be written as,
\begin{equation} \label{eq:bpl_eq_pivot}
    n_{ref}\left[\frac{\xi}{{\xi}_{ref}}\right]^{-\Gamma_{1}}= n_{piv}\left[\frac{\xi}{{\xi}_{piv}}\right]^{-\Gamma_{1}} ~~,
\end{equation}
here,  $n_{ref}$ and  $n_{piv}$ represents the particle densities at reference energy, ${\xi}_{ref}$ and pivot energy, $\xi_{piv}$, respectively. $\Gamma_{1}$ is the index of the particle distribution before the break energy. The pivot energy, {$\xi_{piv}$},  is defined as the energy at which the particle density remains invariant during variations in the spectral index. This implies that the power-law part of the particle spectrum "rotates" around this point, leading to anti-correlated changes in the particle density and the spectral index. The reference energy, {$\xi_{ref}$}, is an arbitrarily chosen energy point used to define the normalization of the particle distribution. In order to obtain the pivot energy and particle density at the pivot energy, Equation \ref{eq:bpl_eq_pivot} can be rearranged such that index  before break energy becomes

%Taking log on both sides,
%\begin{equation}
%    log(n_{ref})-\Gamma_{1}\hspace{0.05cm} log\left[\frac{\xi}{{\xi}_{ref}}\right]=log(n_{piv})- \Gamma_{1}\hspace{0.05cm} log\left[\frac{\xi}{{\xi}_{piv}}\right]
%\end{equation}
%Rearranging the terms, the index before break energy can be obtained as
%\begin{equation}
%    log(n_{ref})=log(n_{piv})-\Gamma_{1} \hspace{0.05cm}log\left[\frac{{\xi}_{ref}}{{\xi}_{piv}}\right]
%\end{equation}

\begin{equation}\label{eq:index_bb}
    \Gamma_{1}=\frac{log(n_{piv})}{log \left[\frac{{\xi}_{ref}}{{\xi}_{piv}}\right]} - \frac{1}{log \left[\frac{{\xi}_{ref}}{{\xi}_{piv} }\right]} {log(n_{ref})} ~~~~~,
\end{equation}
The variation in $\log n$ is expressed as $\Delta \log n = \log(n + \Delta n) - \log n)$. Assuming \(\Delta n \ll n\), the normalized particle density variation becomes
\begin{equation}\label{eq:delta_var}
    \frac{\Delta n}{n}=-\Delta\Gamma_{1}\hspace{0.05cm}log\left[\frac{{\xi}_{ref}}{{\xi}_{piv}}\right] 
\end{equation}
Equation \ref{eq:delta_var} represents a linear equation, with its slope $log\left[\frac{{\xi}_{ref}}{{\xi}_{piv}}\right]$. This slope gives the pivot energy $ \xi_{piv}$ with the known $\xi_{ref}$. 
We conducted linear fits to the variations of the index  $\Delta\Gamma_{1}$ and the normalized particle density variation $\Delta$$n$/$<n>$  for each individual observation and for the combined observation (see Figure \ref{fig:slope-bkn}). The $\xi_{piv}$ values resulting from the linear fits in case of broken powerlaw model for observations S2, S3, S4 and S5 are $0.36\pm 0.04$ $\sqrt{keV}$,    $0.34\pm0.06$ $\sqrt{keV}$, $0.32\pm0.08$ $\sqrt{keV}$ and $0.26\pm0.09$  $\sqrt{keV}$, respectively.

\begin{figure*}
  \centering

  \includegraphics[scale=0.3,angle=-90]{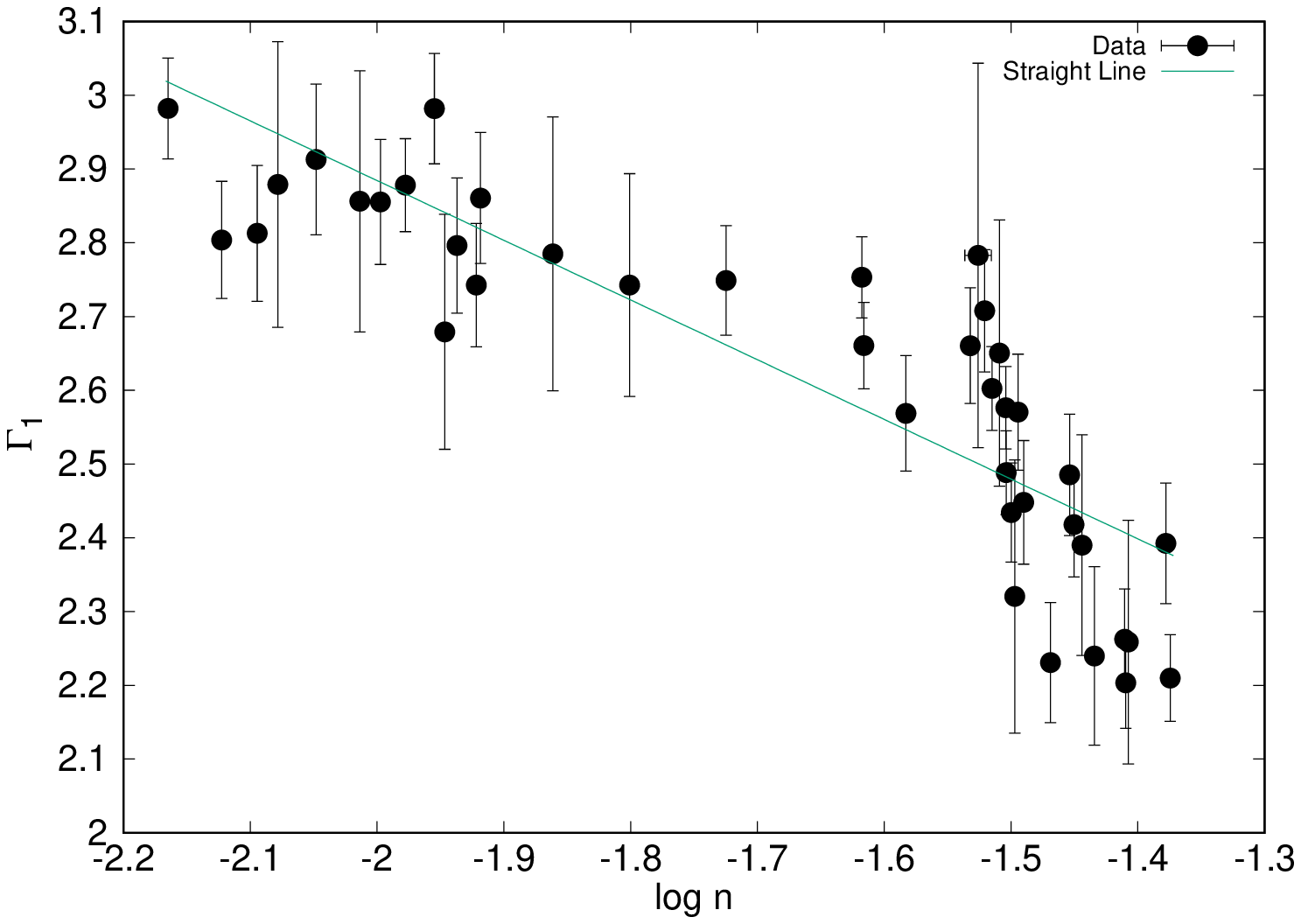}
  \includegraphics[scale=0.3,angle=-90]{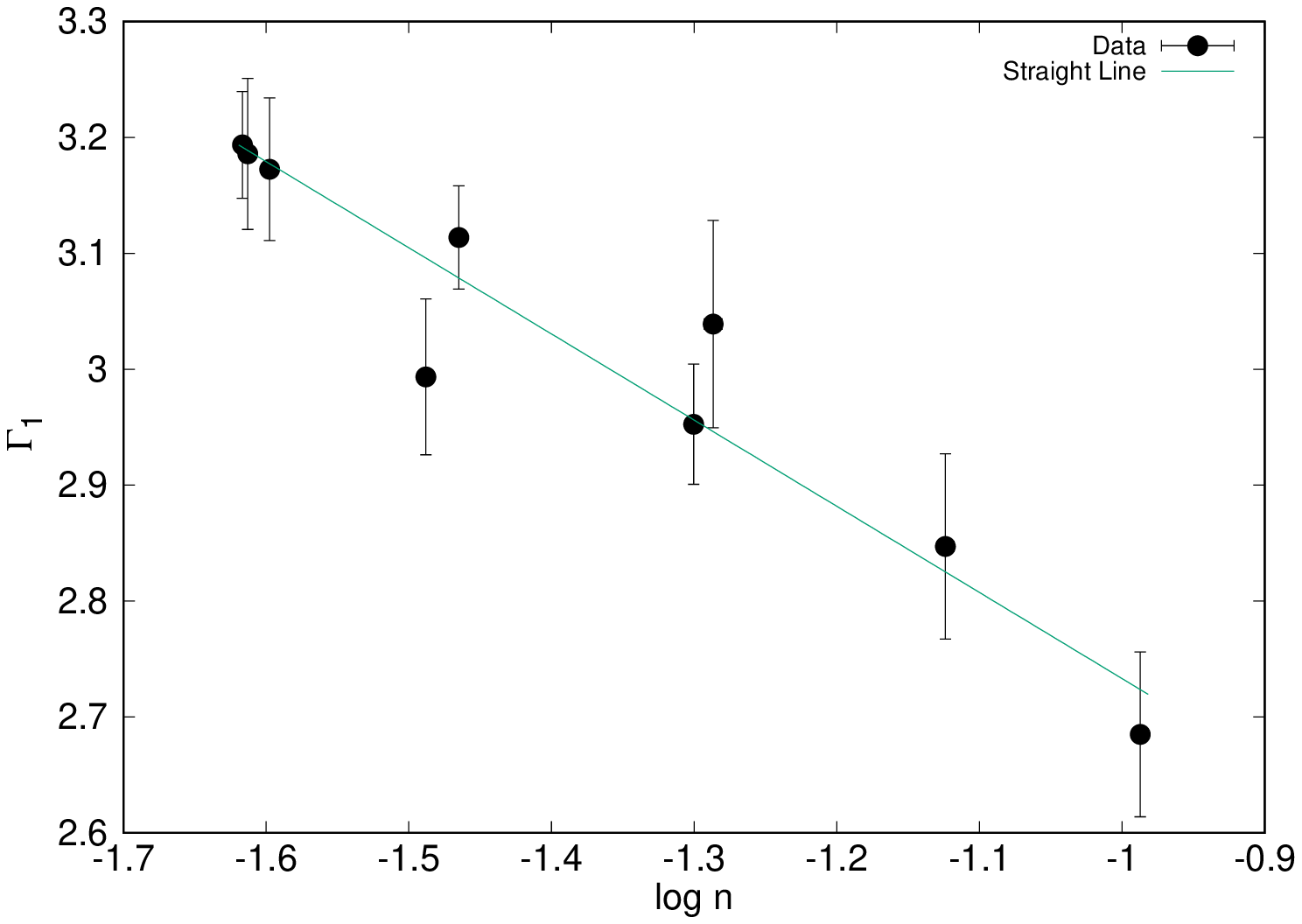} 
  \includegraphics[scale=0.3,angle=-90]{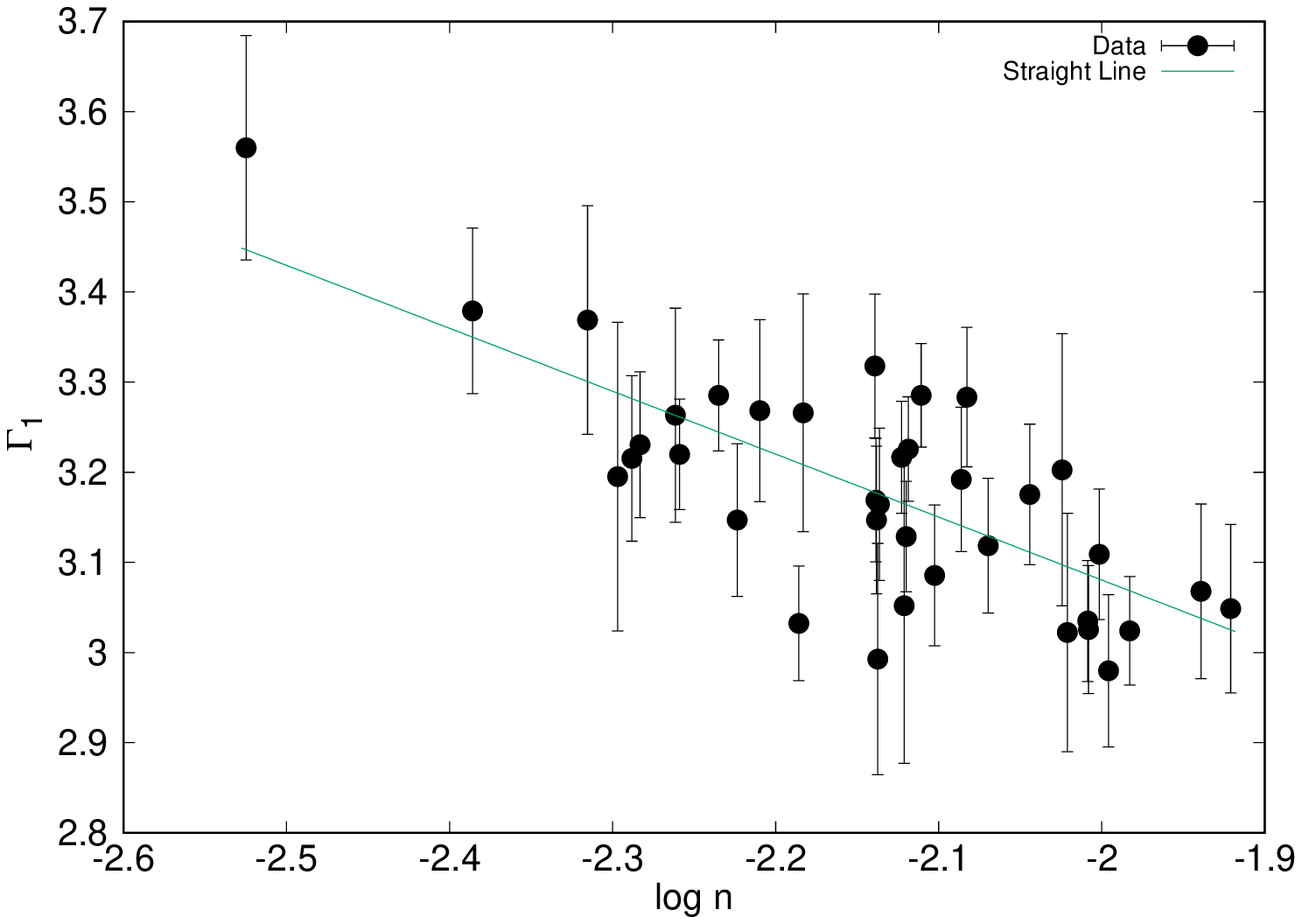} 
  \includegraphics[scale=0.3,angle=-90]{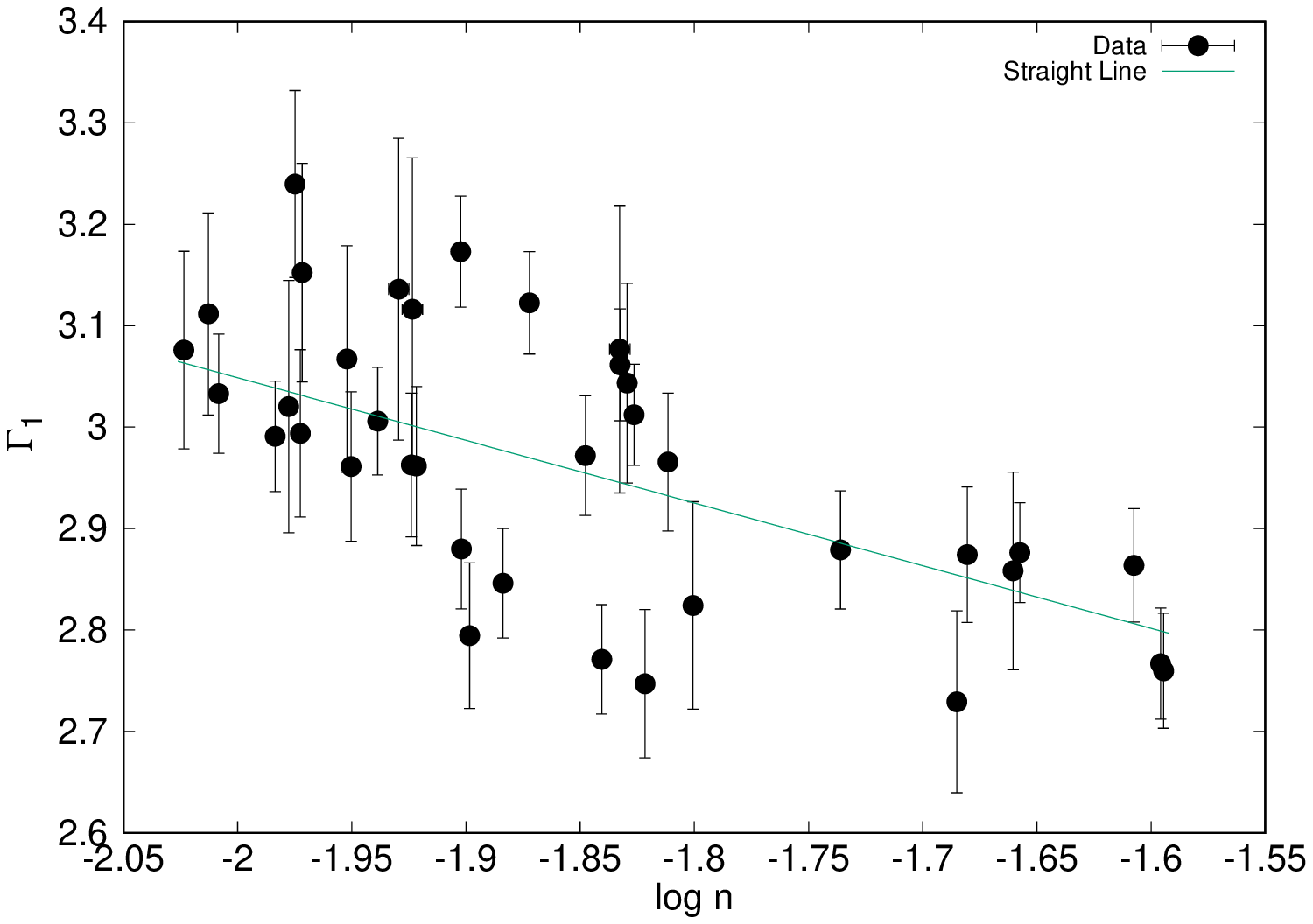} 
   \includegraphics[scale=0.3,angle=-90]{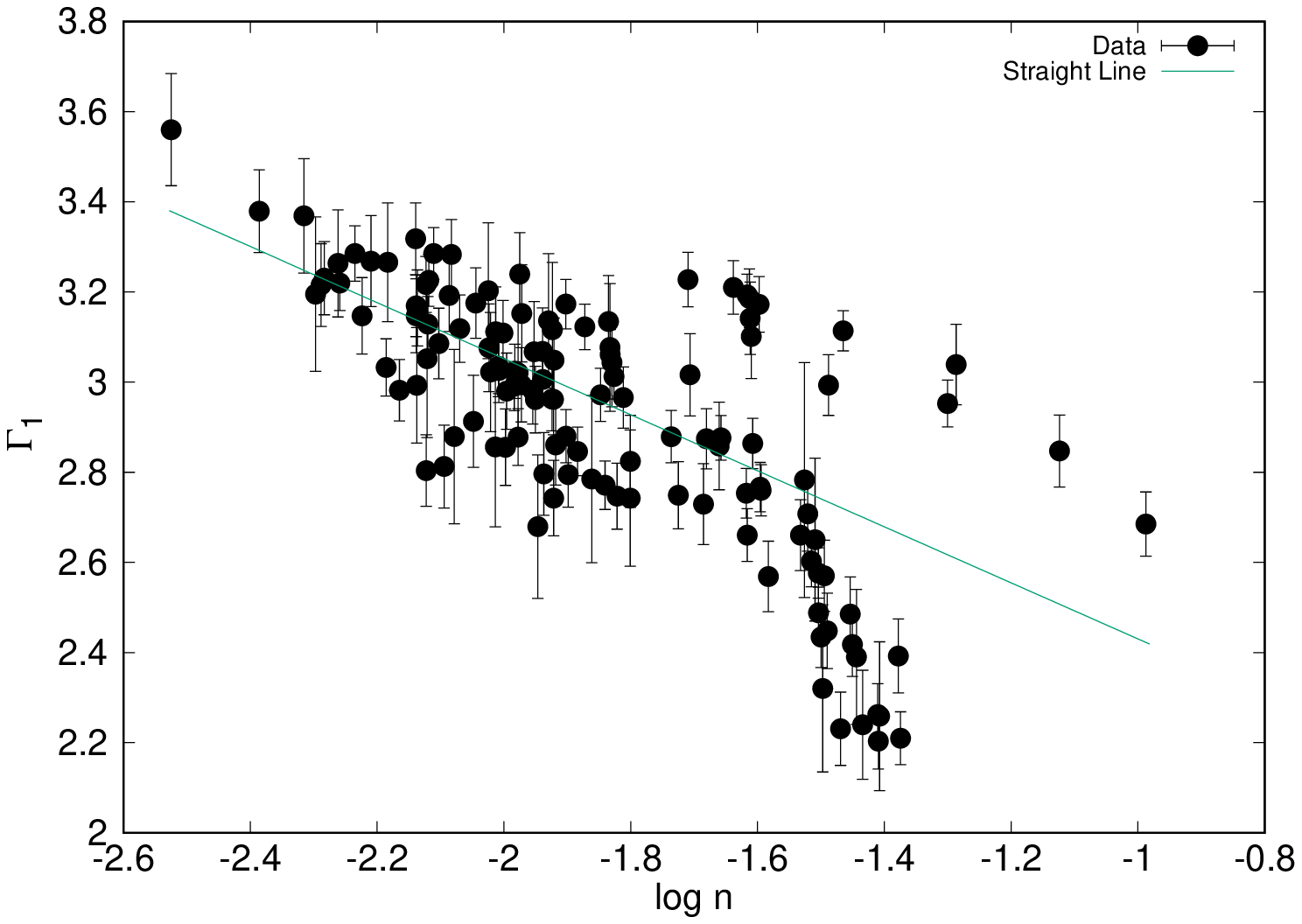}
  \includegraphics[scale=0.3,angle=-90]{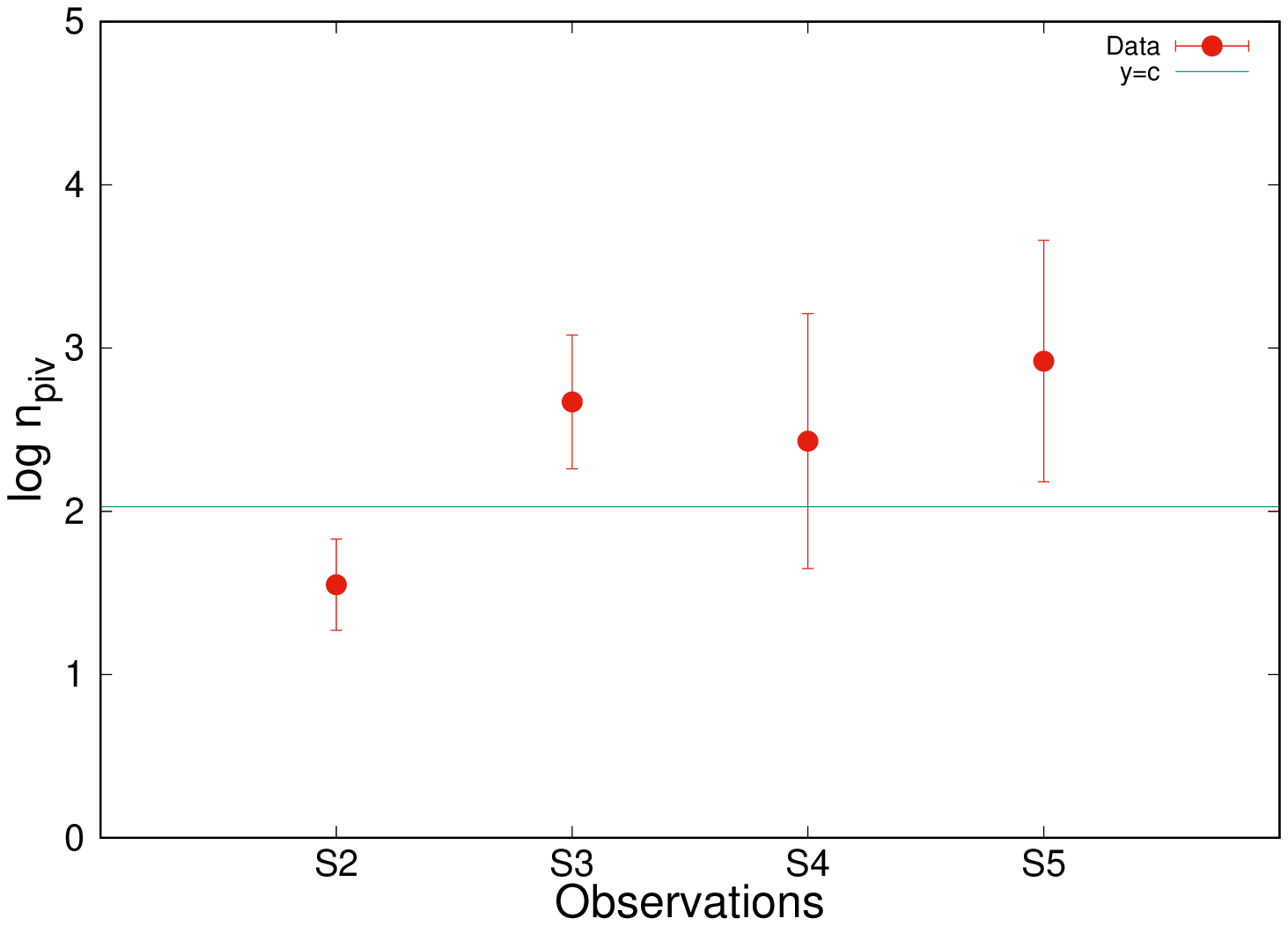}

  \caption{ Liner fit to  the variations between $\Delta\Gamma_{1}$ and $\log n$ at $\xi=\xi_{ref}$ in the BPL model.  The upper left and right panels corresponds to S2 and S3 observations respectively, middle left and right panels corresponds to S4 and S5 observations respectively.  Bottom left panel represents linear fit for the combined observations. The right bottom pannel represents the changes in $\log n_{piv}$ across all observations in the BPL model.}
 
\label{fig:slope2-bkn}
\end{figure*}

We noted that the pivot energy $\xi_{piv}$ remains same  $\xi_{piv} = 0.34\pm 0.02$ $\sqrt{keV}$ for the individual observations and also matches with $\xi_{piv}$ obtained for the combined observation ($\xi_{piv,comb} = 0.35\pm 0.02 \sqrt{keV}$) (see Figure \ref{fig:slope-bkn} ).
The intercept of Equation \ref{eq:index_bb} can be used to obtain $n_{piv}$. Therefore, we performed linear fits to the variations in $\Gamma_{1}$ and $log \hspace{0.05cm} n$ for individual observations  as well as for combined observation (see Figure \ref{fig:slope2-bkn}). The obtained values of  $log \hspace{0.05cm} n_{piv}$ for the individual observations are  shown in Figure \ref{fig:slope2-bkn}.

 To assess the model independence of the same pivot energy in different observations, we employed two models, the log-parabola model and the power-law distribution with maximum electron energy ($\xi$-max model). Both of these models provided a good fit to the data. Detailed descriptions of these models, along with the observations, are provided below.

\begin{figure*}
  \centering

  \includegraphics[scale=0.3,angle=-90]{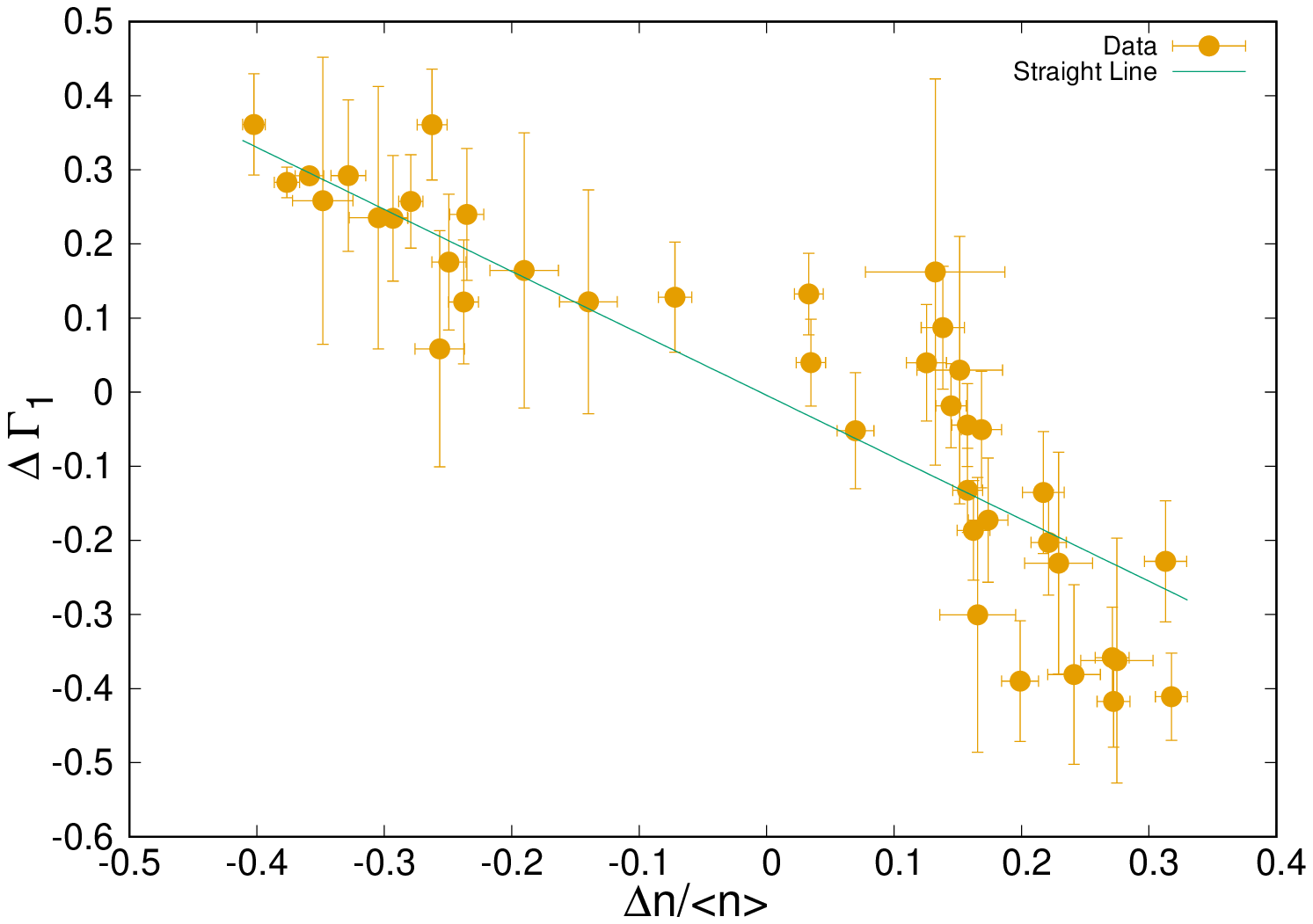}
  \includegraphics[scale=0.3,angle=-90]{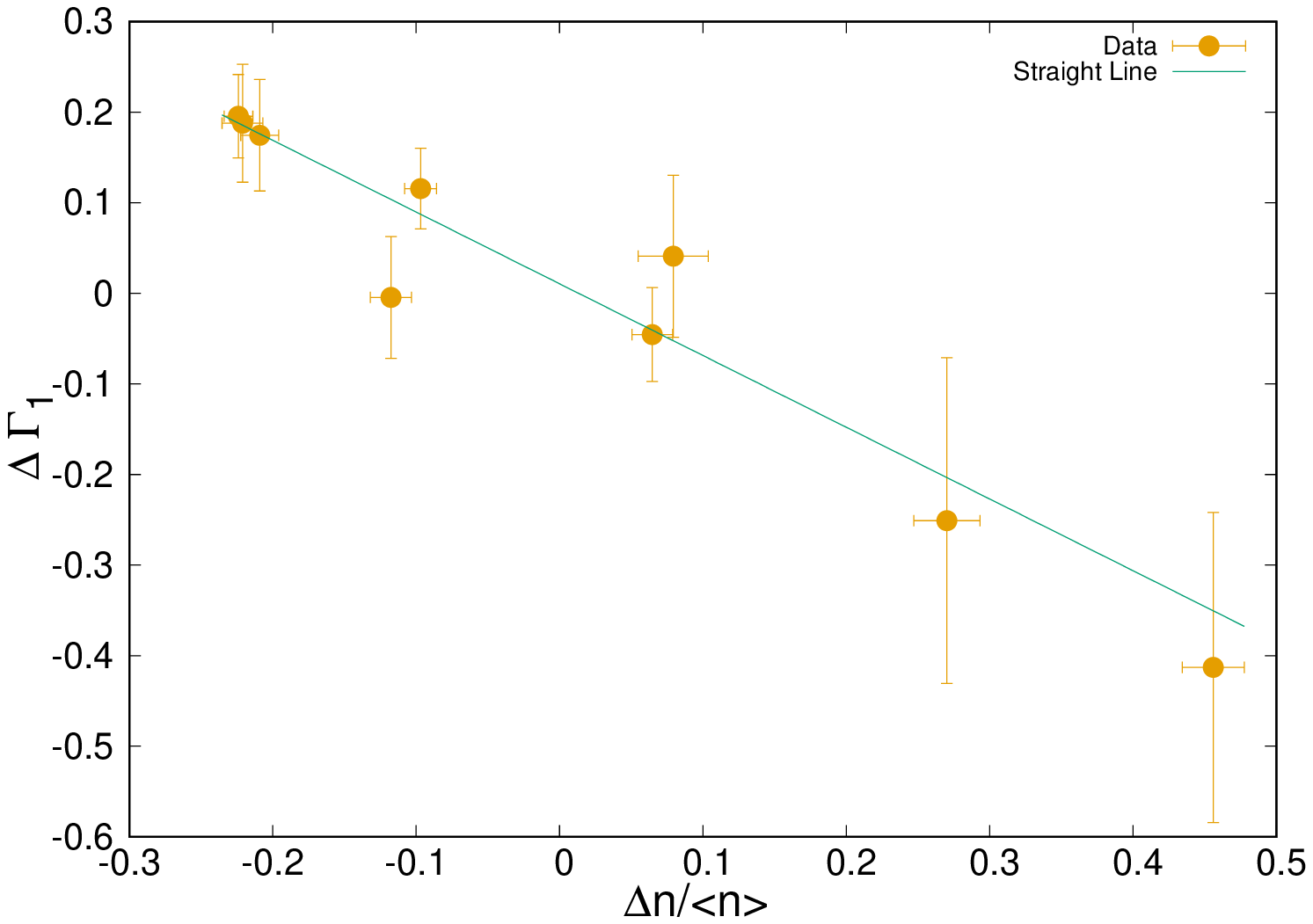}
  \includegraphics[scale=0.3,angle=-90]{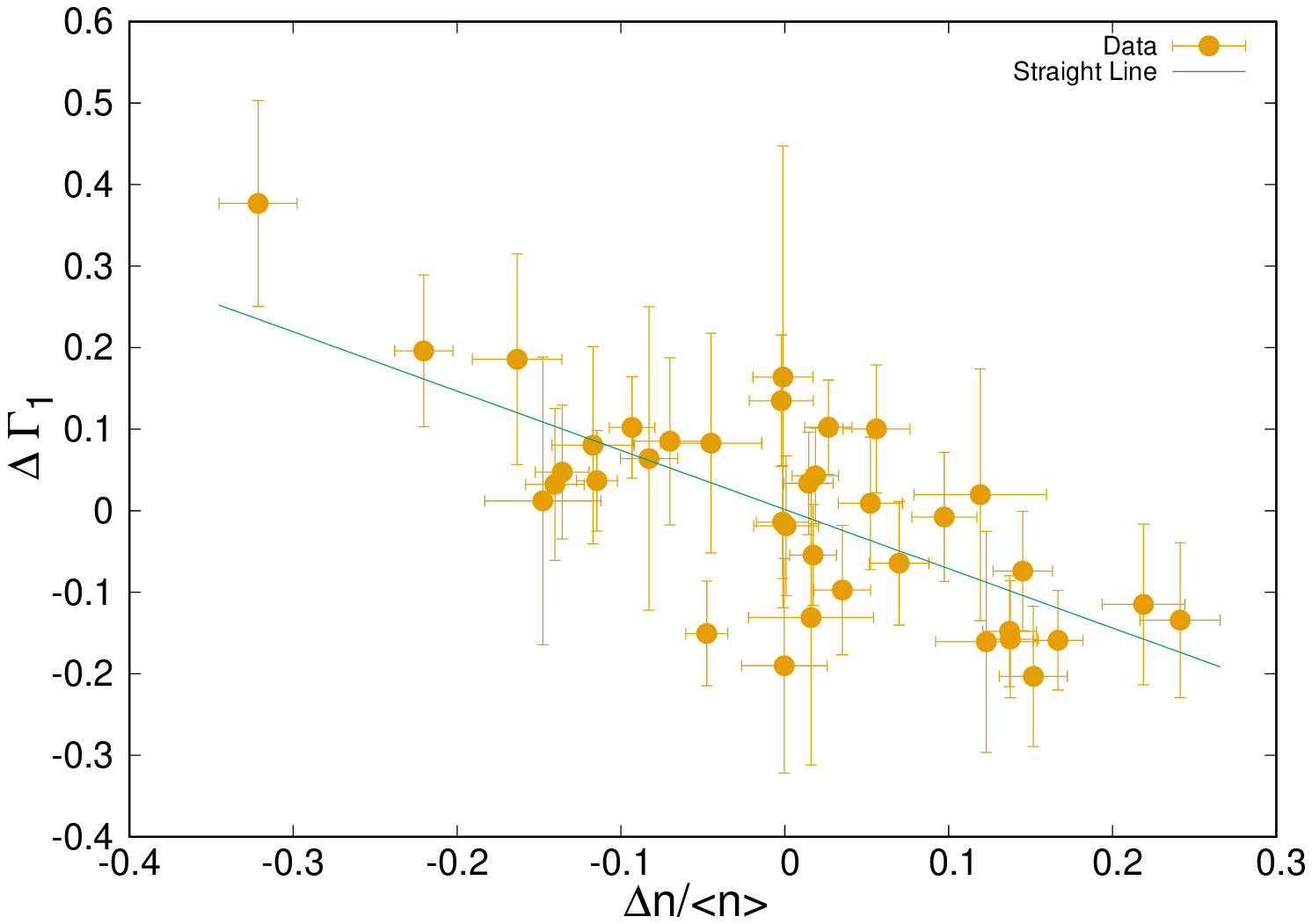}
  \includegraphics[scale=0.3,angle=-90]{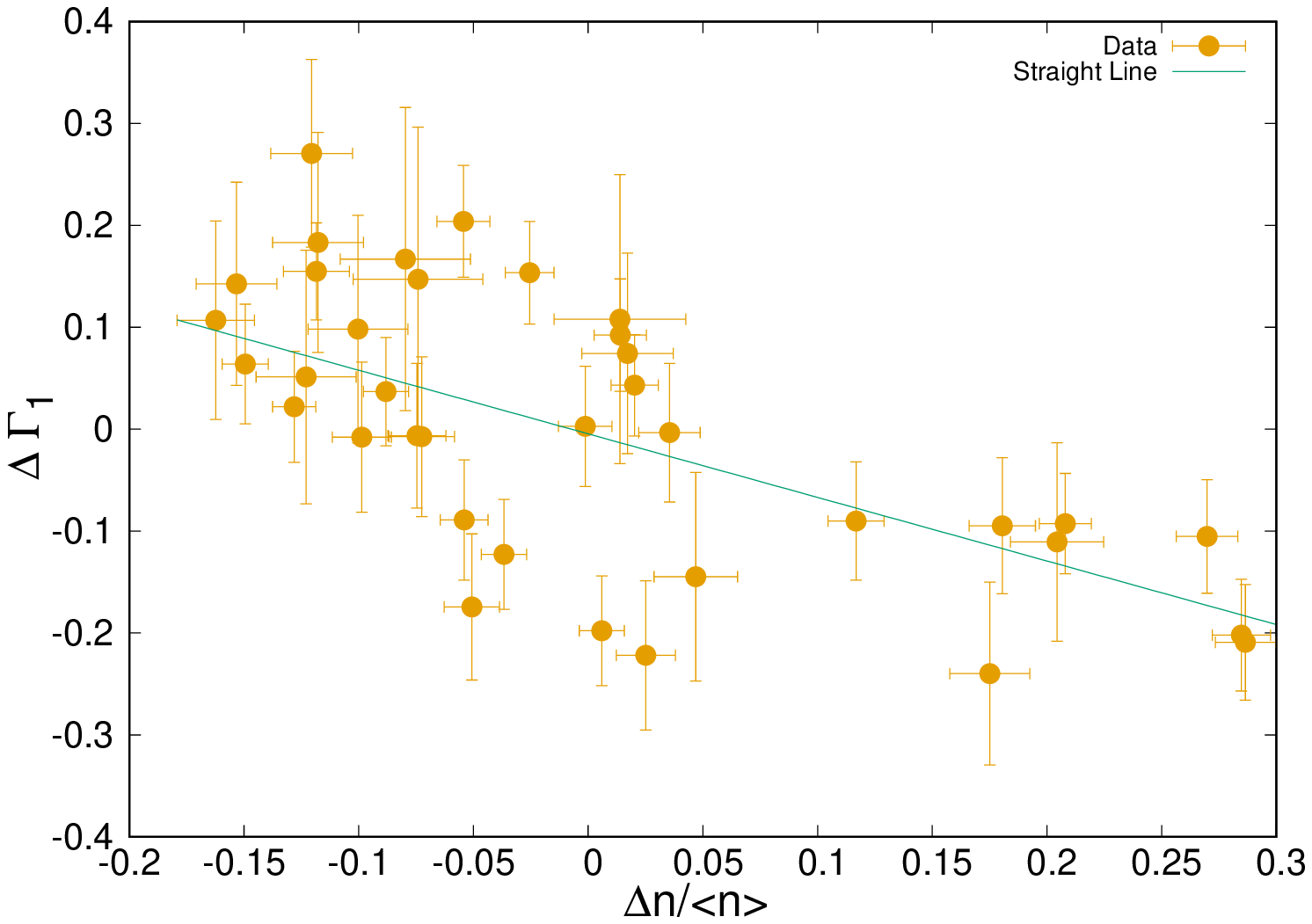}
  \includegraphics[scale=0.3,angle=-90]{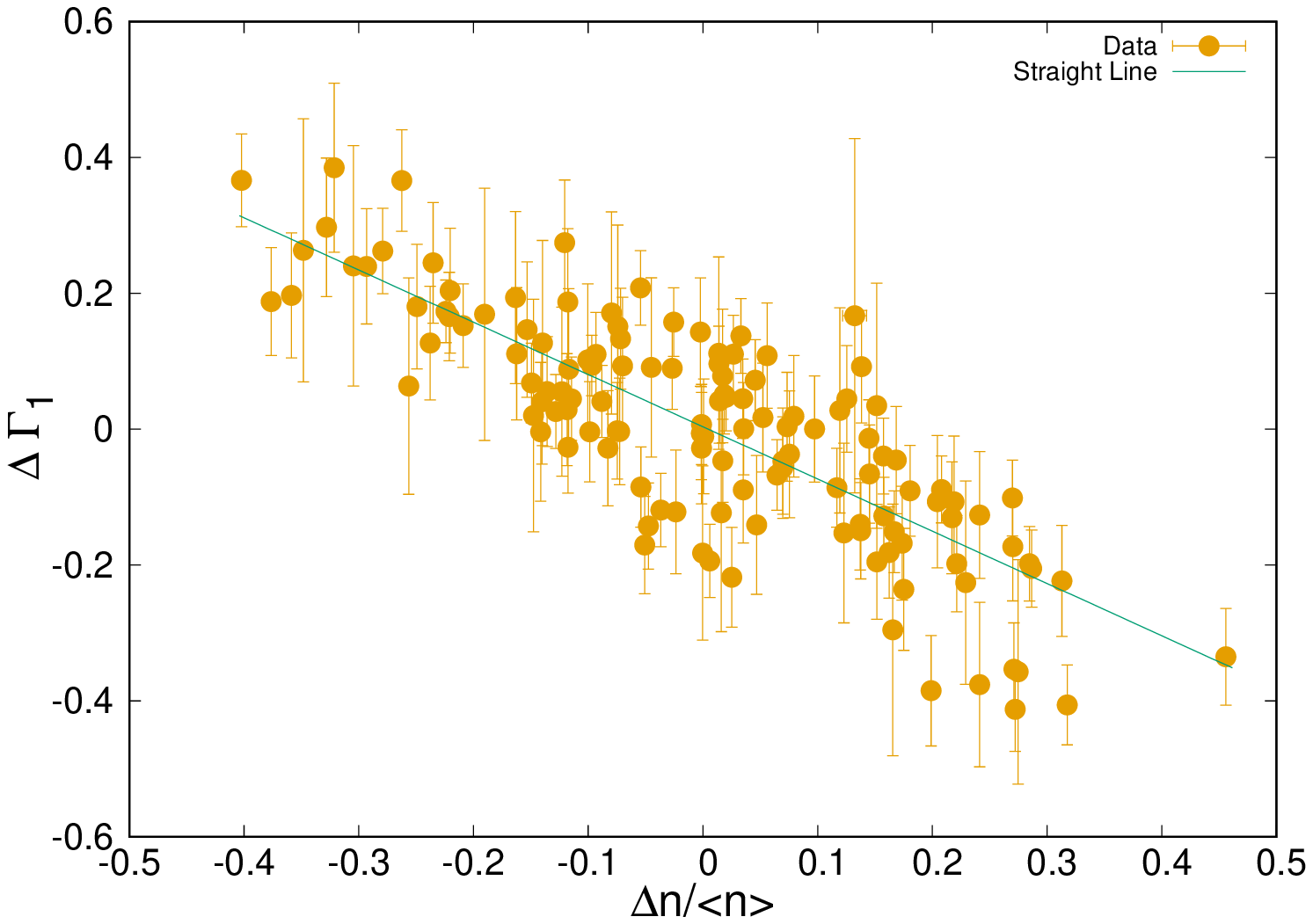}
  \includegraphics[scale=0.3,angle=-90]{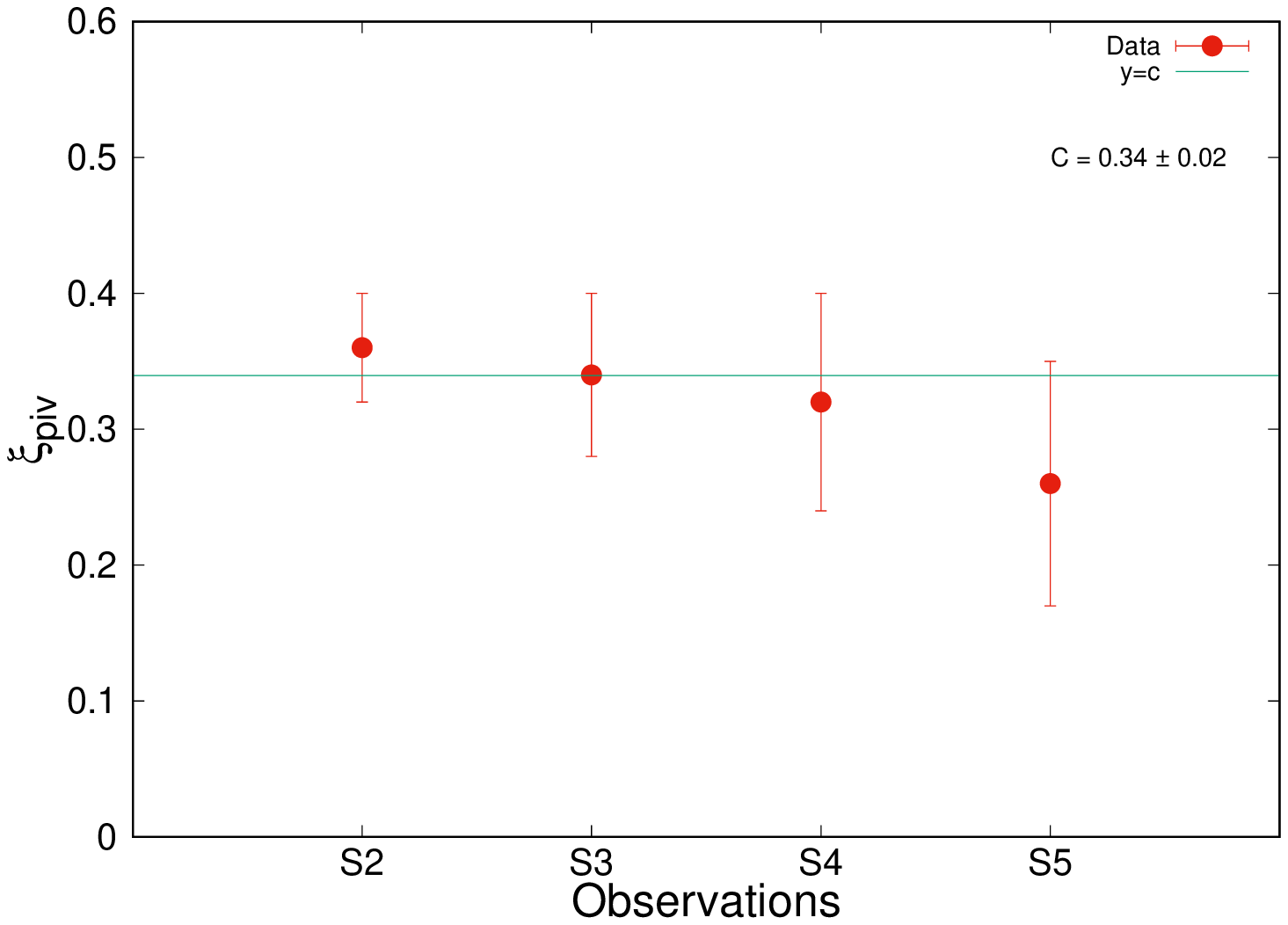}

  \caption{Linear fit  to  the variations between the spectral index ($\Delta\Gamma_{1}$) and the normalized particle density ($\Delta$$n$/$<n>$) at $\xi=\xi_{ref}$ in the  BPL model.  The upper left and right panels corresponds to S2 and S3 observations respectively, middle left and right panels corresponds to S4 and S5 observations respectively.  Bottom left panel represents linear fit for the combined observations. %In case of combined observations, $1/log(\xi_{ref}/\xi_{piv})= 0.77 \pm 0.05$, which means $\xi_{piv}=0.35 \pm 0.03$ $\sqrt{keV}$ (red-chisquare $\sim 0.7$).
  The bottom right plot represents the variation  of $\xi_{piv}$ with respect to the observations in the BPL model. The constant fit results in reduced-$\chi^2\sim1$ and $\xi_{piv}\sim0.34\pm 0.02$ $(\sqrt{keV})$.}
 
\label{fig:slope-bkn}
\end{figure*}

\begin{table}
    \centering
    \caption{Spearman correlation results  between $\Delta$$\Gamma_{1}$ and $\Delta$ $n$/$<n>$ obatined by fitting the synchrotron convolved BPL model to the joint SXT and LAXPC spectra of the selected observations .}
    \scalebox{0.8}{ % Scale the table to 80% of its original size
    
    \begin{tabular}{llll}
    \hline
        Obseravation &Observation ID & $r_{s}$&$P_{s}$  \\ \hline
        S1 & G05\_201T01\_9000000478 &-0.41 &0.24 \\
        S2 &A02\_005T01\_9000000948&-0.95& $3.09 \times 10^{-20}$\\
        S3 & T01\_218T01\_9000001852 &
-0.92 & $5.06 \times 10^{-04}$ \\
       S4 & A05\_015T01\_9000002650 &-0.67 & $4.7 \times 10^{-06}$\\
        
        S5  &A05\_204T01\_9000002856&-0.65& $4.3 \times 10^{-05}$\\
        
        \hline
        
    \end{tabular}
    }
    \label{tab:bkn-src}
\end{table}

\subsubsection{Log-parabola model}
Previous studies on Mrk\,421 has showed that the log-parabola model provides a satisfactory statistical fit to the X-ray spectrum \citep{10.1093/mnras/stac1964, 2021MNRAS.508.5921H} . Therefore, we fitted the combined  X-ray spectra from SXT and LAXPC20 instruments  using \textit{constant$\times$TBabs$\times$Synconv$\otimes$n$(\xi)$} where n($\xi$) is given by a log-parabola form i,e.
\begin{equation}\label{eq:lp_model}
    n(\xi)=K\left(\frac{\xi}{\xi_{ref}}\right)^{-\alpha-\beta  \hspace{0.05cm}log\left(\frac{\xi}{\xi_{ref}}\right)}~~,
\end{equation} 
here, $\alpha$ is the particle spectral index at the reference energy, $\xi_{ref}$,
$\beta$ is the curvature parameter, and $K$ is the normalization of the particle
density at $\xi = \xi_{ref}$.
While performing the spectral fit 
in each time bin, we have fixed the $\xi_{ref}$ at 1 keV for all the observations and  parameters $\alpha$, $\beta$, and the normalization were kept free.
%, where $\mathbb{N}$ is given by 
%\begin{equation}
%N=\frac{\delta^{3}{1+z}}{d_L^2} V\mathbb{A}K
%\end{equation}
The model provides a satisfactory fit to the X-ray spectrum, and the resulting spectral parameters, along with their respective reduced-$\chi^{2}$ values, are presented in Table \ref{tab:t1},\ref{tab:t2},\ref{tab:t3},\ref{tab:t4} and \ref{tab:t5} in the appendix section. 

Table \ref{tab:corr-bp-max} present Spearman correlation results for the log-parabola model fit parameters  i,e. between $\Delta $$\alpha $ and $\Delta$ $\beta$ and between $\Delta $$\alpha $ and $\Delta n/<n>$ . Similar to the broken power-law model, all observations except S1, exhibit a negative correlation between $\Delta $$\alpha $ and the $\Delta n/<n>$, indicating a  harder spectra when the source is brighter. Additionally, for all observations, a negative correlation is evident between $\Delta $$\alpha $ and the $\Delta$ $\beta$. No significant correlation was observed in S1 observation.
Also, we constructed combined plots of all the observations between $\Delta $$\alpha $ and the $\Delta n/<n>$ ; and between  $\Delta $$\alpha $ and the $\Delta$ $\beta$ (see Figure \ref{fig:com-lp} in appendix section). These results show the similar correlation trend and thus reproduce the outcomes observed in individual observations.

Moreover, in order to obtain the pivot energy and particle density at pivot energy in LP model, Equation \ref{eq:lp_model} 
can be written as 

\begin{equation}
   n_{ref}\left(\frac{\xi}{\xi_{ref}}\right)^{-\alpha-\beta\hspace{0.05cm}log\left(\frac{\xi}{\xi_{ref}}\right)}=n_{piv}\left(\frac{\xi}{\xi_{piv}}\right)^{-\alpha^{'}-\beta \hspace{0.05cm}log\left(\frac{\xi}{\xi_{piv}}\right)} ~~,
\end{equation} \label{eq:lp_eq_pivot}
here,  $n_{ref}$ and  $n_{piv}$ represents the particle densities at reference energy, ${\xi}_{ref}$ and pivot energy, $\xi_{piv}$, respectively. 
Following the definition of power law model, the index ${I}$, of LP model can be obtained as:

\begin{equation}
    I=\alpha+2\beta\hspace{0.05cm} log\left(\frac{\xi}{\xi_{ref}}\right)
\end{equation}
Matching the indices in equation(10), we get
\begin{equation}
    \alpha^{'}=\alpha+2\beta\hspace{0.05cm} log\left(\frac{\xi_{piv}}{\xi_{ref}}\right)
\end{equation}
Using $\alpha^{'}$ from equation(12) in equation(10), 
\begin{equation}
\begin{split}
    log (n_{ref})-\left[\alpha + \beta \hspace{0.05cm}log \left(\frac{\xi}{\xi_{ref}}\right)\right]log \hspace{0.05cm}\left(\frac{\xi}{\xi_{ref}}\right)=log (n_{piv})\\
    -\left[\alpha + 2\beta \hspace{0.05cm} log \left(\frac{\xi_{piv}}{\xi_{ref}}\right)+\beta \hspace{0.05cm} log \left(\frac{\xi}{\xi_{ref}}\right)\right]log \left(\frac{\xi}{\xi_{piv}}\right)
\end{split}    
\end{equation}
\begin{equation}
 log (n_{ref})-\alpha \hspace{0.05cm} log\left(\frac{\xi_{piv}}{\xi_{ref}}\right)=log (n_{piv})+\beta\left[log\left(\frac{\xi_{piv}}{\xi_{ref}}\right)\right]^{2}
\end{equation}

The variation in particle density, $\alpha$ and $\beta$ would be related through the equation
\begin{equation}\label{eq:lp_final}
    \frac{\Delta n}{n}-\Delta\alpha \hspace{0.05cm} log\left(\frac{{\xi}_{piv}}{{\xi}_{ref}}\right)= \Delta\beta\left[log\left(\frac{\xi_{piv}}{\xi_{ref}}\right)\right]^{2}    
\end{equation}

\begin{figure*}
  \centering

  \includegraphics[scale=0.38]{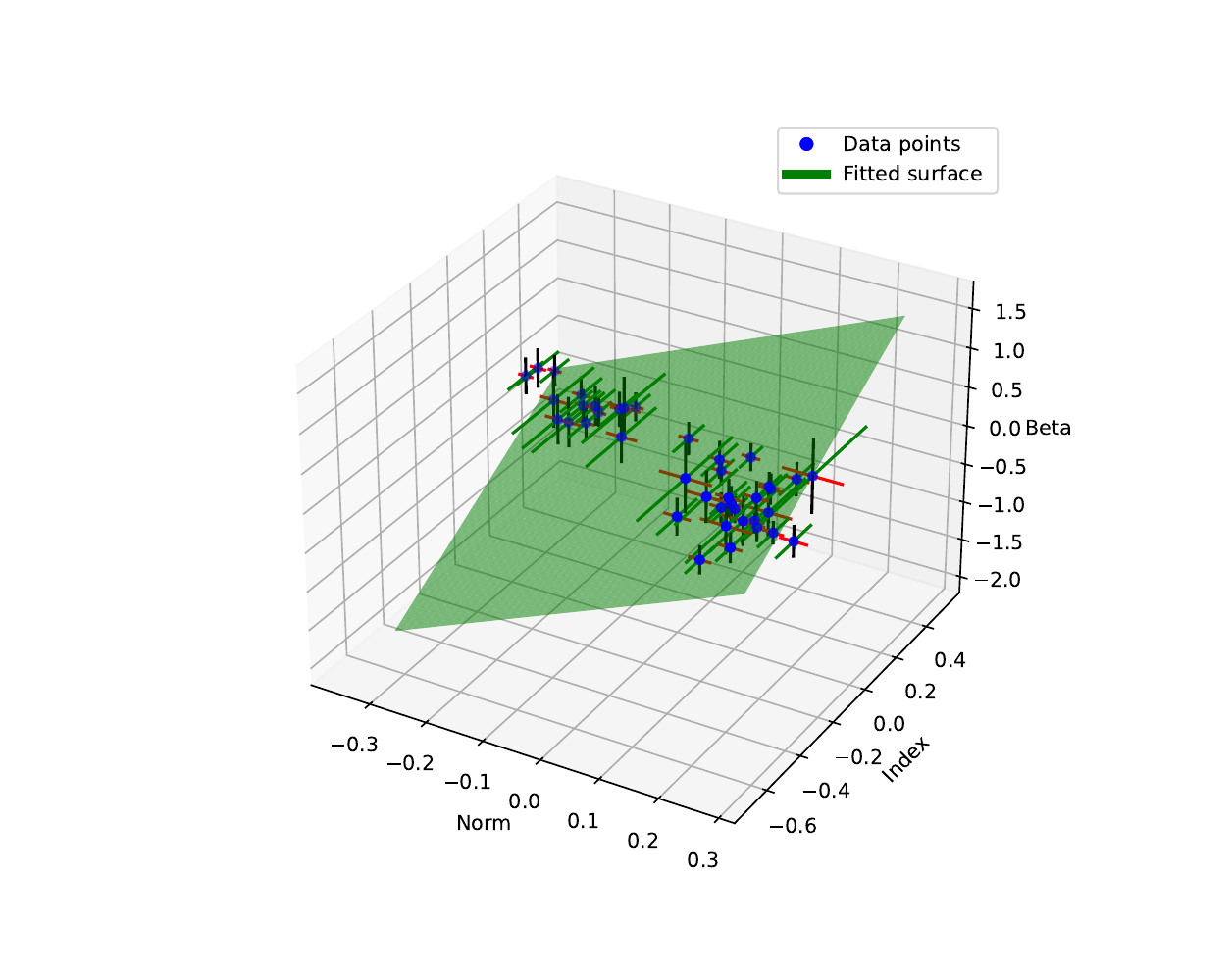}
  \includegraphics[scale=0.38]{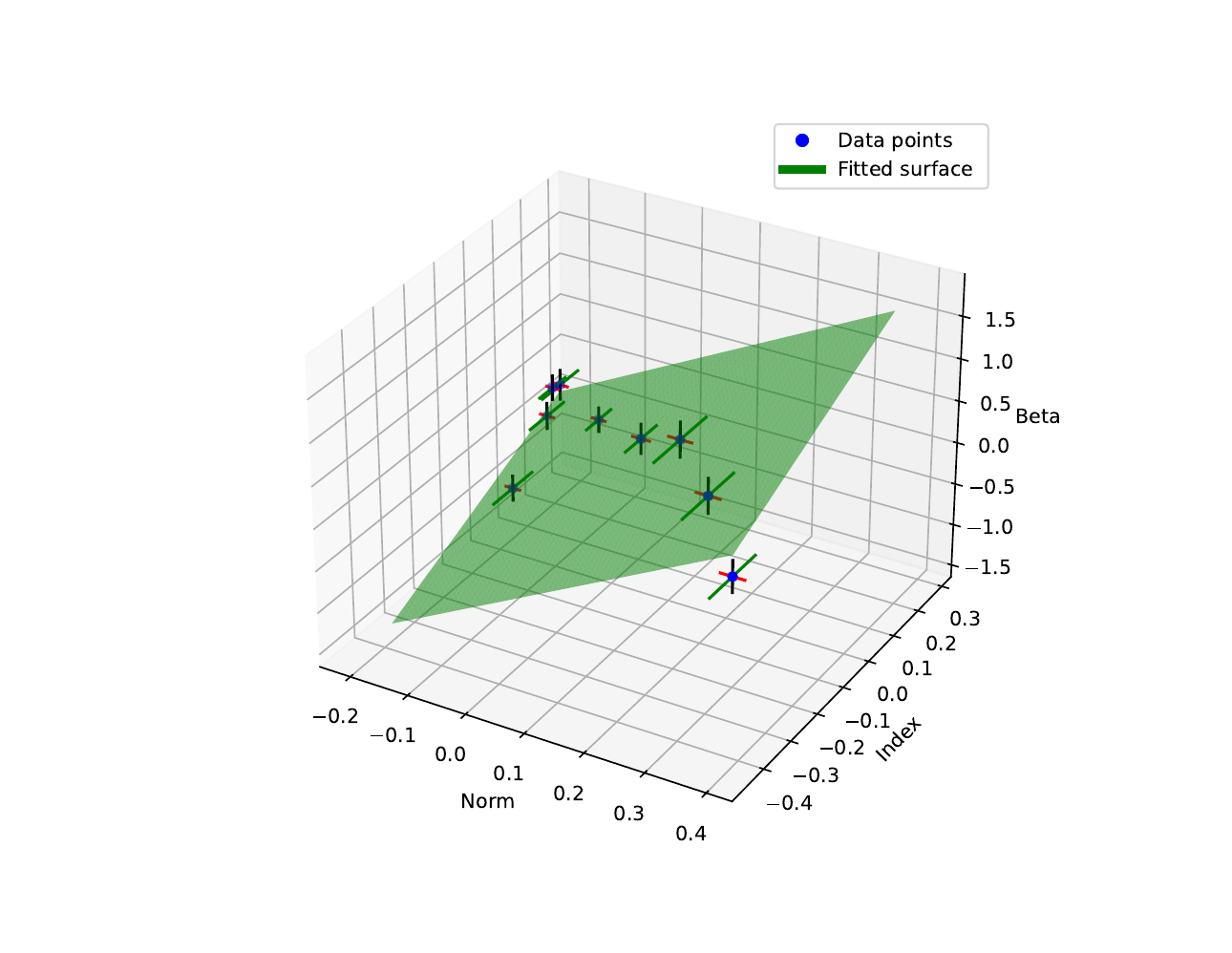}
  \includegraphics[scale=0.38]{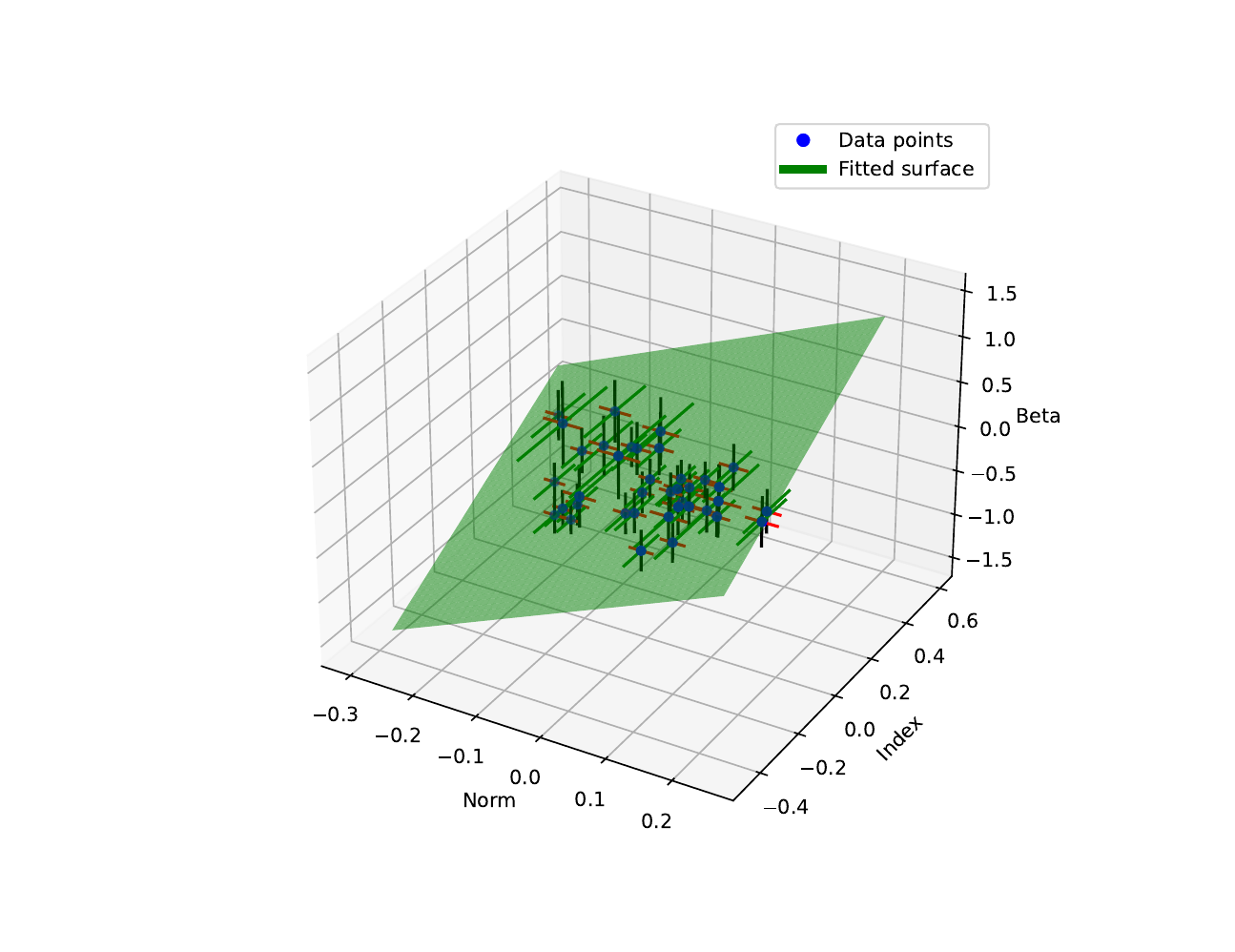}
  \includegraphics[scale=0.38]{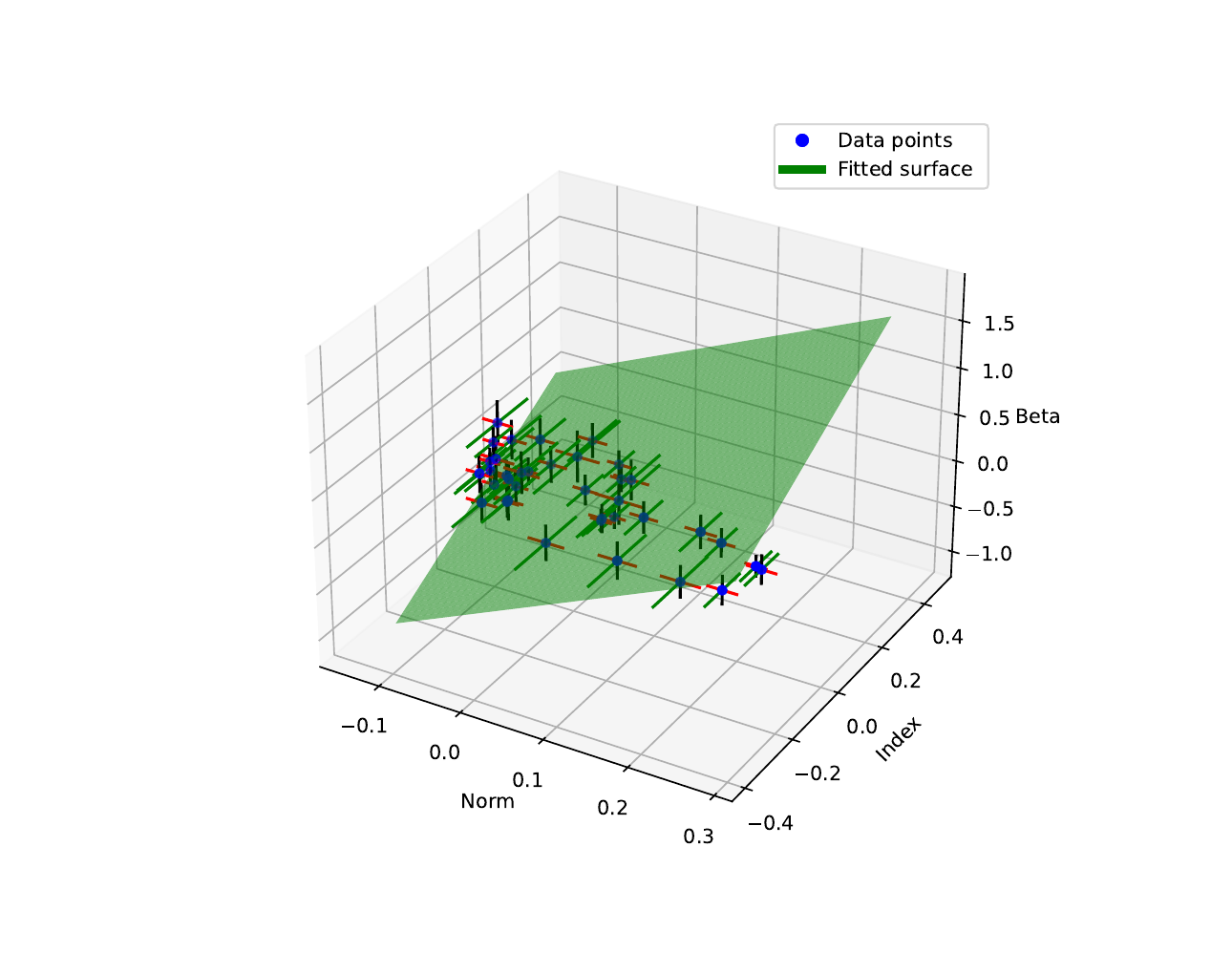}
  \includegraphics[scale=0.38]{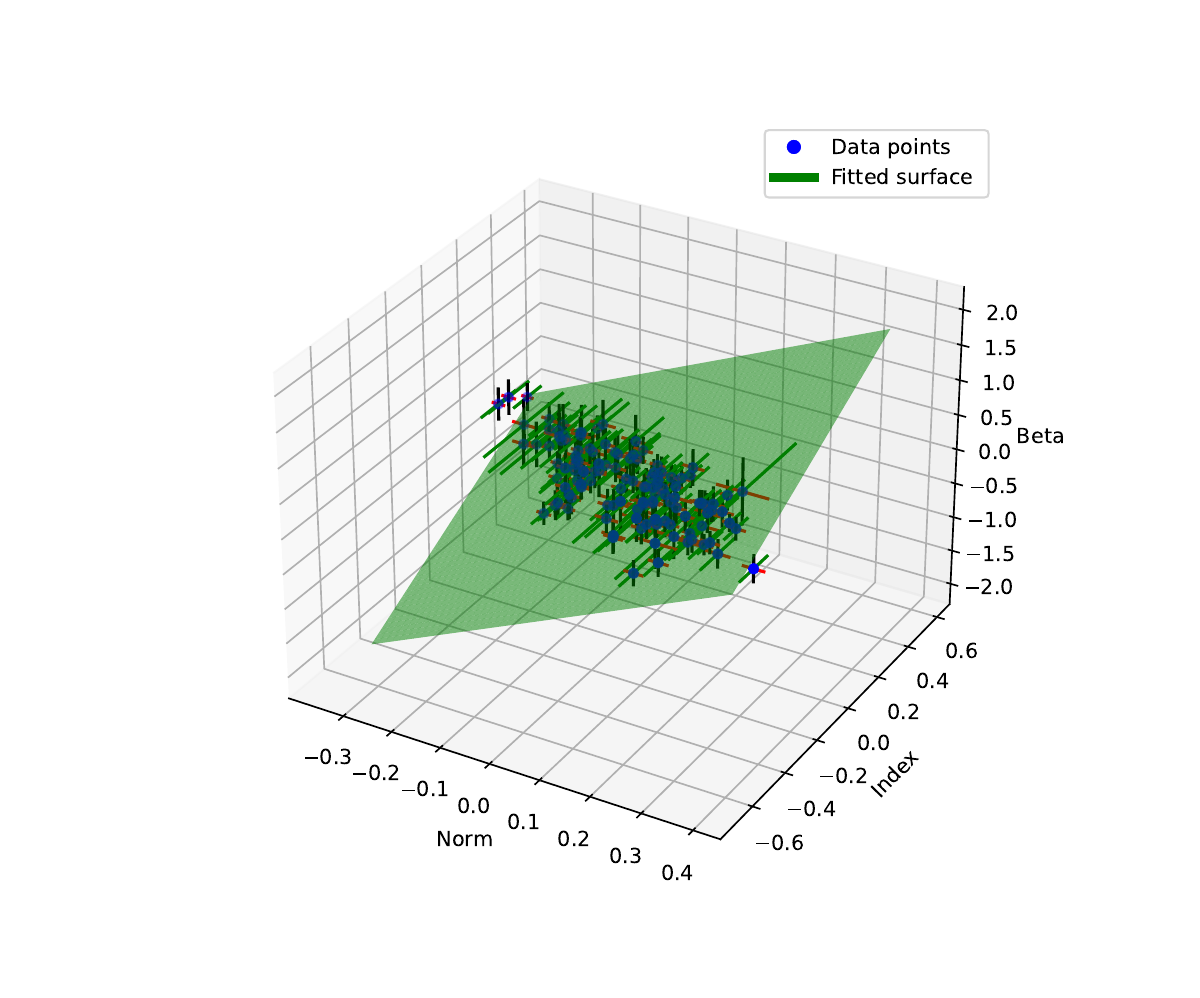}
  \includegraphics[scale=0.3]{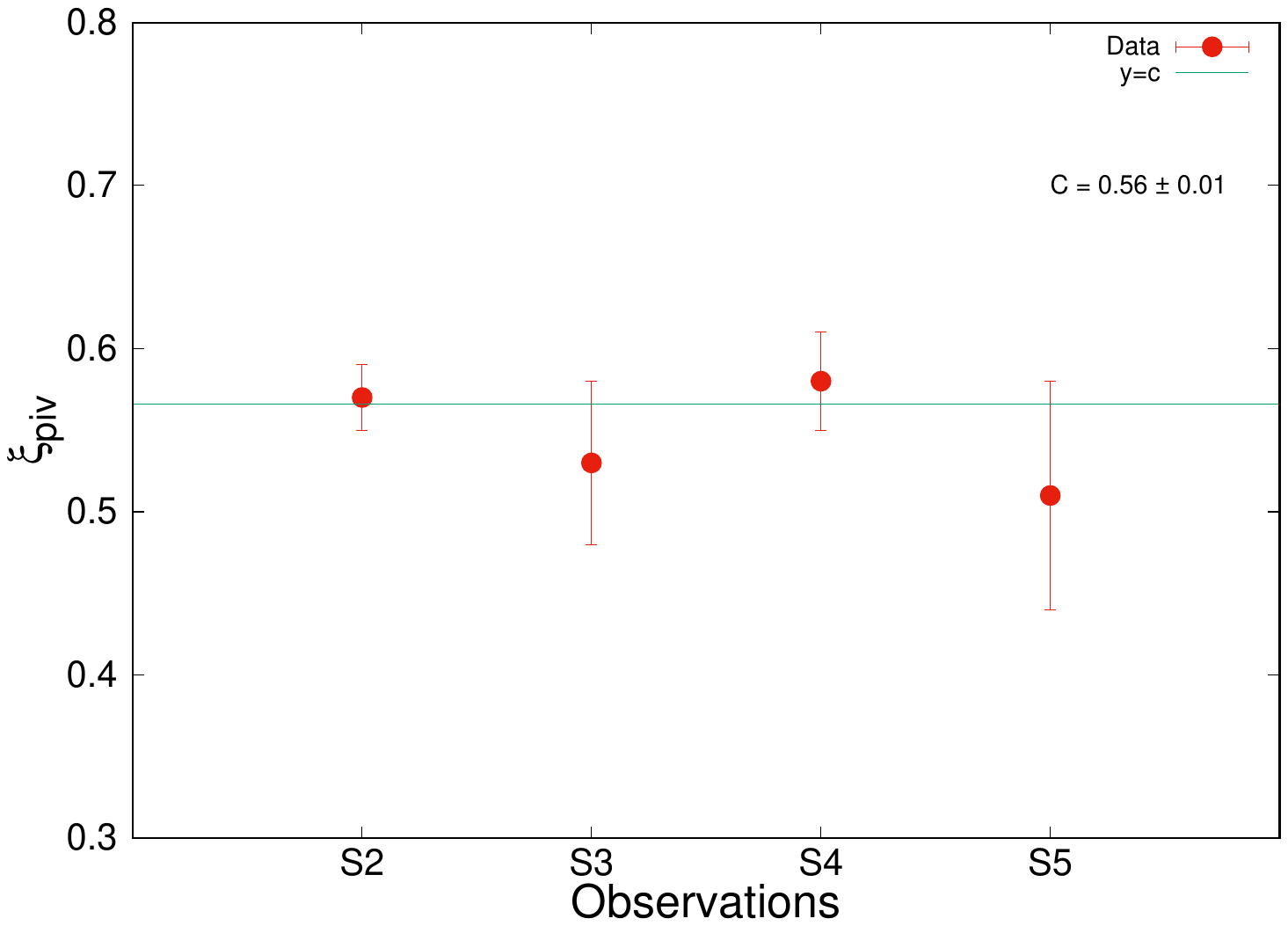}

  \caption{Equation \ref{eq:lp_f}  to  the variations between the spectral index ($\Delta\Gamma_{1}$), ($\Delta\beta$) and  the normalized particle density ($\Delta$$n$/$<n>$) at $\xi=\xi_{ref}$ in the  LP model.  The upper left and right panels corresponds to S2 and S3 observations respectively, middle left and right panels corresponds to S4 and S5 observations respectively.  Bottom left panel represents  fit for the combined observations. The bottom right plot represents the variation  of $\xi_{piv}$ with respect to the observations in the LP model. The constant fit results in reduced-$\chi^2\sim1$ and $\xi_{piv}\sim0.56\pm 0.01$ $(\sqrt{keV})$.}
 
\label{fig:slope-log}
\end{figure*}

This equation can be reformulated as:
\begin{equation}\label{eq:lp_f}
f(x, y) = \frac{1}{t^2}x - \frac{1}{t}y,
\end{equation}

where, x is $\frac{\Delta n}{n}$, y  is $\Delta\alpha$, f(x,y) is $\Delta\beta$ and $t=\log\left(\frac{\xi_{piv}}{\xi_{ref}}\right)$.
Cleary the equation \ref{eq:lp_f} is quadratic in t and hence one expects two values of t. We fitted the above equation to the  $\frac{\Delta n}{n}$, $\Delta\alpha$, and  $\Delta\beta$ values. The fit resulted in two values (positive and negative values) of t. However, we noted that the positive values of t leads to high pivot energy across all observations, contradicting with the observations where pivot energy is often chosen close to the low energy of the observed data. In Figure \ref{fig:slope-log}, we have shown the scatter plot for the values of $\frac{\Delta n}{n}$, $\Delta\alpha$, and  $\Delta\beta$ along with the fit for the negative roots of t.
Upon considering the negative roots, the $\xi_{piv}$ values resulting from the LP model for observations S2, S3, S4, and S5 are $0.57 \pm 0.02$ $\sqrt{keV}$, $0.53 \pm 0.05$ $\sqrt{keV}$, $0.58 \pm 0.03$ $\sqrt{keV}$, and $0.51 \pm 0.07$ $\sqrt{keV}$, respectively. Consequently, the $\xi_{piv}$ value remains consistent across individual observations, the constant fit resulted in value of  $0.56\pm 0.01$ $\sqrt{keV}$ (see Figure \ref{fig:slope-log}). This value aligns with the $\xi_{piv}$ obtained from combined observations, ($\xi_{piv,comb} = 0.56 \pm 0.02$ $\sqrt{keV}$).

\subsubsection{Power-law
distributions with maximum electron energy ($\xi-{max}$)}

We also conducted the braodband X-ray  spectral analysis of Mkn\,421 by considering  the power-law distribution with maximum electron energy. 
Such particle distribution results in regions in which shock-induced particle acceleration leads to radiative energy loss. The loss becomes dominant at higher energies as the energy loss rate is proportional to the square of its energy. The form of such particle energy distribution can be written as (after transforming $\gamma$ to $\xi=\sqrt{C}$$\gamma$)
\begin{equation}\label{eq:xi_max_model}
    n(\xi)=K\left(\frac{\xi}{\xi_{ref}}\right)^{-p}\left(1-\frac{\xi}{\xi_{max}}\right)^{(p-2)}
\end{equation}
%Here, K is determined as $K = Q_0 \tau_{acc} \gamma_0^{p-1}C^{p/2}$, and $\xi_{max}$ is calculated as $\xi_{max} = \gamma_{max} \sqrt{C}$.
where K is particle normalisation, p is the particle spectral index and $\xi_{max}$ corresponds to the maximum Lorentz factor achievable by an electron before energy dissipation occurs. Again the combined X-ray spectrum (0.7-19 keV) in each time bin  is jointly fitted with the model \textit{constant$\times$ TBabs$\times$ Synconv$\otimes$n$(\xi)$}. During the fitting,  
we set $\xi_{ref}$ to a fixed value of 1 keV  and parameters  $p$, $\xi_{max}$, and the normalization were kept free. 
%\begin{equation}
%N=\frac{\delta^{3}{1+z}}{d_L^2} V\mathbb{A}K
%\end{equation}
The model provides a satisfactory fit to the X-ray spectrum, and the resulting spectral parameters, along with their respective reduced-$\chi^{2}$ values, are presented in Table \ref{tab:t1},\ref{tab:t2},\ref{tab:t3},\ref{tab:t4} and \ref{tab:t5} in the appendix section.

\begin{figure*}
  \centering

  \includegraphics[scale=0.30,angle=-90]{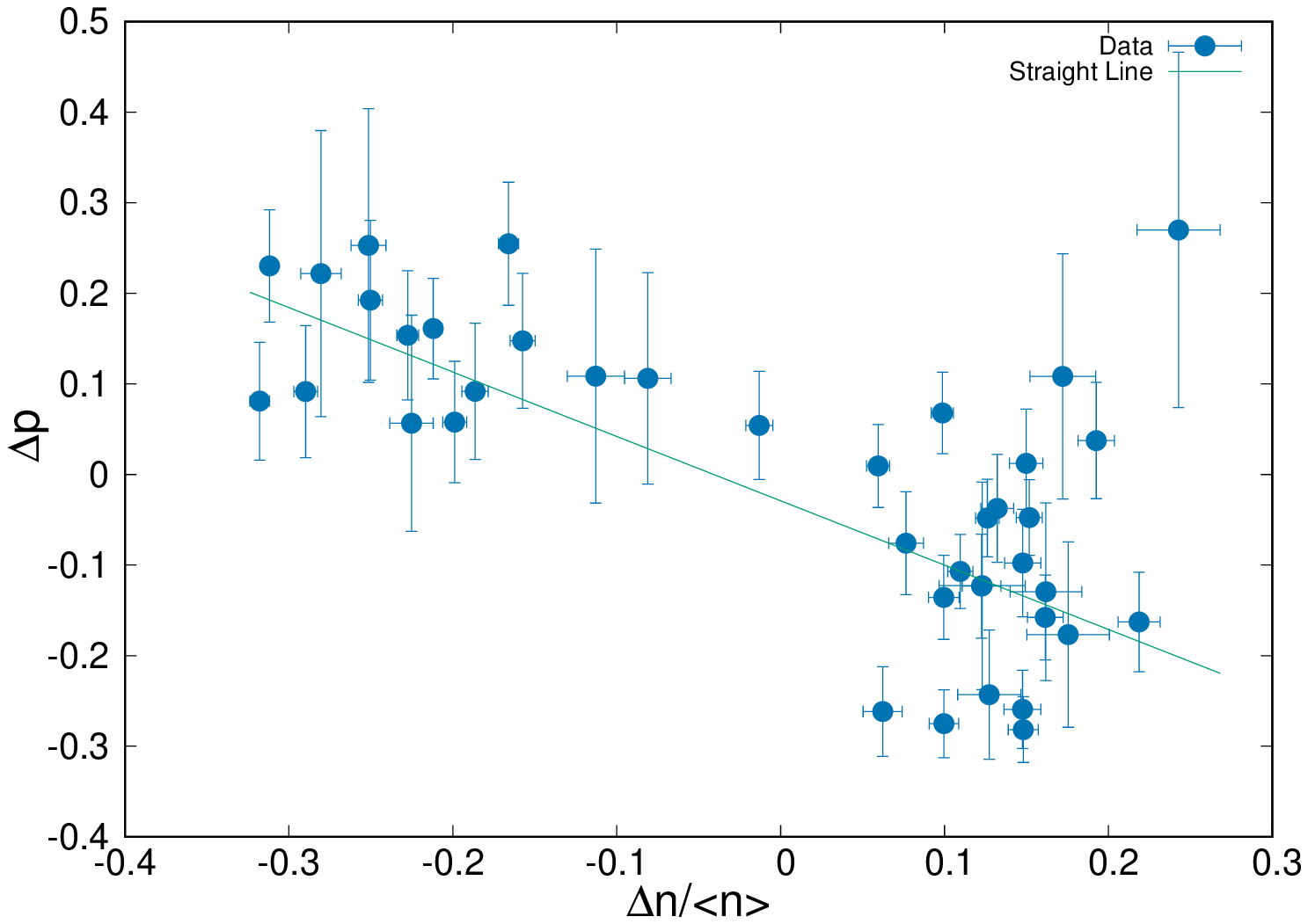}
  \includegraphics[scale=0.30,angle=-90]{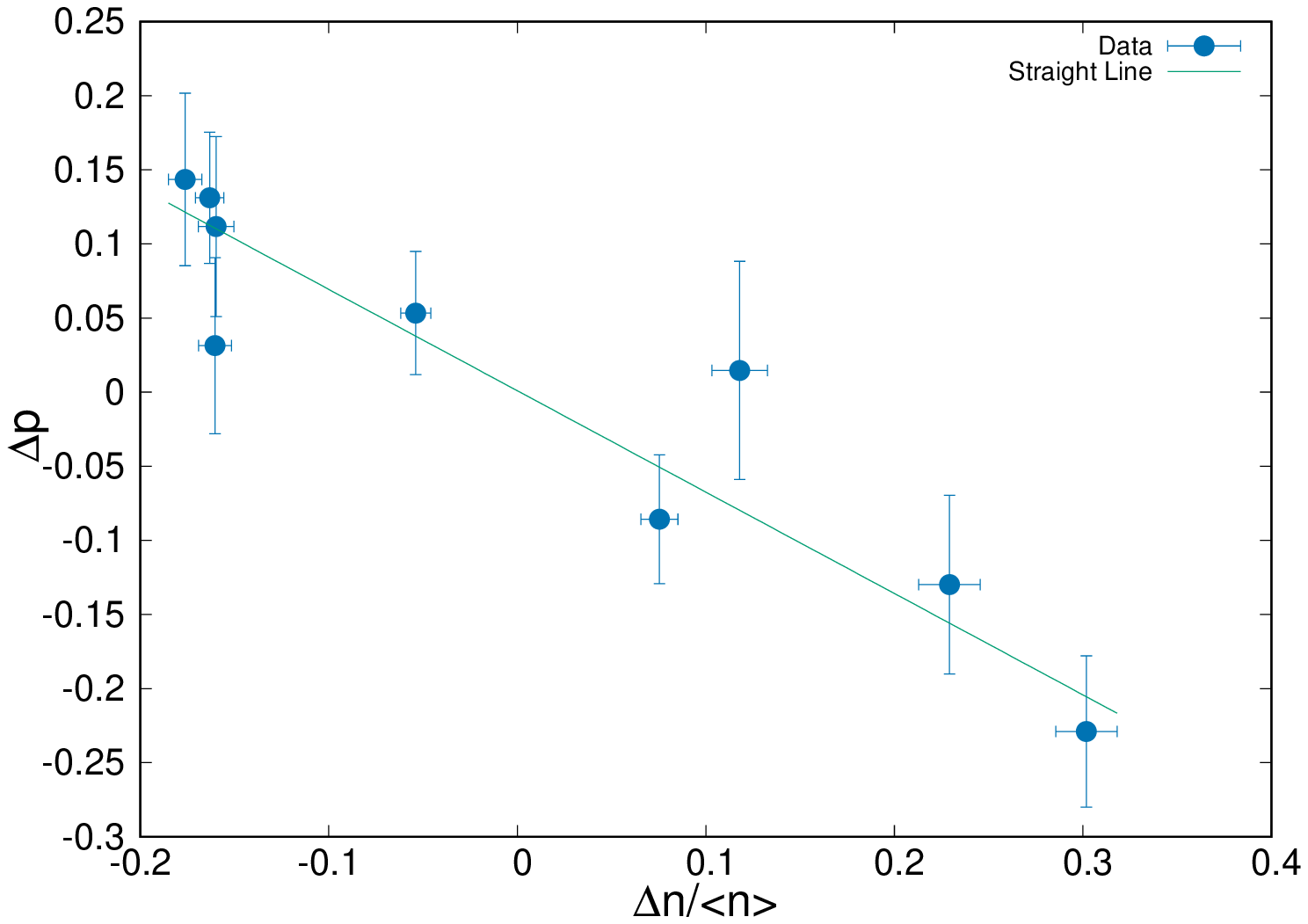} 
  \includegraphics[scale=0.30,angle=-90]{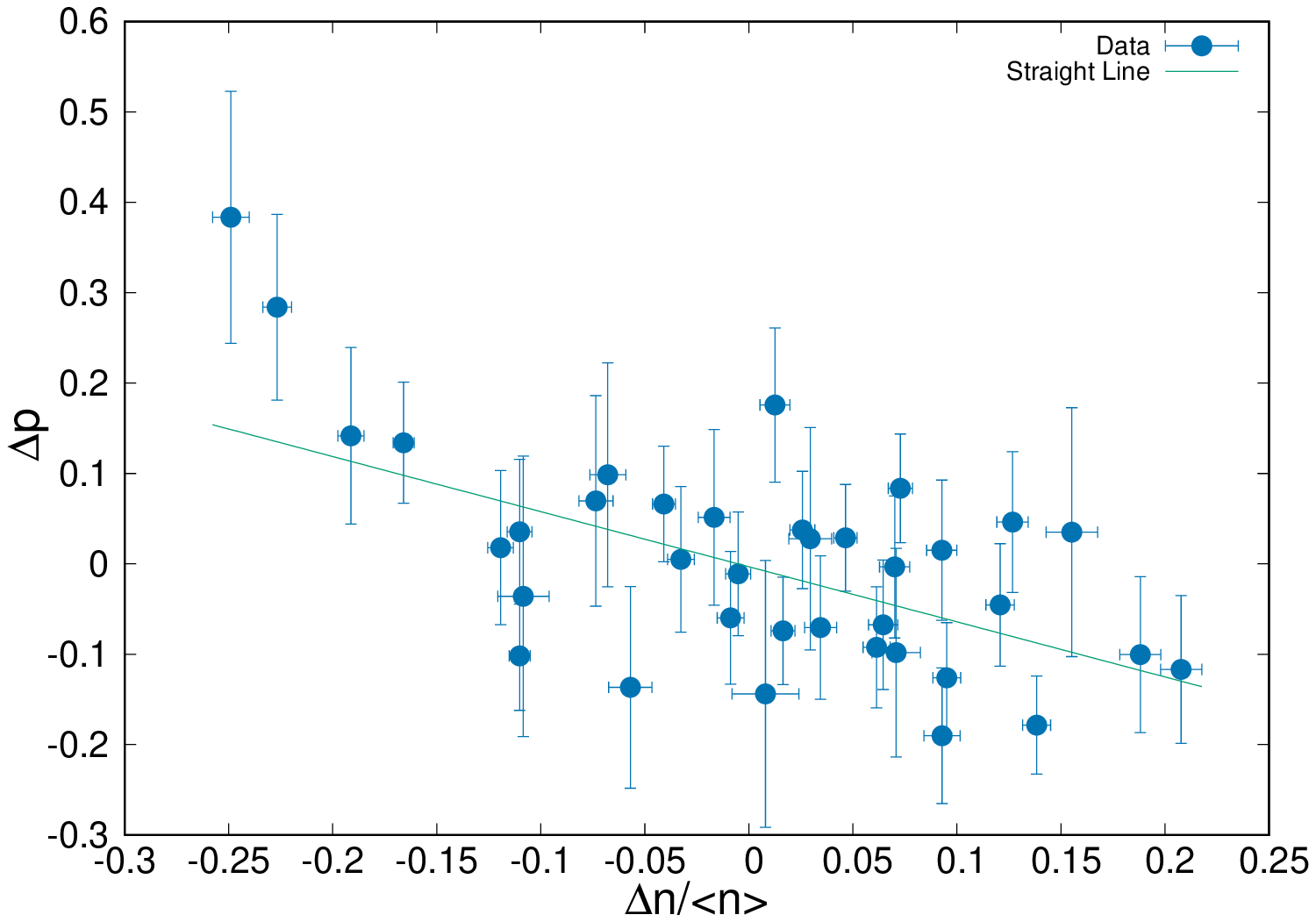} 
  \includegraphics[scale=0.30,angle=-90]{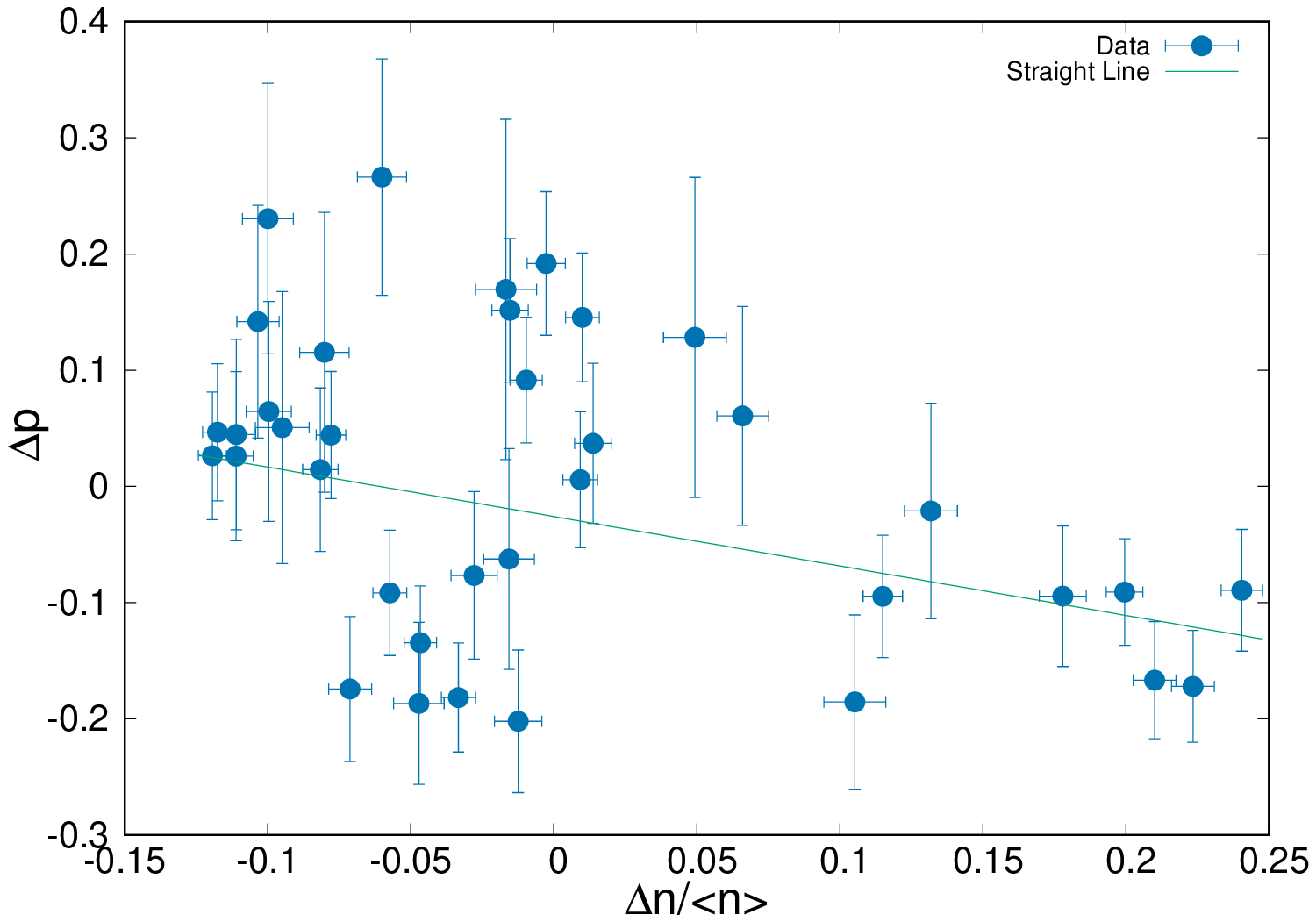} 
  \includegraphics[scale=0.30,angle=-90]{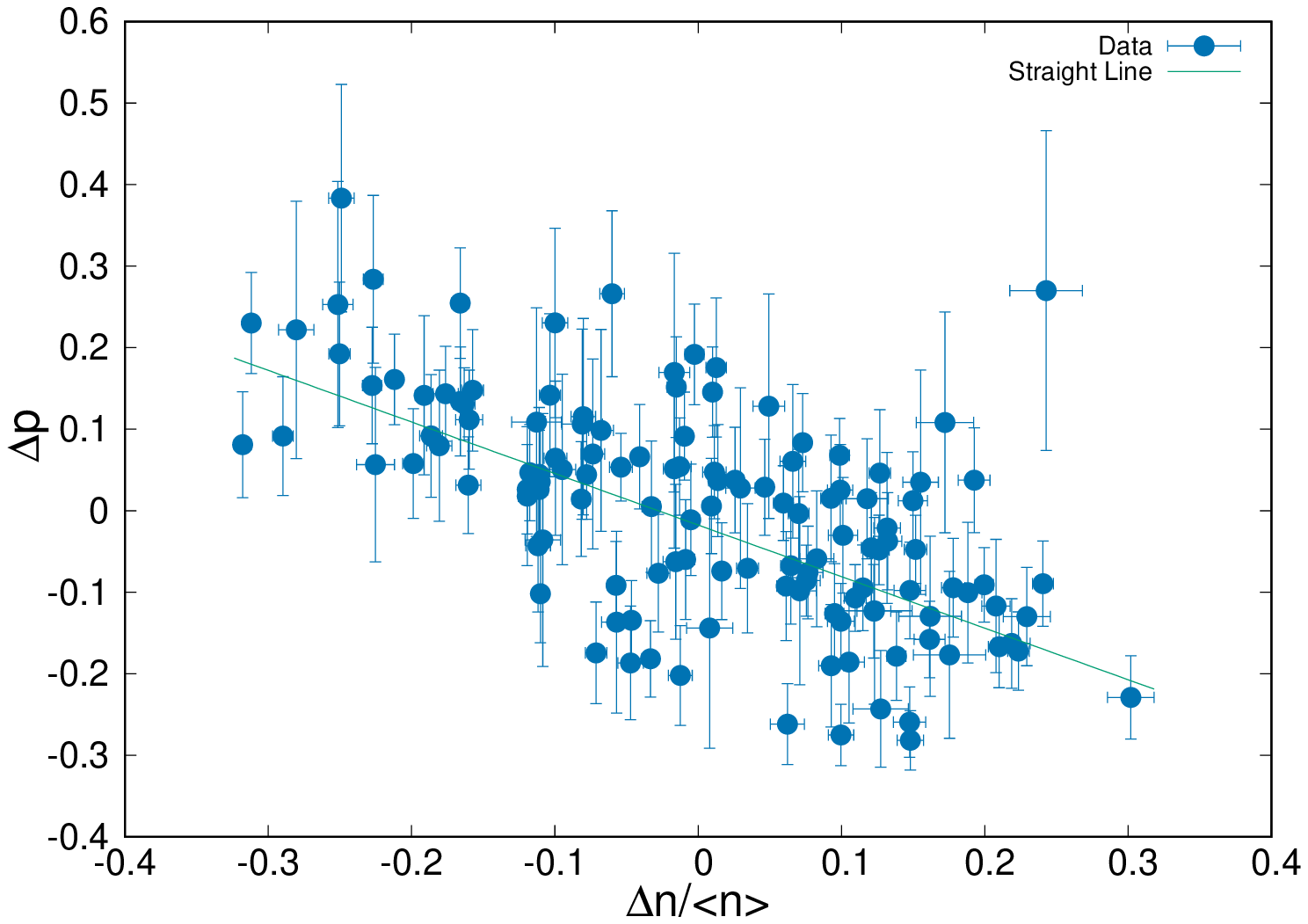}
  \includegraphics[scale=0.30,angle=-90]{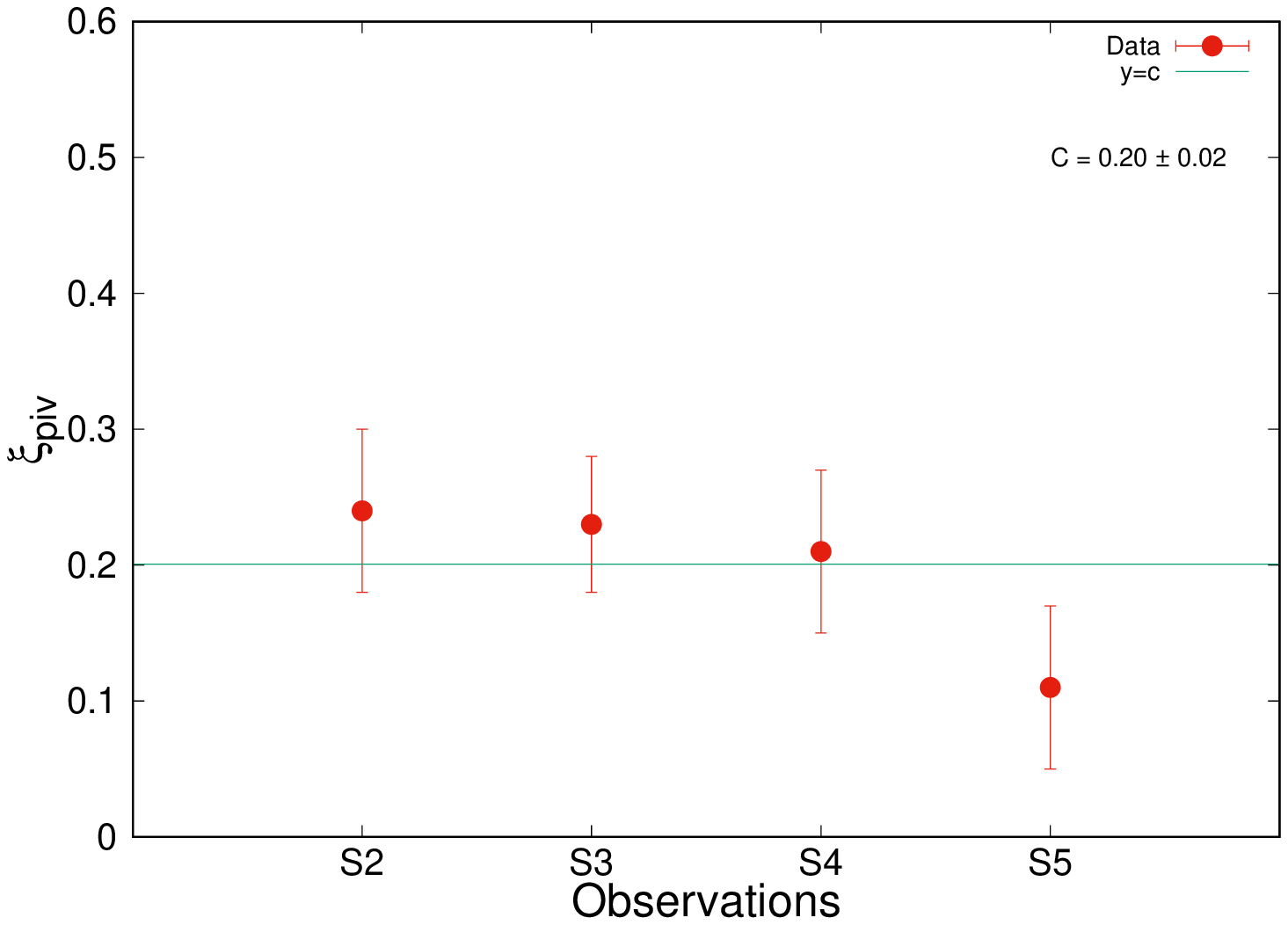}

  \caption{ Linear fit  to  the variations between the spectral index ($\Delta p$) and the normalized particle density ($\Delta$$n$/$<n>$) at $\xi=\xi_{ref}$  in the $\xi-{max}$ model.  The upper left and right panels corresponds to S2 and S3 observations respectively, middle left and right panels corresponds to S4 and S5 observations respectively.  Bottom left panel represents linear fit for the combined observations. %In case of combined observations, $1/log(\xi_{ref}/\xi_{piv})= 0.63 \pm 0.06$, which means $\xi_{piv}=0.21 \pm 0.03$ $\sqrt{keV}$ (red-chisquare $\sim 0.7$).
  The bottom right plot represents the variation  of $\xi_{piv}$ with respect to the observations in the $\xi-{max}$ model. The constant fit results in reduced-$\chi^2\sim1$ and $\xi_{piv}\sim0.20\pm 0.02$ $(\sqrt{keV})$.}
\label{fig:xi-max-slopes}
\end{figure*}

The Spearman correlation results between 
between $\Delta p$ and $\Delta n/<n>$,  and  between $\Delta p$ and $\Delta \xi_{max}$  are shown in Table \ref{tab:corr-bp-max}. No significant correlation was observation S1. 
A moderate anticorrelation is obtained between $\Delta p$ and $\Delta n/<n>$ and a strong positive correlation is observed between $\Delta \xi_{max}$ and $\Delta p$.These results are similar to the combined observation plots (see Figure \ref{fig:combined-gamma} in appendix section).  Further like the BPL model, the $\xi-{max}$ model also shows that
the magnitude of index variation are similar to the normalized particle density relation (see Figure \ref{fig:xi-max-slopes}).

In order to obtain the pivot energy and the particle density at the pivot energy, we can write Equation \ref{eq:xi_max_model} in the form
\begin{equation}
n_{ref}\left(\frac{\xi}{\xi_{ref}}\right)^{-p}=n_{piv}\left(\frac{\xi}{\xi_{piv}}\right)^{-p}
\end{equation}
Here,  $n_{ref}$ and  $n_{piv}$ represents the particle densities at reference energy, ${\xi}_{ref}$ and pivot energy, $\xi_{piv}$, respectively. 
The variation in normalised particle density and  $p$ would be related through the equation,
\begin{equation}
    \frac{\Delta n}{n}=-\Delta p \hspace{0.05cm}log\left[\frac{{\xi}_{ref}}{{\xi}_{piv}}\right]    
\end{equation}
This is an equation of a straight line whose slope is given by $log\left(\frac{{\xi}_{ref}}{{\xi}_{piv}}\right)$.

The linear fits are carried to the variations of the index  $\Delta p$ and the normalized particle density variation $\frac{\Delta(n)}{n}$ for the individual observation and for the combined observation (see Figure \ref{fig:xi-max-slopes}). The $\xi_{piv}$ values resulting from the linear fits in case of $\xi-max$ for observations S2,S3,S4 and S5 are $0.24\pm 0.06$ $\sqrt{keV}$, $0.23\pm0.05$ $\sqrt{keV}$,  $0.21\pm0.06$ $\sqrt{keV}$ and $0.11\pm0.06$ $\sqrt{keV}$, respectively. We noted that the pivot energy $\xi_{piv}$ remains same  ($\xi_{piv} = 0.20 \pm 0.02$ $\sqrt{keV}$) for the individual observations  and also matches with $\xi_{piv}$ obtained for the combined observation (see Figure \ref{fig:xi-max-slopes}). These results are consistent with those derived from the BPL model, reinforcing  the result to be model independent.

\begin{table*}
    \centering
    \caption{Spearman correlation results between the best fit spectral parameters obtained by fitting the joint SXT and LAXPC spectra with the synchrotron convolved LP and $\xi-max$ models.}
    \scalebox{0.8}{ % Scale the table to 80% of its original size
    \begin{tabular}{lllcccccc}
    \hline
       
      Observation &Observation ID &Model&Correlation between&$r_{s}$&$P_{s}$&Correlation between&$r_{s}$&$P_{s}$ \\ \hline
        S1. & G05\_201T01\_9000000478 &&&-0.08 &0.87&&0.08&0.87 \\
        S2 &A02\_005T01\_9000000948&&&-0.83& $4.71 \times 10^{-11}$&&-0.70& $6.10 \times 10^{-07}$\\
       S3& T01\_218T01\_9000001852 &Log-parabola&$\Delta$$\alpha$ $\&$ $\frac{\Delta n}{<n>}$ &-0.86 & $2.49 \times 10^{-03}$ &$\Delta$$\alpha$ $\&$ $\Delta$$\beta$&
-0.65 & $0.05$\\
       S4 & A05\_015T01\_9000002650 &&&-0.63 & $2.31 \times 10^{-05}$&& -0.63& $3.03 \times 10^{-05}$\\
        
       S5 &A05\_204T01\_9000002856&&&-0.51& $0.005$&&-0.73& $3.31 \times 10^{-07}$\\
        
        \hline
        
    %Model & $\xi-{max}$
      % &$\Delta$$p$ $\&$ $\frac{\Delta n}{<n>}$&& $\Delta$$p$ $\&$ $\Delta$$\xi_{max}$\\ \hline
        S No. &Observation ID &Model&Correlation between&$r_{s}$&$P_{s}$&Correlation between&$r_{s}$&$P_{s}$ \\ \hline
        S1. & G05\_201T01\_9000000478 &&&-0.37 &0.46&& 0.31&0.54 \\
        1. &A02\_005T01\_9000000948&&&-0.57 &$0.12 \times 10^{-04}$&&0.73& $1.03 \times 10^{-07}$\\
        2.& T01\_218T01\_9000001852 &$\xi-{max}$&$\Delta$$p$ $\&$ $\frac{\Delta n}{<n>}$&
-0.93 & $2.35 \times 10^{-04}$ &$\Delta$$p$ $\&$ $\Delta$$\xi_{max}$&0.77 & $0.015$ \\
       3.& A05\_015T01\_9000002650 &&&-0.50 & $0.15 \times 10^{-03}$&&0.64 & $2.00 \times 10^{-05}$\\
        
        4. &A05\_204T01\_9000002856&&&-0.33& $0.042$& &0.78& $7.06 \times 10^{-09}$\\
        
        \hline
        
    \end{tabular}
    }
    
    \label{tab:corr-bp-max}
\end{table*}

%\begin{table}
   % \centering
    %\begin{tabular}{l|l|l}
   % \hline
      % Observation ID  & m&c \\
      % \hline
      % A02\_005T01\_9000000948  &$ 0.70 \pm 0.09 $ &  $ 0.02        \pm 0.01$ \\
      % T01\_218T01\_9000001852 &$ 0.68 \pm 0.09$& $0.03 \pm 0.01$ \\
    %A05\_015T01\_9000002650   & $0.62 \pm 0.09$ & $0.02 \pm 0.01$ \\
     
     %A05\_204T01\_9000002856 &$ 0.45 \pm 0.08 $& $0.02 \pm 0.01$ \\
     
     %\hline
    %\end{tabular}
    %\caption{Slope of straight line (gamma max model)}
    %\label{tab:my_label}
%\end{table}

%\begin{table}
   % \centering
    %\begin{tabular}{l|l|l}
    %\hline
      % Observation ID  & m&c \\
      % \hline
      % A02\_005T01\_9000000948  &$ 0.88 \pm 0.07 $ &  $ 0.051        \pm 0.011$ \\
      % T01\_218T01\_9000001852 &$ 0.74 \pm 0.07$& $ 0.001 \pm 0.015$ \\
   % A05\_015T01\_9000002650   & $0.74 \pm 0.12$ & $0.005 \pm 0.003$ \\
     
    % A05\_204T01\_9000002856 &$ 0.62 \pm 0.12 $& $0.020 \pm 0.018$ \\
     
     %\hline
   % \end{tabular}
   % \caption{Slope of straight line (brokenpower)}
   % \label{tab:my_label}
%\end{table}

%\begin{table}

   % \centering
    
   % \begin{tabular}{c|c|}
   % \hline
%$1/log(\xi_{ref}/\xi_{piv})$&$\xi_{piv} (\sqrt{keV})$ \\
     % \hline
     % $0.72\pm 0.01$ & $0.24 \pm 0.05$ \\
      
      %$0.68 \pm 0.01 $  & $0.23 \pm 0.05 $\\
      %$0.62 \pm 0.01$ & $0.20 \pm 0.05$ \\
     % $0.46 \pm 0.01$ & $0.12 \pm 0.04$ \\
     % \hline
    %\end{tabular}
   % \caption{$\xi_{piv}$ values in case of gamma max model.}
    %\label{tab:my_label}
%\end{table}

\section{SUMMARY AND DISCUSSION}
We conducted a detailed X-ray spectral  analysis of  Mkn\,421, utilizing data obtained from AstroSat observations spanning multiple epochs from 2016 to 2019. 
To gain insight into the spectral  characteristics on shorter timescales, we employed a time-resolved spectral analysis approach. For this purpose, we divided the total duration of observation for each individual observation into time intervals of 10 ks. In each of these time bins, we observed that the BPL model for particle distribution provided a good fit statistic to the X-ray spectra in the energy range  0.7-19 keV.
The lower-energy particle index  $\Gamma_{1}$ was observed to exhibit an anticorrelation with the particle density. Furthermore, we found that the variation in the normalized particle density $\frac{\Delta n}{<n>}$   is similar to the  variation in the index $\Delta$$\Gamma_{1}$.   
 Notably, no time lag was detected between particle density and index. In \citet{2021MNRAS.504.5485S}, the variability in the spectral index ($\Gamma$) and particle density is modeled as arising from variations in the acceleration timescale ($\tau_{\text{acc}}$). A sinusoidal perturbation in $\tau_{\text{acc}}$ leads to correlated variations in $\Delta n / n$ and $\Delta \Gamma$, but the relationship between their amplitudes and the presence of a time lag depends  on the relative timescales of the system.  When the variation timescale is comparable to $\tau_{\text{acc}}$, the amplitude of $\Delta \Gamma$ can significantly exceed that of $\Delta n / n$. In this scenario, a notable time lag between the variations in $\Delta n / n$ and $\Delta \Gamma$ is expected, as described by Equation (14) in \citet{2021MNRAS.504.5485S}, which is not observed in our results. However, for slow variations, the phase lag term $\left(\phi_{\text{lag}} / \sin(\phi_{\text{lag}})\right)$ in Equation (14) approaches unity. This leads to the steady-state solution represented by Equation (8) in \citet{2021MNRAS.504.5485S}, where the amplitudes of $\Delta n / n$ is greater than $\Delta \Gamma$. 

Power-law particle distribution pivots over some energy $\xi_{0}$. In case of BPL distribution of electrons,
we found  $\frac{\Delta n}{<n>}$ to be inversely correlated to $\Gamma_{1}$, which suggest the harder when bighter feature typically observed in blazars.  The linear fits applied to the variations of the index $\Gamma_{1}$ and the normalized particle density $\frac{\Delta n}{<n>}$ indicate that the pivot energy $\xi_{\text{piv}}$ remains consistent ($\xi_{\text{piv}} = 0.34 \pm 0.02 \sqrt{keV})$) across individual observations, and matches with the $\xi_{\text{piv}}$ derived from the combined observation. The steady nature of the pivot energy across different observations carries significant implications.  Within context of the convolved  model, the $\xi$ parameter is defined as $\xi=\sqrt{C}\gamma$. Consequently, any model parameter, such as the pivot energy  $\xi_{piv}$, can be linked to $\gamma_{piv}$ as $\gamma_{piv}$ = $\xi_{piv}/\sqrt{C}$, here, $C$ = $1.36 \times {10^{-11}}$$\frac{\delta B}{1+z}$. Given that the estimated $\xi_{piv}$ remains constant,  the product of $\delta^{1/2}$ and $B^{1/2}$ should also remain constant. Thus we are unable to separate them to constrain either $\delta$ or $B$ individually.
Moreover, constant $\xi_{piv}$ suggests that the source's variability stems from index variations and is unaffected by normalization.
 Consequently, factors such as magnetic fields, source region geometry, and other parameters do not contribute to the source's variability. Instead, the variations are primarily linked to the acceleration or escape timescales associated with the emitted particles within the source.

 To assess the model independence of the constancy of the pivot energy, we employed two models, the LP model and the $\xi$-max model.
Besides the BPL model, these alternative models also provide equally good fit to the spectrum. The LP model is characterised by the curved features defined by the parameter, $\beta$. The conventional explanation for the curvature in the spectrum involves considering the energy-dependent acceleration/escape rates process \citep{2004A&A...413..489M, 2021MNRAS.508.5921H}.  The correlation study  between the spectral parameters of LP model shows a negative correlation between $\Delta $$\alpha $ and the $\Delta n/<n>$, indicating a  harder spectra when the source is brighter. Additionally, for all observations, a negative correlation is evident between $\Delta $$\alpha $ and the $\Delta$ $\beta$. No significant correlation was observed in S1 observation. Within the framework of the LP particle distribution, $log\left(\frac{\xi_{piv}}{\xi_{ref}}\right)$ yields two roots, one positive and one negative. The positive root results in a high pivot energy across all observations, which contradicts typical expectations where the pivot energy is often selected close to the low energy of the observed data. The $\xi_{piv}$ values obtained from the negative root for observations S2, S3, S4, and S5 are all consistent with the value $0.56\pm 0.01$ $\sqrt{keV}$. This, once again, indicates the constancy of the pivot energy across different observations.

We further examined the braodband X-ray spectral analysis of Mkn\,421 by considering the power-law distribution with maximum electron
energy. Such particle distribution results in regions in which shock-induced particle acceleration leads to radiative energy loss. The loss becomes dominant at higher energies as the energy loss rate is proportional to the square of its energy,  leading to a departure from a  power-law form at high energies with a maximum Lorentz factor $\gamma_{max}$. We observed that each of the spectra can be adequately represented by this model. The index of the particle distribution at lower energies exhibited a positive correlation with the maximum energy of the particles. However, according to a simplistic interpretation where the observed variation primarily arises from changes in the acceleration time-scale, the correlation should have been negative. A more intricate model, in which the magnetic field is correlated with the acceleration time-scale, could account for the positive correlation observed. 
 By examining the relationship between particle densities at different energy scales 
we established an inverse correlation between $\Delta {n}/<{n}>$ and $\Delta p$, a pattern that aligns with the observed data. Linear fits applied to the variations of the spectral index $\Delta p$ and the normalized particle density $\Delta {n}/<{n}>$ for each individual observation, as well as for the combined observation, enables us to estimate the values of $\xi_{piv}$. Again we noted that $\xi_{piv}$  remains constant ($\xi_{piv} = 0.20\pm 0.02 (\sqrt{keV})$ ) for individual observations,  and are consistent with the $\xi_{piv}$ obtained for the combined observation. These results are consistent with those obtained from the BPL model, underscoring the model independence of the result. This identification enables us to investigate the underlying causes of variability.\\
The observed anti-correlation between the electron density and spectral index, along with the constant pivot Lorentz factor, suggests that these variations may be attributed to changes in the acceleration time scales. However,  the potential influence of other time scales, such as the escape time scale and injection rates can not be ruled out. In this context, \citet{2021MNRAS.508.5921H} investigated the X-ray variability of Mrk 421 by incorporating an energy-dependent escape time scale (EDD). Their study showed that while the EDD and EDA models effectively describe the spectra for individual time segments and explain the correlations among spectral parameters within the model framework, the derived physical parameter estimates challenge the validity of these models in their simplest form. Since the log-parabola model is essentially an approximation of the such model, the similar correlation trends observed between the fitting parameters in their study and ours can be used to challenge the hypothesis of energy-dependent escape/acceleration time scales. Additionally, the constant pivot energy observed accross the observations indicate that their is minor feedback of B and $\delta$ on the particle spectrum.  Moderate changes in $\delta$ and $B$ would not affect the constant pivot energy directly, but they may amplify or suppress the observed flux variability. Thus, while external variations cannot be entirely ruled out, the pivot energy’s stability points toward intrinsic energy-dependent processes as the primary drivers of variability.  In conclusion, the models like EDD and EDA are simplified, with analytical solutions, but the actual physical processes are likely more intricate. Developing and implementing advanced methods to test non-analytical models is crucial. For instance, both acceleration and diffusion time-scales may exhibit energy dependence, which would be more physically realistic \citep{2004A&A...413..489M,2006A&A...448..861M, tramacere2007signatures}. Stochastic acceleration might play a significant role in shaping the particle distribution.\\

%\documentclass{article}

% Define the \aap command to expand to "Astronomy & Astrophysics"

\newcommand{\aap}{Astronomy \& Astrophysics}
\newcommand{\apj}{The Astrophysical Journal}
\newcommand{\ssr}{Space Science Reviews}
\newcommand{\mnras}{Monthly Notices of the Royal Astronomical Society}
\newcommand{\apjl}{The Astrophysical Journal Letters}
\newcommand{\pasp}{Publications of the Astronomical Society of the Pacific}
\newcommand{\apjs}{Astrophysical Journal Supplement Series}
\newcommand{\araa}{Annual Review of Astronomy and Astrophysics}
\newcommand{\aj}{AJ}

\section{Acknowledgements}
 SAD is thankful to the MOMA for the MANF fellowship (No.F.82-27/2019(SA-III)). ZS is supported by the Department of Science and Technology, Govt. of India, under the INSPIRE Faculty grant (DST/INSPIRE/04/2020/002319). SAD, ZS and  NI express  gratitude to the Inter-University Centre for Astronomy and Astrophysics (IUCAA) in Pune, India, for the support and facilities provided.

\bibliographystyle{elsarticle-harv} 
%\biboptions{authoryear}
\bibliography{sample631}

%% else use the following coding to input the bibitems directly in the
%% TeX file.

%%\begin{thebibliography}{00}

%% \bibitem[Author(year)]{label}
%% For example:

%% \bibitem[Aladro et al.(2015)]{Aladro15} Aladro, R., Martín, S., Riquelme, D., et al. 2015, \aas, 579, A101

%%\end{thebibliography}
 
 \section{Appendix}

%\section{Best fit parameters}
%The best fit spectral parameters for all observations using different models are given below
\begin{table*}

     \centering
     \caption{The best fit parameter values obtained by fitting  time-resolved broadband X-ray spectra of S1 observation with the snchrotron convolved BPL/LP/$\xi-{max}$ models.} 
     \scalebox{0.87}{ % Scale the table to 80% of its original size
     \begin{tabular}{rlllllllllllll} \hline
     && &Broken&&&&Log&&&&$\xi-{max}$\\
      &&&Power law&&&&parabola&&&&model&\\
      \hline
       &Time&$\Gamma_{1}$&$\Gamma_{2}$&norm (n)&$\frac{\chi^{2}}{dof}$&$\alpha$&$\beta$&norm (n)&$\frac{\chi^{2}}{dof}$&p&$\xi_{max}$&{norm} (n)&$\frac{\chi^{2}}{dof}$\\

&&&&$(\times 10^{-2})$&&&&$(\times 10^{-2})$&&&&$(\times 10^{-2})$&\\
\hline
&5000&$3.13^{+ 0.10}_{- 0.10}$& $ 4.32^{+ 0.11}_{- 0.11}$& $ 15.96^{+ 0.41}_{- 0.41}$& $ 1.03$	&$3.19^{+ 0.10}_{- 0.10}$& $ 1.17^{+ 0.20}_{- 0.18}$& $ 34.27^{+ 0.66}_{- 0.65}$& $ 1.04$	&$3.08^{+ 0.10}_{- 0.09}$& $ 6.14^{+ 0.86}_{- 0.64}$& $ 38.67^{+ 0.88}_{- 0.88}$& $ 1.05$\\
&15000&$3.02^{+ 0.09}_{- 0.09}$& $ 4.13^{+ 0.11}_{- 0.11}$& $ 18.15^{+ 0.39}_{- 0.39}$& $ 1.00$	&$3.02^{+ 0.09}_{- 0.09}$& $ 1.11^{+ 0.18}_{- 0.17}$& $ 37.93^{+ 0.67}_{- 0.66}$& $ 1.02$	&$2.96^{+ 0.08}_{- 0.08}$& $ 6.24^{+ 0.85}_{- 0.64}$& $ 41.93^{+ 0.88}_{- 0.87}$& $ 1.03$\\
&25000&$3.23^{+ 0.06}_{- 0.06}$& $ 4.87^{+ 0.13}_{- 0.12}$& $ 18.09^{+ 0.3}_{- 0.3}$& $ 1.09$	&$3.29^{+ 0.06}_{- 0.06}$& $ 1.58^{+ 0.17}_{- 0.16}$& $ 40.73^{+ 0.61}_{- 0.60}$& $ 1.14$	&$3.05^{+ 0.06}_{- 0.06}$& $ 4.76^{+ 0.35}_{- 0.29}$& $ 47.73^{+ 0.82}_{- 0.82}$& $ 1.12$\\
&35000&$3.21^{+ 0.06}_{- 0.06}$& $ 5.06^{+ 0.14}_{- 0.13}$& $ 19.44^{+ 0.32}_{- 0.32}$& $ 1.16$	&$3.28^{+ 0.06}_{- 0.06}$& $ 1.78^{+ 0.18}_{- 0.17}$& $ 44.07^{+ 0.67}_{- 0.66}$& $ 1.22$	&$3.03^{+ 0.06}_{- 0.06}$& $ 4.44^{+ 0.30}_{- 0.25}$& $ 51.88^{+ 0.89}_{- 0.88}$& $ 1.20$\\
&45000&$3.14^{+ 0.08}_{- 0.08}$& $ 5.03^{+ 0.15}_{- 0.14}$& $ 19.96^{+ 0.42}_{- 0.42}$& $ 0.94$	&$3.15^{+ 0.09}_{- 0.09}$& $ 1.96^{+ 0.22}_{- 0.21}$& $ 45.09^{+ 0.80}_{- 0.79}$& $ 0.98$	&$2.97^{+ 0.07}_{- 0.07}$& $ 4.35^{+ 0.32}_{- 0.27}$& $ 51.97^{+ 1.04}_{- 1.02}$& $ 0.97$\\
&55000&$3.10^{+ 0.10}_{- 0.09}$& $ 4.94^{+ 0.19}_{- 0.17}$& $ 19.99^{+ 0.49}_{- 0.49}$& $ 0.94$	&$3.13^{+ 0.10}_{- 0.10}$& $ 1.84^{+ 0.26}_{- 0.25}$& $ 44.41^{+ 0.95}_{- 0.92}$& $ 1.00$	&$2.94^{+ 0.09}_{- 0.08}$& $ 4.37^{+ 0.42}_{- 0.34}$& $ 51.10^{+ 1.21}_{- 1.19}$& $ 0.96$\\
\hline
     
     \end{tabular}
     }
     \label{tab:t1}      
\end{table*}

\begin{table*}

     \centering
     \caption{The best fit parameter values obtained by fitting  time-resolved broadband X-ray spectra of S2 observation with the snchrotron convolved BPL/LP/$\xi-{max}$ models.} 
     \scalebox{0.87}{ % Scale the table to 80% of its original size
     \begin{tabular}{rlllllllllllll} \hline
     && &Broken&&&&Log&&&&$\xi-{max}$\\
      &&&Power law&&&&parabola&&&&model&\\
      \hline
       &Time&$\Gamma_{1}$&$\Gamma_{2}$&norm (n)&$\frac{\chi^{2}}{dof}$&$\alpha$&$\beta$&norm (n)&$\frac{\chi^{2}}{dof}$&p&$\xi_{max}$&{norm} (n)&$\frac{\chi^{2}}{dof}$\\

&&&&$(\times 10^{-2})$&&&&$(\times 10^{-2})$&&&&$(\times 10^{-2})$&\\
\hline
&5000&$2.90^{+ 0.08}_{- 0.08}$& $ 4.70^{+ 0.16}_{- 0.15}$& $ 11.97^{+ 0.19}_{- 0.19}$& $ 1.03$  	&$2.74^{+ 0.10}_{- 0.09}$& $ 1.81^{+ 0.22}_{- 0.21}$& $ 24.89^{+ 0.42}_{- 0.42}$& $ 1.06$  	&$2.70^{+ 0.07}_{- 0.07}$& $ 4.39^{+ 0.33}_{- 0.28}$& $ 26.95^{+ 0.60}_{- 0.59}$& $ 1.05$\\
&15000&$2.91^{+ 0.09}_{- 0.09}$& $ 4.84^{+ 0.16}_{- 0.15}$& $ 12.31^{+ 0.22}_{- 0.22}$& $ 0.93$  	&$2.73^{+ 0.12}_{- 0.11}$& $ 1.98^{+ 0.24}_{- 0.22}$& $ 25.87^{+ 0.46}_{- 0.46}$& $ 0.95$  	&$2.71^{+ 0.08}_{- 0.07}$& $ 4.27^{+ 0.31}_{- 0.26}$& $ 28.06^{+ 0.72}_{- 0.70}$& $ 0.96$\\
&25000&$2.98^{+ 0.07}_{- 0.07}$& $ 4.57^{+ 0.14}_{- 0.14}$& $ 11.48^{+ 0.17}_{- 0.17}$& $ 1.04$  	&$2.95^{+ 0.08}_{- 0.08}$& $ 1.49^{+ 0.19}_{- 0.18}$& $ 24.47^{+ 0.38}_{- 0.37}$& $ 1.06$  	&$2.85^{+ 0.06}_{- 0.06}$& $ 4.98^{+ 0.43}_{- 0.35}$& $ 27.18^{+ 0.49}_{- 0.50}$& $ 1.07$\\
&35000&$2.91^{+ 0.10}_{- 0.10}$& $ 4.42^{+ 0.16}_{- 0.15}$& $ 12.90^{+ 0.26}_{- 0.26}$& $ 1.14$  	&$2.90^{+ 0.12}_{- 0.11}$& $ 1.38^{+ 0.23}_{- 0.22}$& $ 26.85^{+ 0.52}_{- 0.51}$& $ 1.18$  	&$2.81^{+ 0.09}_{- 0.09}$& $ 5.13^{+ 0.59}_{- 0.46}$& $ 29.61^{+ 0.74}_{- 0.73}$& $ 1.16$\\
&45000&$2.88^{+ 0.20}_{- 0.19}$& $ 4.32^{+ 0.16}_{- 0.15}$& $ 12.51^{+ 0.47}_{- 0.46}$& $ 0.78$  	&$2.82^{+ 0.23}_{- 0.22}$& $ 1.40^{+ 0.35}_{- 0.32}$& $ 25.87^{+ 0.84}_{- 0.84}$& $ 0.78$  	&$2.84^{+ 0.17}_{- 0.16}$& $ 5.49^{+ 1.10}_{- 0.73}$& $ 28.42^{+ 1.23}_{- 1.11}$& $ 0.81$\\
&55000&$2.86^{+ 0.18}_{- 0.18}$& $ 4.11^{+ 0.15}_{- 0.15}$& $ 13.35^{+ 0.45}_{- 0.44}$& $ 0.95$  	&$2.80^{+ 0.21}_{- 0.19}$& $ 1.22^{+ 0.32}_{- 0.29}$& $ 27.18^{+ 0.82}_{- 0.82}$& $ 0.95$  	&$2.87^{+ 0.16}_{- 0.15}$& $ 6.27^{+ 1.54}_{- 0.96}$& $ 29.57^{+ 1.06}_{- 1.01}$& $ 1.00$\\
&65000&$2.74^{+ 0.09}_{- 0.08}$& $ 4.35^{+ 0.13}_{- 0.13}$& $ 14.64^{+ 0.23}_{- 0.23}$& $ 0.95$  	&$2.70^{+ 0.10}_{- 0.10}$& $ 1.52^{+ 0.20}_{- 0.19}$& $ 29.50^{+ 0.50}_{- 0.49}$& $ 0.98$  	&$2.67^{+ 0.07}_{- 0.07}$& $ 4.84^{+ 0.41}_{- 0.34}$& $ 31.65^{+ 0.73}_{- 0.71}$& $ 0.97$\\
&75000&$2.80^{+ 0.09}_{- 0.09}$& $ 4.55^{+ 0.17}_{- 0.16}$& $ 14.41^{+ 0.26}_{- 0.26}$& $ 0.80$  	&$2.74^{+ 0.11}_{- 0.11}$& $ 1.67^{+ 0.24}_{- 0.23}$& $ 29.68^{+ 0.56}_{- 0.55}$& $ 0.81$  	&$2.71^{+ 0.08}_{- 0.08}$& $ 4.61^{+ 0.43}_{- 0.35}$& $ 32.14^{+ 0.80}_{- 0.78}$& $ 0.81$\\
&85000&$2.88^{+ 0.06}_{- 0.06}$& $ 4.31^{+ 0.13}_{- 0.12}$& $ 13.84^{+ 0.18}_{- 0.18}$& $ 1.01$  	&$2.87^{+ 0.07}_{- 0.07}$& $ 1.29^{+ 0.17}_{- 0.16}$& $ 28.44^{+ 0.41}_{- 0.40}$& $ 1.05$  	&$2.78^{+ 0.06}_{- 0.06}$& $ 5.24^{+ 0.45}_{- 0.37}$& $ 31.13^{+ 0.53}_{- 0.53}$& $ 1.02$\\
&95000&$2.86^{+ 0.09}_{- 0.08}$& $ 4.39^{+ 0.13}_{- 0.13}$& $ 13.57^{+ 0.22}_{- 0.22}$& $ 1.01$  	&$2.81^{+ 0.10}_{- 0.10}$& $ 1.46^{+ 0.20}_{- 0.19}$& $ 28.01^{+ 0.47}_{- 0.47}$& $ 1.02$  	&$2.77^{+ 0.07}_{- 0.07}$& $ 5.09^{+ 0.47}_{- 0.38}$& $ 30.51^{+ 0.66}_{- 0.65}$& $ 1.04$\\
&105000&$2.68^{+ 0.17}_{- 0.16}$& $ 4.39^{+ 0.15}_{- 0.14}$& $ 14.28^{+ 0.38}_{- 0.37}$& $ 0.85$  	&$2.62^{+ 0.20}_{- 0.18}$& $ 1.66^{+ 0.31}_{- 0.28}$& $ 28.54^{+ 0.77}_{- 0.79}$& $ 0.86$  	&$2.67^{+ 0.13}_{- 0.12}$& $ 4.81^{+ 0.60}_{- 0.46}$& $ 30.61^{+ 1.33}_{- 1.19}$& $ 0.89$\\
&115000&$2.98^{+ 0.08}_{- 0.07}$& $ 4.34^{+ 0.13}_{- 0.12}$& $ 14.16^{+ 0.23}_{- 0.22}$& $ 0.91$  	&$2.98^{+ 0.08}_{- 0.08}$& $ 1.23^{+ 0.17}_{- 0.17}$& $ 29.72^{+ 0.46}_{- 0.45}$& $ 0.94$  	&$2.87^{+ 0.07}_{- 0.07}$& $ 5.53^{+ 0.55}_{- 0.44}$& $ 32.94^{+ 0.61}_{- 0.62}$& $ 0.93$\\
&125000&$2.86^{+ 0.09}_{- 0.09}$& $ 4.48^{+ 0.14}_{- 0.14}$& $ 14.68^{+ 0.26}_{- 0.26}$& $ 0.87$  	&$2.81^{+ 0.11}_{- 0.10}$& $ 1.55^{+ 0.21}_{- 0.20}$& $ 30.50^{+ 0.54}_{- 0.53}$& $ 0.89$  	&$2.76^{+ 0.08}_{- 0.07}$& $ 4.89^{+ 0.45}_{- 0.36}$& $ 33.28^{+ 0.76}_{- 0.75}$& $ 0.89$\\
&135000&$2.78^{+ 0.19}_{- 0.19}$& $ 4.68^{+ 0.18}_{- 0.17}$& $ 15.55^{+ 0.53}_{- 0.52}$& $ 1.03$  	&$2.68^{+ 0.24}_{- 0.22}$& $ 1.89^{+ 0.38}_{- 0.34}$& $ 32.25^{+ 0.99}_{- 1.01}$& $ 1.06$  	&$2.73^{+ 0.15}_{- 0.14}$& $ 4.51^{+ 0.60}_{- 0.44}$& $ 35.04^{+ 1.74}_{- 1.54}$& $ 1.05$\\
&145000&$2.74^{+ 0.16}_{- 0.15}$& $ 4.46^{+ 0.15}_{- 0.14}$& $ 16.52^{+ 0.44}_{- 0.44}$& $ 1.04$  	&$2.53^{+ 0.20}_{- 0.18}$& $ 1.88^{+ 0.32}_{- 0.29}$& $ 33.79^{+ 0.93}_{- 0.95}$& $ 1.04$  	&$2.72^{+ 0.12}_{- 0.12}$& $ 4.84^{+ 0.60}_{- 0.45}$& $ 36.30^{+ 1.42}_{- 1.29}$& $ 1.09$\\
&155000&$2.75^{+ 0.08}_{- 0.07}$& $ 4.53^{+ 0.14}_{- 0.13}$& $ 17.82^{+ 0.25}_{- 0.25}$& $ 0.86$  	&$2.70^{+ 0.09}_{- 0.09}$& $ 1.68^{+ 0.20}_{- 0.19}$& $ 36.31^{+ 0.57}_{- 0.57}$& $ 0.91$  	&$2.67^{+ 0.06}_{- 0.06}$& $ 4.56^{+ 0.33}_{- 0.28}$& $ 38.99^{+ 0.82}_{- 0.81}$& $ 0.90$\\
&165000&$2.66^{+ 0.06}_{- 0.06}$& $ 4.37^{+ 0.12}_{- 0.11}$& $ 19.87^{+ 0.22}_{- 0.22}$& $ 1.05$  	&$2.61^{+ 0.07}_{- 0.07}$& $ 1.61^{+ 0.17}_{- 0.16}$& $ 39.46^{+ 0.53}_{- 0.52}$& $ 1.09$  	&$2.63^{+ 0.05}_{- 0.05}$& $ 4.71^{+ 0.30}_{- 0.26}$& $ 41.85^{+ 0.70}_{- 0.69}$& $ 1.08$\\
&175000&$2.57^{+ 0.08}_{- 0.08}$& $ 4.71^{+ 0.15}_{- 0.14}$& $ 20.54^{+ 0.28}_{- 0.28}$& $ 0.95$  	&$2.46^{+ 0.10}_{- 0.10}$& $ 2.11^{+ 0.22}_{- 0.21}$& $ 40.93^{+ 0.68}_{- 0.67}$& $ 0.98$  	&$2.54^{+ 0.06}_{- 0.06}$& $ 4.08^{+ 0.24}_{- 0.20}$& $ 42.52^{+ 1.05}_{- 1.02}$& $ 0.98$\\
&185000&$2.75^{+ 0.06}_{- 0.05}$& $ 4.52^{+ 0.12}_{- 0.11}$& $ 19.83^{+ 0.22}_{- 0.22}$& $ 0.99$  	&$2.70^{+ 0.07}_{- 0.06}$& $ 1.67^{+ 0.16}_{- 0.15}$& $ 40.41^{+ 0.52}_{- 0.51}$& $ 1.03$  	&$2.68^{+ 0.05}_{- 0.05}$& $ 4.65^{+ 0.27}_{- 0.24}$& $ 43.40^{+ 0.68}_{- 0.68}$& $ 1.03$\\
&195000&$2.60^{+ 0.06}_{- 0.06}$& $ 4.48^{+ 0.11}_{- 0.11}$& $ 21.98^{+ 0.23}_{- 0.23}$& $ 1.05$  	&$2.52^{+ 0.07}_{- 0.07}$& $ 1.81^{+ 0.16}_{- 0.16}$& $ 43.48^{+ 0.57}_{- 0.56}$& $ 1.10$  	&$2.57^{+ 0.04}_{- 0.04}$& $ 4.42^{+ 0.23}_{- 0.20}$& $ 45.49^{+ 0.79}_{- 0.78}$& $ 1.08$\\
&205000&$2.71^{+ 0.08}_{- 0.08}$& $ 4.50^{+ 0.13}_{- 0.12}$& $ 21.85^{+ 0.33}_{- 0.33}$& $ 1.05$  	&$2.62^{+ 0.10}_{- 0.10}$& $ 1.76^{+ 0.20}_{- 0.19}$& $ 44.24^{+ 0.73}_{- 0.73}$& $ 1.09$  	&$2.65^{+ 0.07}_{- 0.06}$& $ 4.59^{+ 0.33}_{- 0.27}$& $ 47.11^{+ 1.11}_{- 1.07}$& $ 1.08$\\
&215000&$2.66^{+ 0.08}_{- 0.08}$& $ 4.32^{+ 0.11}_{- 0.11}$& $ 21.61^{+ 0.30}_{- 0.30}$& $ 1.02$  	&$2.59^{+ 0.10}_{- 0.09}$& $ 1.62^{+ 0.18}_{- 0.17}$& $ 42.94^{+ 0.69}_{- 0.68}$& $ 1.03$  	&$2.63^{+ 0.06}_{- 0.06}$& $ 4.80^{+ 0.34}_{- 0.29}$& $ 45.42^{+ 1.02}_{- 0.99}$& $ 1.04$\\
&225000&$2.65^{+ 0.19}_{- 0.18}$& $ 4.06^{+ 0.12}_{- 0.12}$& $ 22.11^{+ 0.66}_{- 0.64}$& $ 1.01$  	&$2.55^{+ 0.22}_{- 0.21}$& $ 1.41^{+ 0.31}_{- 0.28}$& $ 43.23^{+ 1.41}_{- 1.48}$& $ 1.02$  	&$2.73^{+ 0.14}_{- 0.14}$& $ 5.78^{+ 1.05}_{- 0.72}$& $ 46.30^{+ 2.01}_{- 1.79}$& $ 1.05$\\
&235000&$2.57^{+ 0.08}_{- 0.08}$& $ 3.88^{+ 0.11}_{- 0.10}$& $ 22.44^{+ 0.30}_{- 0.30}$& $ 1.08$  	&$2.53^{+ 0.09}_{- 0.09}$& $ 1.22^{+ 0.17}_{- 0.16}$& $ 42.70^{+ 0.71}_{- 0.71}$& $ 1.10$  	&$2.58^{+ 0.06}_{- 0.06}$& $ 5.53^{+ 0.52}_{- 0.42}$& $ 44.72^{+ 0.99}_{- 0.97}$& $ 1.11$\\
&245000&$2.49^{+ 0.08}_{- 0.08}$& $ 3.87^{+ 0.11}_{- 0.10}$& $ 23.37^{+ 0.31}_{- 0.31}$& $ 1.00$  	&$2.46^{+ 0.09}_{- 0.09}$& $ 1.39^{+ 0.13}_{- 0.13}$& $ 44.44^{+ 0.59}_{- 0.60}$& $ 1.38$  	&$2.52^{+ 0.06}_{- 0.06}$& $ 5.32^{+ 0.48}_{- 0.39}$& $ 45.33^{+ 1.10}_{- 1.06}$& $ 1.03$\\
&255000&$2.58^{+ 0.06}_{- 0.06}$& $ 3.97^{+ 0.10}_{- 0.10}$& $ 22.22^{+ 0.23}_{- 0.23}$& $ 1.00$  	&$2.53^{+ 0.07}_{- 0.06}$& $ 1.30^{+ 0.15}_{- 0.14}$& $ 42.51^{+ 0.55}_{- 0.55}$& $ 1.03$  	&$2.57^{+ 0.04}_{- 0.04}$& $ 5.25^{+ 0.38}_{- 0.32}$& $ 44.48^{+ 0.72}_{- 0.70}$& $ 1.02$\\
&265000&$2.49^{+ 0.06}_{- 0.06}$& $ 4.23^{+ 0.11}_{- 0.10}$& $ 22.23^{+ 0.22}_{- 0.22}$& $ 0.94$  	&$2.41^{+ 0.07}_{- 0.07}$& $ 1.67^{+ 0.16}_{- 0.15}$& $ 42.47^{+ 0.56}_{- 0.55}$& $ 0.97$  	&$2.51^{+ 0.04}_{- 0.04}$& $ 4.62^{+ 0.26}_{- 0.22}$& $ 43.83^{+ 0.77}_{- 0.76}$& $ 0.97$\\
&275000&$2.43^{+ 0.07}_{- 0.07}$& $ 4.30^{+ 0.11}_{- 0.11}$& $ 22.32^{+ 0.25}_{- 0.25}$& $ 0.92$  	&$2.34^{+ 0.09}_{- 0.08}$& $ 1.84^{+ 0.17}_{- 0.17}$& $ 42.43^{+ 0.63}_{- 0.63}$& $ 0.95$  	&$2.48^{+ 0.05}_{- 0.05}$& $ 4.46^{+ 0.24}_{- 0.21}$& $ 43.43^{+ 0.93}_{- 0.91}$& $ 0.96$\\
&285000&$2.45^{+ 0.09}_{- 0.08}$& $ 4.39^{+ 0.12}_{- 0.12}$& $ 22.54^{+ 0.30}_{- 0.30}$& $ 0.97$  	&$2.33^{+ 0.11}_{- 0.10}$& $ 1.95^{+ 0.21}_{- 0.20}$& $ 43.17^{+ 0.76}_{- 0.77}$& $ 1.00$  	&$2.49^{+ 0.06}_{- 0.06}$& $ 4.37^{+ 0.27}_{- 0.23}$& $ 44.34^{+ 1.18}_{- 1.14}$& $ 1.01$\\
&295000&$2.39^{+ 0.08}_{- 0.08}$& $ 4.19^{+ 0.11}_{- 0.10}$& $ 25.21^{+ 0.32}_{- 0.32}$& $ 0.97$  	&$2.31^{+ 0.10}_{- 0.10}$& $ 1.75^{+ 0.19}_{- 0.18}$& $ 47.29^{+ 0.82}_{- 0.83}$& $ 1.00$  	&$2.45^{+ 0.06}_{- 0.05}$& $ 4.54^{+ 0.28}_{- 0.24}$& $ 48.14^{+ 1.28}_{- 1.24}$& $ 1.00$\\
&305000&$2.39^{+ 0.16}_{- 0.15}$& $ 4.30^{+ 0.13}_{- 0.12}$& $ 23.60^{+ 0.51}_{- 0.51}$& $ 0.99$  	&$2.27^{+ 0.19}_{- 0.18}$& $ 1.92^{+ 0.30}_{- 0.28}$& $ 44.52^{+ 1.30}_{- 1.36}$& $ 1.01$  	&$2.49^{+ 0.10}_{- 0.10}$& $ 4.49^{+ 0.42}_{- 0.33}$& $ 45.89^{+ 2.19}_{- 2.02}$& $ 1.03$\\
&315000&$2.26^{+ 0.17}_{- 0.17}$& $ 4.44^{+ 0.13}_{- 0.13}$& $ 24.48^{+ 0.54}_{- 0.54}$& $ 0.82$  	&$2.09^{+ 0.22}_{- 0.21}$& $ 2.26^{+ 0.34}_{- 0.31}$& $ 45.37^{+ 1.52}_{- 1.64}$& $ 0.83$  	&$2.44^{+ 0.11}_{- 0.10}$& $ 4.24^{+ 0.36}_{- 0.30}$& $ 46.43^{+ 2.52}_{- 2.33}$& $ 0.90$\\
&325000&$2.23^{+ 0.08}_{- 0.08}$& $ 4.41^{+ 0.12}_{- 0.12}$& $ 23.02^{+ 0.28}_{- 0.28}$& $ 1.05$  	&$2.09^{+ 0.11}_{- 0.10}$& $ 2.21^{+ 0.21}_{- 0.20}$& $ 42.37^{+ 0.77}_{- 0.78}$& $ 1.08$  	&$2.36^{+ 0.05}_{- 0.05}$& $ 4.04^{+ 0.21}_{- 0.17}$& $ 41.96^{+ 1.19}_{- 1.16}$& $ 1.11$\\
&335000&$2.20^{+ 0.06}_{- 0.06}$& $ 3.96^{+ 0.10}_{- 0.10}$& $ 24.43^{+ 0.25}_{- 0.25}$& $ 0.86$  	&$2.11^{+ 0.08}_{- 0.08}$& $ 1.70^{+ 0.16}_{- 0.16}$& $ 43.59^{+ 0.64}_{- 0.64}$& $ 0.90$  	&$2.34^{+ 0.04}_{- 0.04}$& $ 4.58^{+ 0.26}_{- 0.22}$& $ 43.44^{+ 0.89}_{- 0.88}$& $ 0.89$\\
&345000&$2.21^{+ 0.06}_{- 0.06}$& $ 4.27^{+ 0.11}_{- 0.10}$& $ 25.30^{+ 0.24}_{- 0.24}$& $ 0.95$  	&$2.06^{+ 0.08}_{- 0.08}$& $ 2.08^{+ 0.17}_{- 0.17}$& $ 45.95^{+ 0.65}_{- 0.65}$& $ 0.98$  	&$2.34^{+ 0.04}_{- 0.04}$& $ 4.16^{+ 0.17}_{- 0.17}$& $ 45.35^{+ 0.92}_{- 0.91}$& $ 1.00$\\
&355000&$2.78^{+ 0.28}_{- 0.26}$& $ 4.47^{+ 0.15}_{- 0.15}$& $ 21.74^{+ 1.10}_{- 1.05}$& $ 0.72$  	&$2.69^{+ 0.34}_{- 0.30}$& $ 1.69^{+ 0.47}_{- 0.41}$& $ 44.59^{+ 1.89}_{- 1.95}$& $ 0.73$  	&$2.89^{+ 0.21}_{- 0.20}$& $ 5.50^{+ 1.30}_{- 0.81}$& $ 49.09^{+ 2.53}_{- 2.27}$& $ 0.82$\\
&365000&$2.42^{+ 0.07}_{- 0.07}$& $ 4.62^{+ 0.12}_{- 0.11}$& $ 23.45^{+ 0.27}_{- 0.27}$& $ 0.93$  	&$2.28^{+ 0.09}_{- 0.09}$& $ 2.23^{+ 0.19}_{- 0.19}$& $ 45.25^{+ 0.70}_{- 0.70}$& $ 0.95$  	&$2.46^{+ 0.05}_{- 0.05}$& $ 4.05^{+ 0.20}_{- 0.17}$& $ 45.88^{+ 1.08}_{- 1.06}$& $ 1.00$\\
&375000&$2.26^{+ 0.07}_{- 0.07}$& $ 4.89^{+ 0.13}_{- 0.12}$& $ 24.41^{+ 0.25}_{- 0.25}$& $ 0.83$  	&$2.02^{+ 0.10}_{- 0.09}$& $ 2.79^{+ 0.21}_{- 0.20}$& $ 46.24^{+ 0.73}_{- 0.73}$& $ 0.87$  	&$2.36^{+ 0.04}_{- 0.04}$& $ 3.66^{+ 0.14}_{- 0.13}$& $ 45.33^{+ 1.12}_{- 1.10}$& $ 0.88$\\
&385000&$2.24^{+ 0.13}_{- 0.12}$& $ 4.81^{+ 0.15}_{- 0.14}$& $ 23.83^{+ 0.39}_{- 0.40}$& $ 0.86$  	&$2.02^{+ 0.17}_{- 0.16}$& $ 2.72^{+ 0.30}_{- 0.28}$& $ 44.86^{+ 1.14}_{- 1.19}$& $ 0.86$  	&$2.37^{+ 0.08}_{- 0.07}$& $ 3.74^{+ 0.22}_{- 0.19}$& $ 44.53^{+ 1.93}_{- 1.85}$& $ 0.95$\\
&395000&$2.32^{+ 0.19}_{- 0.19}$& $ 4.84^{+ 0.16}_{- 0.15}$& $ 22.38^{+ 0.58}_{- 0.57}$& $ 0.90$  	&$2.09^{+ 0.26}_{- 0.24}$& $ 2.69^{+ 0.41}_{- 0.37}$& $ 42.80^{+ 1.55}_{- 1.69}$& $ 0.89$  	&$2.49^{+ 0.12}_{- 0.11}$& $ 3.95^{+ 0.34}_{- 0.28}$& $ 44.36^{+ 2.62}_{- 2.42}$& $ 1.02$\\
\hline
     
     \end{tabular}
     }
     \label{tab:t2}      
\end{table*}

\begin{table*}

     \centering
     \caption{The best fit parameter values obtained by fitting  time-resolved broadband X-ray spectra of S3 observation with the snchrotron convolved BPL/LP/$\xi-{max}$ models.}
     \scalebox{0.87}{ % Scale the table to 80% of its original size 
     \begin{tabular}{rlllllllllllll} \hline
     && &Broken&&&&Log&&&&$\xi-{max}$\\
      &&&Power law&&&&parabola&&&&model&\\
      \hline
       &Time&$\Gamma_{1}$&$\Gamma_{2}$&norm (n)&$\frac{\chi^{2}}{dof}$&$\alpha$&$\beta$&norm (n)&$\frac{\chi^{2}}{dof}$&p&$\xi_{max}$&{norm} (n)&$\frac{\chi^{2}}{dof}$\\

&&&&$(\times 10^{-2})$&&&&$(\times 10^{-2})$&&&&$(\times 10^{-2})$&\\
\hline
&5000&$2.58^{+ 0.07}_{- 0.07}$& $ 4.94^{+ 0.12}_{- 0.11}$& $ 37.26^{+ 0.55}_{- 0.55}$& $ 1.16$	&$2.69^{+ 0.08}_{- 0.08}$& $ 2.30^{+ 0.19}_{- 0.18}$& $ 76.88^{+ 1.11}_{- 1.09}$& $ 1.23$	&$2.65^{+ 0.05}_{- 0.05}$& $ 3.90^{+ 0.18}_{- 0.16}$& $ 81.75^{+ 1.58}_{- 1.63}$& $ 1.22$\\
&15000&$2.75^{+ 0.08}_{- 0.08}$& $ 5.28^{+ 0.13}_{- 0.12}$& $ 32.50^{+ 0.59}_{- 0.59}$& $ 0.78$	&$2.85^{+ 0.10}_{- 0.09}$& $ 2.52^{+ 0.21}_{- 0.20}$& $ 70.64^{+ 1.11}_{- 1.09}$& $ 0.82$	&$2.75^{+ 0.06}_{- 0.06}$& $ 3.78^{+ 0.18}_{- 0.16}$& $ 77.20^{+ 1.59}_{- 1.63}$& $ 0.85$\\
&25000&$3.04^{+ 0.09}_{- 0.09}$& $ 5.19^{+ 0.14}_{- 0.14}$& $ 27.62^{+ 0.63}_{- 0.63}$& $ 1.04$	&$3.11^{+ 0.10}_{- 0.10}$& $ 2.13^{+ 0.22}_{- 0.21}$& $ 61.53^{+ 1.12}_{- 1.11}$& $ 1.05$	&$2.89^{+ 0.08}_{- 0.07}$& $ 4.08^{+ 0.26}_{- 0.22}$&  $70.21^{+ 1.47}_{- 1.47}$& $ 1.13$\\
&35000&$2.95^{+ 0.05}_{- 0.05}$& $ 5.37^{+ 0.14}_{- 0.14}$& $ 27.24^{+ 0.36}_{- 0.36}$& $ 1.03$	&$3.02^{+ 0.06}_{- 0.06}$& $ 2.33^{+ 0.18}_{- 0.17}$& $ 60.03^{+ 0.82}_{- 0.80}$& $ 1.11$	&$2.79^{+ 0.04}_{- 0.04}$& $ 3.74^{+ 0.16}_{- 0.15}$& $ 67.54^{+ 0.99}_{- 0.98}$& $ 1.08$\\
&45000&$3.11^{+ 0.04}_{- 0.04}$& $ 5.03^{+ 0.12}_{- 0.12}$& $ 23.11^{+ 0.29}_{- 0.29}$& $ 1.04$	&$3.19^{+ 0.05}_{- 0.04}$& $ 1.77^{+ 0.15}_{- 0.14}$& $ 51.27^{+ 0.65}_{- 0.63}$& $ 1.14$	&$2.93^{+ 0.04}_{- 0.04}$& $ 4.31^{+ 0.23}_{- 0.20}$& $ 59.42^{+ 0.81}_{- 0.80}$& $ 1.11$\\
&55000&$3.19^{+ 0.07}_{- 0.07}$& $ 5.06^{+ 0.15}_{- 0.14}$& $ 19.93^{+ 0.36}_{- 0.36}$& $ 1.06$	&$3.29^{+ 0.07}_{- 0.07}$& $ 1.71^{+ 0.19}_{- 0.18}$& $ 44.83^{+ 0.74}_{- 0.72}$& $ 1.12$	&$2.99^{+ 0.06}_{- 0.06}$& $ 4.38^{+ 0.31}_{- 0.26}$& $ 52.77^{+ 0.96}_{- 0.95}$& $ 1.10$\\
&65000&$3.19^{+ 0.05}_{- 0.05}$& $ 5.06^{+ 0.12}_{- 0.12}$& $ 19.86^{+ 0.26}_{- 0.26}$& $ 0.89$	&$3.26^{+ 0.05}_{- 0.05}$& $ 1.76^{+ 0.15}_{- 0.15}$& $ 44.91^{+ 0.57}_{- 0.56}$& $ 0.95$	&$3.01^{+ 0.04}_{- 0.04}$& $ 4.47^{+ 0.25}_{- 0.22}$& $ 52.55^{+ 0.76}_{- 0.75}$& $ 0.94$\\
&75000&$3.17^{+ 0.06}_{- 0.06}$& $ 4.72^{+ 0.12}_{- 0.12}$& $ 20.24^{+ 0.34}_{- 0.34}$& $ 1.04$	&$3.24^{+ 0.06}_{- 0.06}$& $ 1.45^{+ 0.16}_{- 0.16}$& $ 44.76^{+ 0.67}_{- 0.66}$& $ 1.07$	&$3.02^{+ 0.06}_{- 0.06}$& $ 5.01^{+ 0.39}_{- 0.33}$& $ 51.73^{+ 0.89}_{- 0.89}$& $ 1.09$\\
&85000&$2.99^{+ 0.07}_{- 0.07}$& $ 4.32^{+ 0.10}_{- 0.10}$& $ 22.58^{+ 0.37}_{- 0.37}$& $ 1.01$	&$3.02^{+ 0.07}_{- 0.07}$& $ 1.28^{+ 0.15}_{- 0.15}$& $ 47.46^{+ 0.69}_{- 0.68}$& $ 1.04$	&$2.91^{+ 0.06}_{- 0.06}$& $ 5.51^{+ 0.48}_{- 0.39}$& $ 52.73^{+ 0.89}_{- 0.87}$& $ 1.04$\\
\hline
     
     \end{tabular}
     }
     \label{tab:t3}      
\end{table*}

\begin{figure}
    \centering
    \begin{subfigure}{1.15\linewidth}
    \includegraphics[width=1\textwidth]{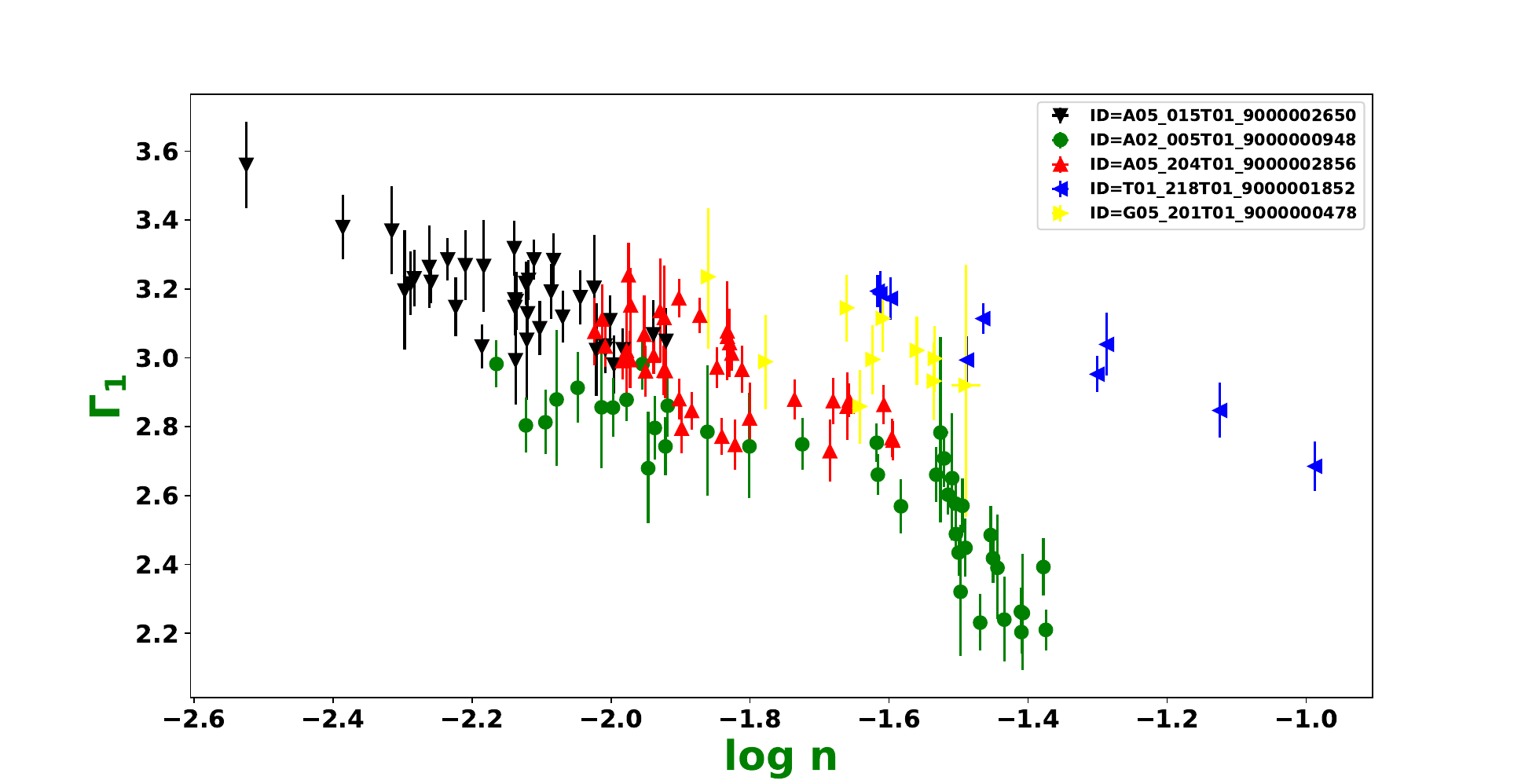}
    \end{subfigure}
    \begin{subfigure}{1.15\linewidth}
    \includegraphics[width=1\textwidth]{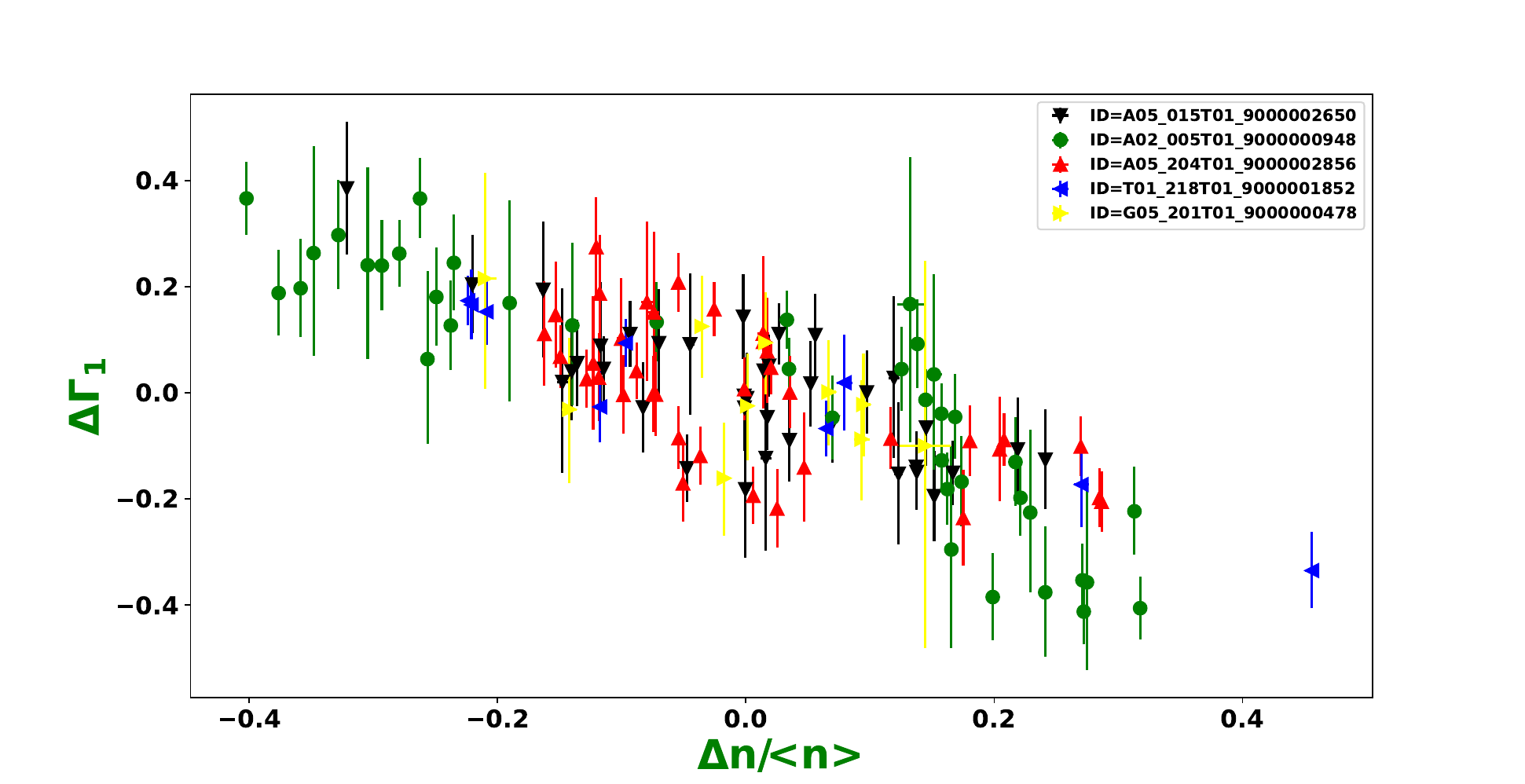}
    \end{subfigure}
    
    \caption{ Variation of  spectral index $\Gamma_{1}$ and  particle density $<n>$ in BPL model. The upper pannel represents relationship between  $\Gamma_{1}$ and $log\hspace{0.05cm}n$. The lower pannel represents  correlation  bewteen index ($\Delta\Gamma_{1}$) and the normalized particle density ($\Delta$$n$/$<n>$) at $\xi=\xi_{ref}$. }
    \label{fig:com-bkn}
\end{figure}
 
 \begin{table*}

     \centering
     \caption{The best fit parameter values obtained by fitting  time-resolved broadband X-ray spectra of S4 observation with the snchrotron convolved BPL/LP/$\xi-{max}$ models.} 
     \scalebox{0.87}{ % Scale the table to 80% of its original size
     \begin{tabular}{rlllllllllllll} \hline
     && &Broken&&&&Log&&&&$\xi-{max}$\\
      &&&Power law&&&&parabola&&&&model&\\
      \hline
       &Time&$\Gamma_{1}$&$\Gamma_{2}$&norm (n)&$\frac{\chi^{2}}{dof}$&$\alpha$&$\beta$&norm (n)&$\frac{\chi^{2}}{dof}$&p&$\xi_{max}$&{norm} (n)&$\frac{\chi^{2}}{dof}$\\

&&&&$(\times 10^{-2})$&&&&$(\times 10^{-2})$&&&&$(\times 10^{-2})$&\\
\hline
&5000&$3.35^{+ 0.08}_{- 0.08}$& $ 5.02^{+ 0.21}_{- 0.23}$& $ 11.79^{+ 0.22}_{- 0.22}$& $ 0.98$  	&$3.04^{+ 0.1}_{- 0.1}$& $ 1.6^{+ 0.23}_{- 0.22}$& $ 26.55^{+ 0.49}_{- 0.48}$& $ 0.99$  	&$2.92^{+ 0.08}_{- 0.08}$& $ 4.87^{+ 0.5}_{- 0.39}$& $ 29.99^{+ 0.65}_{- 0.66}$& $ 1.00$\\
&15000&$3.25^{+ 0.09}_{- 0.08}$& $ 4.89^{+ 0.22}_{- 0.24}$& $ 10.82^{+ 0.21}_{- 0.21}$& $ 0.94$  	&$3.09^{+ 0.1}_{- 0.1}$& $ 1.41^{+ 0.23}_{- 0.22}$& $ 24.04^{+ 0.46}_{- 0.45}$& $ 0.94$  	&$2.94^{+ 0.09}_{- 0.09}$& $ 5.08^{+ 0.61}_{- 0.47}$& $ 27.30^{+ 0.62}_{- 0.63}$& $ 0.95$\\
&25000&$3.03^{+ 0.06}_{- 0.06}$& $ 5.09^{+ 0.19}_{- 0.21}$& $ 11.24^{+ 0.15}_{- 0.15}$& $ 1.03$  	&$2.94^{+ 0.08}_{- 0.08}$& $ 1.72^{+ 0.2}_{- 0.19}$& $ 24.74^{+ 0.38}_{- 0.37}$& $ 1.04$  	&$2.82^{+ 0.06}_{- 0.06}$& $ 4.54^{+ 0.34}_{- 0.28}$& $ 27.59^{+ 0.5}_{- 0.51}$& $ 1.04$\\
&35000&$3.22^{+ 0.06}_{- 0.06}$& $ 4.42^{+ 0.15}_{- 0.15}$& $ 10.45^{+ 0.15}_{- 0.15}$& $ 1.10$  	&$3.17^{+ 0.07}_{- 0.07}$& $ 1^{+ 0.16}_{- 0.15}$& $ 23.01^{+ 0.34}_{- 0.33}$& $ 1.11$  	&$3.05^{+ 0.07}_{- 0.07}$& $ 6.55^{+ 0.86}_{- 0.65}$& $ 25.86^{+ 0.5}_{- 0.5}$& $ 1.12$\\
&45000&$3.22^{+ 0.09}_{- 0.09}$& $ 4.42^{+ 0.18}_{- 0.19}$& $ 10.14^{+ 0.21}_{- 0.21}$& $ 1.04$  	&$3.17^{+ 0.11}_{- 0.1}$& $ 1.01^{+ 0.21}_{- 0.2}$& $ 22.31^{+ 0.43}_{- 0.43}$& $ 1.03$  	&$3.06^{+ 0.1}_{- 0.1}$& $ 6.59^{+ 1.21}_{- 0.84}$& $ 25.07^{+ 0.63}_{- 0.62}$& $ 1.06$\\
&55000&$3.38^{+ 0.09}_{- 0.09}$& $ 4.56^{+ 0.19}_{- 0.20}$& $ 9.20^{+ 0.21}_{- 0.21}$& $ 1.16$  	&$3.3^{+ 0.11}_{- 0.1}$& $ 1.05^{+ 0.22}_{- 0.2}$& $ 21.16^{+ 0.42}_{- 0.42}$& $ 1.13$  	&$3.2^{+ 0.1}_{- 0.11}$& $ 6.86^{+ 1.36}_{- 0.92}$& $ 23.97^{+ 0.69}_{- 0.66}$& $ 1.17$\\
&65000&$3.56^{+ 0.13}_{- 0.12}$& $ 5.21^{+ 0.29}_{- 0.32}$& $ 8.01^{+ 0.28}_{- 0.28}$& $ 1.06$  	&$3.51^{+ 0.14}_{- 0.14}$& $ 1.32^{+ 0.3}_{- 0.28}$& $ 19.55^{+ 0.55}_{- 0.54}$& $ 1.04$  	&$3.3^{+ 0.14}_{- 0.15}$& $ 5.7^{+ 1.15}_{- 0.78}$& $ 23.28^{+ 0.88}_{- 0.82}$& $ 1.07$\\
&75000&$3.20^{+ 0.18}_{- 0.17}$& $ 6.26^{+ 0.48}_{- 0.57}$& $ 10.06^{+ 0.42}_{- 0.41}$& $ 0.95$  	&$3.02^{+ 0.24}_{- 0.22}$& $ 2.46^{+ 0.5}_{- 0.45}$& $ 24.08^{+ 0.81}_{- 0.79}$& $ 0.93$  	&$2.88^{+ 0.16}_{- 0.16}$& $ 3.83^{+ 0.5}_{- 0.38}$& $ 27.64^{+ 1.23}_{- 1.15}$& $ 0.96$\\
&85000&$2.99^{+ 0.13}_{- 0.13}$& $ 5.53^{+ 0.30}_{- 0.33}$& $ 11.80^{+ 0.31}_{- 0.30}$& $ 0.94$  	&$2.84^{+ 0.17}_{- 0.16}$& $ 2.16^{+ 0.35}_{- 0.33}$& $ 26.33^{+ 0.65}_{- 0.64}$& $ 0.93$  	&$2.78^{+ 0.11}_{- 0.12}$& $ 4.1^{+ 0.43}_{- 0.34}$& $ 29.23^{+ 1.04}_{- 0.98}$& $ 0.97$\\
&95000&$3.17^{+ 0.07}_{- 0.07}$& $ 5.39^{+ 0.24}_{- 0.27}$& $ 11.78^{+ 0.19}_{- 0.19}$& $ 0.92$  	&$3.07^{+ 0.09}_{- 0.08}$& $ 1.81^{+ 0.23}_{- 0.22}$& $ 27.00^{+ 0.47}_{- 0.46}$& $ 0.93$  	&$2.91^{+ 0.07}_{- 0.07}$& $ 4.43^{+ 0.37}_{- 0.31}$& $ 30.84^{+ 0.61}_{- 0.62}$& $ 0.94$\\
&105000&$3.13^{+ 0.06}_{- 0.06}$& $ 5.82^{+ 0.26}_{- 0.29}$& $ 12.00^{+ 0.17}_{- 0.17}$& $ 1.21$  	&$3.03^{+ 0.08}_{- 0.08}$& $ 2.07^{+ 0.22}_{- 0.21}$& $ 27.64^{+ 0.44}_{- 0.43}$& $ 1.30$  	&$2.85^{+ 0.06}_{- 0.06}$& $ 4.01^{+ 0.26}_{- 0.22}$& $ 31.51^{+ 0.57}_{- 0.58}$& $ 1.23$\\
&115000&$3.16^{+ 0.09}_{- 0.08}$& $ 6.32^{+ 0.36}_{- 0.41}$& $ 11.81^{+ 0.23}_{- 0.23}$& $ 1.05$  	&$3.04^{+ 0.11}_{- 0.11}$& $ 2.43^{+ 0.3}_{- 0.28}$& $ 27.97^{+ 0.56}_{- 0.55}$& $ 1.04$  	&$2.85^{+ 0.08}_{- 0.08}$& $ 3.75^{+ 0.28}_{- 0.23}$& $ 32.07^{+ 0.77}_{- 0.77}$& $ 1.06$\\
&125000&$3.29^{+ 0.06}_{- 0.06}$& $ 5.76^{+ 0.26}_{- 0.28}$& $ 10.70^{+ 0.16}_{- 0.16}$& $ 1.11$  	&$3.21^{+ 0.08}_{- 0.07}$& $ 1.94^{+ 0.22}_{- 0.21}$& $ 25.40^{+ 0.42}_{- 0.41}$& $ 1.09$  	&$2.99^{+ 0.06}_{- 0.07}$& $ 4.28^{+ 0.32}_{- 0.27}$& $ 29.73^{+ 0.55}_{- 0.56}$& $ 1.12$\\
&135000&$3.23^{+ 0.08}_{- 0.08}$& $ 5.61^{+ 0.27}_{- 0.29}$& $ 10.20^{+ 0.20}_{- 0.19}$& $ 0.97$  	&$3.14^{+ 0.1}_{- 0.1}$& $ 1.92^{+ 0.25}_{- 0.24}$& $ 23.85^{+ 0.45}_{- 0.44}$& $ 0.97$  	&$2.96^{+ 0.08}_{- 0.08}$& $ 4.35^{+ 0.38}_{- 0.31}$& $ 27.59^{+ 0.6}_{- 0.61}$& $ 0.99$\\
&145000&$3.26^{+ 0.12}_{- 0.12}$& $ 5.74^{+ 0.34}_{- 0.39}$& $ 10.42^{+ 0.30}_{- 0.29}$& $ 1.09$  	&$3.15^{+ 0.15}_{- 0.14}$& $ 2.01^{+ 0.34}_{- 0.32}$& $ 24.69^{+ 0.62}_{- 0.60}$& $ 1.06$  	&$2.99^{+ 0.12}_{- 0.12}$& $ 4.32^{+ 0.52}_{- 0.4}$& $ 28.72^{+ 0.82}_{- 0.84}$& $ 1.11$\\
&155000&$3.37^{+ 0.13}_{- 0.13}$& $ 6.17^{+ 0.38}_{- 0.44}$& $ 9.87^{+ 0.32}_{- 0.32}$& $ 1.09$  	&$3.25^{+ 0.16}_{- 0.15}$& $ 2.18^{+ 0.36}_{- 0.33}$& $ 24.30^{+ 0.65}_{- 0.64}$& $ 1.08$  	&$3.02^{+ 0.12}_{- 0.13}$& $ 4.06^{+ 0.45}_{- 0.35}$& $ 28.90^{+ 0.86}_{- 0.87}$& $ 1.10$\\
&165000&$3.27^{+ 0.13}_{- 0.13}$& $ 6.33^{+ 0.41}_{- 0.47}$& $ 11.27^{+ 0.37}_{- 0.36}$& $ 1.02$  	&$3.12^{+ 0.18}_{- 0.17}$& $ 2.41^{+ 0.39}_{- 0.36}$& $ 27.35^{+ 0.73}_{- 0.72}$& $ 0.98$  	&$2.95^{+ 0.12}_{- 0.13}$& $ 3.89^{+ 0.41}_{- 0.32}$& $ 31.91^{+ 1.03}_{- 1.01}$& $ 1.03$\\
&175000&$3.12^{+ 0.08}_{- 0.07}$& $ 5.75^{+ 0.28}_{- 0.32}$& $ 12.62^{+ 0.21}_{- 0.21}$& $ 1.00$  	&$2.99^{+ 0.1}_{- 0.09}$& $ 2.16^{+ 0.26}_{- 0.24}$& $ 29.12^{+ 0.53}_{- 0.52}$& $ 0.96$  	&$2.85^{+ 0.07}_{- 0.07}$& $ 4.06^{+ 0.31}_{- 0.26}$& $ 33.00^{+ 0.7}_{- 0.71}$& $ 1.02$\\
&185000&$2.98^{+ 0.09}_{- 0.08}$& $ 5.94^{+ 0.32}_{- 0.37}$& $ 13.59^{+ 0.25}_{- 0.24}$& $ 1.28$  	&$2.78^{+ 0.12}_{- 0.11}$& $ 2.5^{+ 0.3}_{- 0.29}$& $ 30.87^{+ 0.62}_{- 0.60}$& $ 1.26$  	&$2.73^{+ 0.07}_{- 0.08}$& $ 3.76^{+ 0.28}_{- 0.23}$& $ 33.88^{+ 0.87}_{- 0.85}$& $ 1.30$\\
&195000&$3.02^{+ 0.06}_{- 0.06}$& $ 6.35^{+ 0.30}_{- 0.33}$& $ 13.77^{+ 0.18}_{- 0.18}$& $ 1.20$  	&$2.88^{+ 0.08}_{- 0.08}$& $ 2.59^{+ 0.24}_{- 0.23}$& $ 31.75^{+ 0.50}_{- 0.49}$& $ 1.23$  	&$2.74^{+ 0.05}_{- 0.06}$& $ 3.61^{+ 0.19}_{- 0.17}$& $ 35.29^{+ 0.67}_{- 0.67}$& $ 1.22$\\
&205000&$3.11^{+ 0.07}_{- 0.07}$& $ 5.45^{+ 0.21}_{- 0.22}$& $ 13.51^{+ 0.21}_{- 0.21}$& $ 1.14$  	&$2.99^{+ 0.09}_{- 0.09}$& $ 1.98^{+ 0.22}_{- 0.21}$& $ 30.79^{+ 0.50}_{- 0.49}$& $ 1.09$  	&$2.87^{+ 0.07}_{- 0.07}$& $ 4.34^{+ 0.31}_{- 0.26}$& $ 34.74^{+ 0.68}_{- 0.68}$& $ 1.17$\\
&215000&$3.32^{+ 0.08}_{- 0.08}$& $ 4.95^{+ 0.21}_{- 0.22}$& $ 11.78^{+ 0.23}_{- 0.23}$& $ 1.22$  	&$3.26^{+ 0.09}_{- 0.09}$& $ 1.35^{+ 0.21}_{- 0.2}$& $ 27.15^{+ 0.50}_{- 0.49}$& $ 1.20$  	&$3.1^{+ 0.09}_{- 0.09}$& $ 5.47^{+ 0.68}_{- 0.52}$& $ 31.39^{+ 0.72}_{- 0.72}$& $ 1.24$\\
&225000&$3.09^{+ 0.08}_{- 0.08}$& $ 5.19^{+ 0.20}_{- 0.22}$& $ 12.21^{+ 0.21}_{- 0.20}$& $ 1.01$  	&$3^{+ 0.1}_{- 0.09}$& $ 1.75^{+ 0.22}_{- 0.21}$& $ 27.29^{+ 0.46}_{- 0.45}$& $ 1.02$  	&$2.86^{+ 0.07}_{- 0.08}$& $ 4.53^{+ 0.37}_{- 0.31}$& $ 30.73^{+ 0.65}_{- 0.65}$& $ 1.03$\\
&235000&$3.27^{+ 0.10}_{- 0.10}$& $ 5.86^{+ 0.30}_{- 0.33}$& $ 10.97^{+ 0.27}_{- 0.26}$& $ 1.24$  	&$3.13^{+ 0.13}_{- 0.12}$& $ 2.14^{+ 0.29}_{- 0.28}$& $ 26.30^{+ 0.56}_{- 0.55}$& $ 1.21$  	&$2.97^{+ 0.1}_{- 0.1}$& $ 4.19^{+ 0.39}_{- 0.31}$& $ 30.48^{+ 0.76}_{- 0.77}$& $ 1.26$\\
&245000&$3.05^{+ 0.18}_{- 0.17}$& $ 6.37^{+ 0.44}_{- 0.52}$& $ 11.99^{+ 0.45}_{- 0.44}$& $ 0.94$  	&$2.84^{+ 0.25}_{- 0.23}$& $ 2.71^{+ 0.5}_{- 0.45}$& $ 27.93^{+ 0.86}_{- 0.84}$& $ 0.93$  	&$2.78^{+ 0.15}_{- 0.16}$& $ 3.64^{+ 0.4}_{- 0.31}$& $ 31.25^{+ 1.61}_{- 1.43}$& $ 0.96$\\
&255000&$3.02^{+ 0.14}_{- 0.13}$& $ 5.52^{+ 0.28}_{- 0.31}$& $ 13.25^{+ 0.37}_{- 0.36}$& $ 0.93$  	&$2.88^{+ 0.18}_{- 0.17}$& $ 2.12^{+ 0.34}_{- 0.32}$& $ 29.72^{+ 0.73}_{- 0.72}$& $ 0.91$  	&$2.82^{+ 0.12}_{- 0.12}$& $ 4.2^{+ 0.45}_{- 0.35}$& $ 33.20^{+ 1.16}_{- 1.09}$& $ 0.96$\\
&265000&$3.03^{+ 0.07}_{- 0.07}$& $ 5.10^{+ 0.20}_{- 0.22}$& $ 13.42^{+ 0.20}_{- 0.20}$& $ 1.24$  	&$2.91^{+ 0.09}_{- 0.09}$& $ 1.76^{+ 0.22}_{- 0.2}$& $ 29.58^{+ 0.49}_{- 0.48}$& $ 1.25$  	&$2.83^{+ 0.07}_{- 0.07}$& $ 4.57^{+ 0.37}_{- 0.31}$& $ 32.90^{+ 0.65}_{- 0.66}$& $ 1.26$\\
&285000&$3.19^{+ 0.08}_{- 0.08}$& $ 5.62^{+ 0.27}_{- 0.30}$& $ 12.42^{+ 0.23}_{- 0.23}$& $ 1.00$  	&$3.09^{+ 0.1}_{- 0.1}$& $ 1.98^{+ 0.26}_{- 0.24}$& $ 28.87^{+ 0.55}_{- 0.53}$& $ 0.98$  	&$2.92^{+ 0.08}_{- 0.08}$& $ 4.26^{+ 0.37}_{- 0.3}$& $ 33.17^{+ 0.72}_{- 0.74}$& $ 1.02$\\
&295000&$3.23^{+ 0.06}_{- 0.06}$& $ 5.60^{+ 0.23}_{- 0.25}$& $ 12.02^{+ 0.17}_{- 0.17}$& $ 1.15$  	&$3.14^{+ 0.07}_{- 0.07}$& $ 1.91^{+ 0.2}_{- 0.19}$& $ 28.06^{+ 0.43}_{- 0.42}$& $ 1.13$  	&$2.95^{+ 0.06}_{- 0.06}$& $ 4.34^{+ 0.3}_{- 0.26}$& $ 32.44^{+ 0.56}_{- 0.57}$& $ 1.17$\\
&305000&$3.04^{+ 0.07}_{- 0.07}$& $ 5.72^{+ 0.23}_{- 0.25}$& $ 13.42^{+ 0.19}_{- 0.19}$& $ 1.16$  	&$2.87^{+ 0.09}_{- 0.09}$& $ 2.3^{+ 0.23}_{- 0.22}$& $ 30.57^{+ 0.48}_{- 0.47}$& $ 1.08$  	&$2.79^{+ 0.06}_{- 0.06}$& $ 4^{+ 0.24}_{- 0.21}$& $ 33.95^{+ 0.67}_{- 0.66}$& $ 1.19$\\
&315000&$3.07^{+ 0.10}_{- 0.10}$& $ 5.70^{+ 0.27}_{- 0.29}$& $ 14.38^{+ 0.30}_{- 0.30}$& $ 1.14$  	&$2.96^{+ 0.12}_{- 0.12}$& $ 2.28^{+ 0.24}_{- 0.23}$& $ 33.06^{+ 0.54}_{- 0.54}$& $ 1.37$  	&$2.82^{+ 0.09}_{- 0.09}$& $ 4.05^{+ 0.32}_{- 0.27}$& $ 36.83^{+ 0.99}_{- 0.96}$& $ 1.16$\\
&325000&$3.05^{+ 0.09}_{- 0.09}$& $ 5.83^{+ 0.25}_{- 0.27}$& $ 14.65^{+ 0.29}_{- 0.29}$& $ 1.16$  	&$2.86^{+ 0.13}_{- 0.12}$& $ 2.4^{+ 0.27}_{- 0.26}$& $ 33.63^{+ 0.63}_{- 0.62}$& $ 1.11$  	&$2.8^{+ 0.08}_{- 0.08}$& $ 3.95^{+ 0.27}_{- 0.23}$& $ 37.44^{+ 0.99}_{- 0.96}$& $ 1.19$\\
&335000&$3.20^{+ 0.15}_{- 0.15}$& $ 5.80^{+ 0.31}_{- 0.34}$& $ 13.21^{+ 0.48}_{- 0.47}$& $ 1.27$  	&$2.86^{+ 0.22}_{- 0.2}$& $ 2.5^{+ 0.41}_{- 0.38}$& $ 31.90^{+ 0.88}_{- 0.87}$& $ 1.20$  	&$2.96^{+ 0.14}_{- 0.14}$& $ 4.23^{+ 0.5}_{- 0.39}$& $ 35.81^{+ 1.24}_{- 1.18}$& $ 1.30$\\
&345000&$3.28^{+ 0.08}_{- 0.08}$& $ 5.95^{+ 0.31}_{- 0.35}$& $ 12.46^{+ 0.24}_{- 0.24}$& $ 0.91$  	&$3.22^{+ 0.1}_{- 0.09}$& $ 2.02^{+ 0.26}_{- 0.24}$& $ 29.69^{+ 0.58}_{- 0.56}$& $ 0.92$  	&$2.97^{+ 0.08}_{- 0.08}$& $ 4.12^{+ 0.34}_{- 0.28}$& $ 34.93^{+ 0.76}_{- 0.77}$& $ 0.92$\\
&355000&$3.22^{+ 0.06}_{- 0.06}$& $ 5.37^{+ 0.24}_{- 0.26}$& $ 11.97^{+ 0.18}_{- 0.18}$& $ 1.17$  	&$3.16^{+ 0.07}_{- 0.07}$& $ 1.67^{+ 0.21}_{- 0.2}$& $ 27.44^{+ 0.45}_{- 0.44}$& $ 1.18$  	&$2.96^{+ 0.06}_{- 0.07}$& $ 4.56^{+ 0.39}_{- 0.32}$& $ 31.80^{+ 0.6}_{- 0.61}$& $ 1.19$\\
&365000&$3.18^{+ 0.08}_{- 0.08}$& $ 5.34^{+ 0.25}_{- 0.27}$& $ 12.95^{+ 0.23}_{- 0.23}$& $ 1.04$  	&$3.08^{+ 0.1}_{- 0.09}$& $ 1.78^{+ 0.24}_{- 0.23}$& $ 29.63^{+ 0.55}_{- 0.54}$& $ 1.01$  	&$2.94^{+ 0.08}_{- 0.08}$& $ 4.55^{+ 0.43}_{- 0.35}$& $ 33.87^{+ 0.72}_{- 0.74}$& $ 1.06$\\
&375000&$3.29^{+ 0.06}_{- 0.06}$& $ 5.54^{+ 0.22}_{- 0.24}$& $ 12.11^{+ 0.17}_{- 0.17}$& $ 1.08$  	&$3.22^{+ 0.07}_{- 0.07}$& $ 1.79^{+ 0.2}_{- 0.19}$& $ 28.50^{+ 0.43}_{- 0.43}$& $ 1.07$  	&$3.00^{+ 0.06}_{- 0.06}$& $ 4.5^{+ 0.33}_{- 0.28}$& $ 33.26^{+ 0.58}_{- 0.59}$& $ 1.10$\\
\hline
      
     \end{tabular}
     }
     
     \label{tab:t4}   
\end{table*}

\begin{table*}

     \centering
     \caption{The best fit parameter values obtained by fitting  time-resolved broadband X-ray spectra of S5 observation with the snchrotron convolved BPL/LP/$\xi-{max}$ models.} 
     \scalebox{0.87}{ % Scale the table to 80% of its original size
     \begin{tabular}{rlllllllllllll} \hline
     && &Broken&&&&Log&&&&$\xi-{max}$\\
      &&&Power law&&&&parabola&&&&model&\\
      \hline
       &Time&$\Gamma_{1}$&$\Gamma_{2}$&norm (n)&$\frac{\chi^{2}}{dof}$&$\alpha$&$\beta$&norm (n)&$\frac{\chi^{2}}{dof}$&p&$\xi_{max}$&{norm} (n)&$\frac{\chi^{2}}{dof}$\\

&&&&$(\times 10^{-2})$&&&&$(\times 10^{-2})$&&&&$(\times 10^{-2})$&\\
\hline
&5000&$3.12^{+ 0.08}_{- 0.08}$& $ 4.26^{+ 0.17}_{- 0.16}$& $ 13.91^{+ 0.23}_{- 0.23}$& $ 0.97$	&$2.87^{+ 0.10}_{- 0.10}$& $ 1.09^{+ 0.19}_{- 0.18}$& $ 29.27^{+ 0.52}_{- 0.52}$& $ 0.96$	&$2.87^{+ 0.09}_{- 0.08}$& $ 6.44^{+ 0.96}_{- 0.70}$& $ 31.83^{+ 0.67}_{- 0.68}$& $ 0.99$\\
&15000&$2.96^{+ 0.07}_{- 0.07}$& $ 4.08^{+ 0.14}_{- 0.13}$& $ 14.22^{+ 0.20}_{- 0.20}$& $ 1.00$	&$2.88^{+ 0.09}_{- 0.09}$& $ 0.94^{+ 0.16}_{- 0.15}$& $ 29.41^{+ 0.47}_{- 0.46}$& $ 1.00$	&$2.85^{+ 0.08}_{- 0.07}$& $ 6.90^{+ 0.98}_{- 0.72}$& $ 31.83^{+ 0.60}_{- 0.61}$& $ 1.02$\\
&25000&$3.12^{+ 0.15}_{- 0.15}$& $ 4.49^{+ 0.22}_{- 0.20}$& $ 14.61^{+ 0.46}_{- 0.45}$& $ 0.97$	&$2.98^{+ 0.19}_{- 0.18}$& $ 1.19^{+ 0.29}_{- 0.27}$& $ 31.78^{+ 0.86}_{- 0.86}$& $ 0.93$	&$3.00^{+ 0.16}_{- 0.15}$& $ 6.44^{+ 1.50}_{- 0.95}$& $ 35.20^{+ 1.07}_{- 1.08}$& $ 1.01$\\
&35000&$3.14^{+ 0.15}_{- 0.15}$& $ 4.48^{+ 0.21}_{- 0.20}$& $ 14.52^{+ 0.46}_{- 0.45}$& $ 0.97$	&$2.57^{+ 0.11}_{- 0.11}$& $ 1.66^{+ 0.21}_{- 0.19}$& $ 32.16^{+ 0.57}_{- 0.58}$& $ 1.20$	&$2.64^{+ 0.07}_{- 0.07}$& $ 4.85^{+ 0.39}_{- 0.32}$& $ 34.11^{+ 0.89}_{- 0.85}$& $ 1.23$\\
&45000&$2.88^{+ 0.06}_{- 0.06}$& $ 4.63^{+ 0.15}_{- 0.15}$& $ 14.93^{+ 0.17}_{- 0.17}$& $ 1.17$	&$2.75^{+ 0.08}_{- 0.07}$& $ 1.47^{+ 0.16}_{- 0.15}$& $ 31.25^{+ 0.43}_{- 0.43}$& $ 1.15$	&$2.74^{+ 0.06}_{- 0.05}$& $ 5.15^{+ 0.38}_{- 0.32}$& $ 33.75^{+ 0.59}_{- 0.58}$& $ 1.20$\\
&55000&$2.75^{+ 0.07}_{- 0.07}$& $ 4.85^{+ 0.19}_{- 0.18}$& $ 16.18^{+ 0.20}_{- 0.20}$& $ 1.18$	&$2.55^{+ 0.10}_{- 0.10}$& $ 1.83^{+ 0.20}_{- 0.19}$& $ 33.41^{+ 0.55}_{- 0.55}$& $ 1.14$	&$2.63^{+ 0.06}_{- 0.06}$& $ 4.59^{+ 0.33}_{- 0.27}$& $ 35.35^{+ 0.83}_{- 0.81}$& $ 1.22$\\
&65000&$2.77^{+ 0.05}_{- 0.05}$& $ 4.65^{+ 0.15}_{- 0.15}$& $ 15.87^{+ 0.15}_{- 0.16}$& $ 1.11$	&$2.62^{+ 0.07}_{- 0.07}$& $ 1.58^{+ 0.16}_{- 0.15}$& $ 32.58^{+ 0.43}_{- 0.43}$& $ 1.12$	&$2.65^{+ 0.05}_{- 0.05}$& $ 4.88^{+ 0.31}_{- 0.27}$& $ 34.61^{+ 0.59}_{- 0.58}$& $ 1.15$\\
&75000&$2.85^{+ 0.05}_{- 0.05}$& $ 4.75^{+ 0.17}_{- 0.16}$& $ 15.20^{+ 0.16}_{- 0.16}$& $ 1.04$	&$2.71^{+ 0.07}_{- 0.07}$& $ 1.56^{+ 0.16}_{- 0.16}$& $ 31.77^{+ 0.43}_{- 0.42}$& $ 1.06$	&$2.69^{+ 0.05}_{- 0.05}$& $ 4.86^{+ 0.33}_{- 0.28}$& $ 34.13^{+ 0.57}_{- 0.57}$& $ 1.07$\\
&85000&$2.96^{+ 0.08}_{- 0.08}$& $ 5.15^{+ 0.25}_{- 0.23}$& $ 14.64^{+ 0.23}_{- 0.23}$& $ 0.85$	&$2.84^{+ 0.10}_{- 0.10}$& $ 1.69^{+ 0.22}_{- 0.21}$& $ 31.71^{+ 0.56}_{- 0.55}$& $ 0.89$	&$2.75^{+ 0.07}_{- 0.07}$& $ 4.54^{+ 0.38}_{- 0.31}$& $ 34.80^{+ 0.81}_{- 0.79}$& $ 0.86$\\
&95000&$2.79^{+ 0.07}_{- 0.07}$& $ 4.88^{+ 0.20}_{- 0.19}$& $ 14.98^{+ 0.19}_{- 0.19}$& $ 1.10$	&$2.66^{+ 0.10}_{- 0.09}$& $ 1.68^{+ 0.20}_{- 0.19}$& $ 31.10^{+ 0.50}_{- 0.50}$& $ 1.12$	&$2.65^{+ 0.06}_{- 0.06}$& $ 4.62^{+ 0.34}_{- 0.29}$& $ 33.25^{+ 0.75}_{- 0.73}$& $ 1.13$\\
&105000&$3.02^{+ 0.13}_{- 0.12}$& $ 4.45^{+ 0.19}_{- 0.18}$& $ 13.84^{+ 0.35}_{- 0.34}$& $ 0.88$	&$2.87^{+ 0.16}_{- 0.15}$& $ 1.26^{+ 0.25}_{- 0.23}$& $ 29.59^{+ 0.71}_{- 0.70}$& $ 0.86$	&$2.88^{+ 0.12}_{- 0.12}$& $ 5.95^{+ 0.98}_{- 0.69}$& $ 32.40^{+ 0.94}_{- 0.91}$& $ 0.89$\\
&115000&$3.11^{+ 0.10}_{- 0.10}$& $ 4.47^{+ 0.18}_{- 0.17}$& $ 13.36^{+ 0.28}_{- 0.28}$& $ 1.36$	&$2.98^{+ 0.12}_{- 0.12}$& $ 1.19^{+ 0.21}_{- 0.20}$& $ 29.10^{+ 0.58}_{- 0.57}$& $ 1.33$	&$2.97^{+ 0.10}_{- 0.10}$& $ 6.33^{+ 1.00}_{- 0.72}$& $ 32.10^{+ 0.74}_{- 0.75}$& $ 1.39$\\
&125000&$3.08^{+ 0.10}_{- 0.10}$& $ 4.83^{+ 0.24}_{- 0.22}$& $ 13.22^{+ 0.27}_{- 0.27}$& $ 1.02$	&$2.93^{+ 0.13}_{- 0.12}$& $ 1.48^{+ 0.24}_{- 0.23}$& $ 29.06^{+ 0.59}_{- 0.58}$& $ 0.98$	&$2.89^{+ 0.10}_{- 0.09}$& $ 5.30^{+ 0.67}_{- 0.51}$& $ 32.23^{+ 0.79}_{- 0.80}$& $ 1.04$\\
&135000&$3.03^{+ 0.06}_{- 0.06}$& $ 4.51^{+ 0.16}_{- 0.15}$& $ 13.42^{+ 0.16}_{- 0.16}$& $ 1.03$	&$2.94^{+ 0.07}_{- 0.07}$& $ 1.20^{+ 0.16}_{- 0.15}$& $ 28.68^{+ 0.40}_{- 0.39}$& $ 1.02$	&$2.88^{+ 0.06}_{- 0.06}$& $ 5.82^{+ 0.56}_{- 0.45}$& $ 31.59^{+ 0.52}_{- 0.53}$& $ 1.06$\\
&145000&$2.99^{+ 0.06}_{- 0.05}$& $ 4.30^{+ 0.14}_{- 0.13}$& $ 13.76^{+ 0.15}_{- 0.15}$& $ 1.24$	&$2.91^{+ 0.07}_{- 0.07}$& $ 1.07^{+ 0.14}_{- 0.14}$& $ 28.86^{+ 0.38}_{- 0.37}$& $ 1.23$	&$2.86^{+ 0.06}_{- 0.05}$& $ 6.23^{+ 0.63}_{- 0.50}$& $ 31.53^{+ 0.49}_{- 0.50}$& $ 1.27$\\
&155000&$2.96^{+ 0.07}_{- 0.07}$& $ 4.16^{+ 0.15}_{- 0.14}$& $ 14.60^{+ 0.20}_{- 0.20}$& $ 1.08$	&$2.88^{+ 0.09}_{- 0.08}$& $ 0.99^{+ 0.17}_{- 0.16}$& $ 30.29^{+ 0.48}_{- 0.47}$& $ 1.08$	&$2.84^{+ 0.07}_{- 0.07}$& $ 6.57^{+ 0.88}_{- 0.66}$& $ 32.88^{+ 0.62}_{- 0.63}$& $ 1.10$\\
&165000&$3.01^{+ 0.05}_{- 0.05}$& $ 4.27^{+ 0.14}_{- 0.14}$& $ 14.39^{+ 0.16}_{- 0.15}$& $ 1.11$	&$2.94^{+ 0.06}_{- 0.06}$& $ 1.01^{+ 0.14}_{- 0.14}$& $ 30.18^{+ 0.39}_{- 0.39}$& $ 1.12$	&$2.87^{+ 0.06}_{- 0.05}$& $ 6.41^{+ 0.69}_{- 0.54}$& $ 33.01^{+ 0.52}_{- 0.52}$& $ 1.13$\\
&175000&$2.97^{+ 0.06}_{- 0.06}$& $ 4.38^{+ 0.16}_{- 0.15}$& $ 15.76^{+ 0.18}_{- 0.18}$& $ 1.08$	&$2.88^{+ 0.07}_{- 0.07}$& $ 1.16^{+ 0.16}_{- 0.15}$& $ 33.11^{+ 0.47}_{- 0.46}$& $ 1.08$	&$2.83^{+ 0.06}_{- 0.06}$& $ 5.94^{+ 0.60}_{- 0.48}$& $ 36.13^{+ 0.61}_{- 0.61}$& $ 1.11$\\
&185000&$3.04^{+ 0.10}_{- 0.10}$& $ 4.58^{+ 0.19}_{- 0.17}$& $ 16.05^{+ 0.32}_{- 0.31}$& $ 1.02$	&$2.92^{+ 0.12}_{- 0.12}$& $ 1.30^{+ 0.22}_{- 0.20}$& $ 34.61^{+ 0.67}_{- 0.67}$& $ 0.99$	&$2.89^{+ 0.10}_{- 0.09}$& $ 5.72^{+ 0.74}_{- 0.56}$& $ 38.16^{+ 0.90}_{- 0.90}$& $ 1.04$\\
&195000&$3.08^{+ 0.14}_{- 0.14}$& $ 4.25^{+ 0.18}_{- 0.17}$& $ 16.00^{+ 0.46}_{- 0.45}$& $ 1.42$	&$2.89^{+ 0.18}_{- 0.17}$& $ 1.11^{+ 0.27}_{- 0.25}$& $ 34.35^{+ 0.91}_{- 0.91}$& $ 1.38$	&$2.96^{+ 0.15}_{- 0.14}$& $ 6.91^{+ 1.67}_{- 1.05}$& $ 37.57^{+ 1.10}_{- 1.10}$& $ 1.46$\\
&205000&$2.73^{+ 0.09}_{- 0.09}$& $ 4.38^{+ 0.15}_{- 0.14}$& $ 18.54^{+ 0.28}_{- 0.27}$& $ 1.25$	&$2.58^{+ 0.12}_{- 0.11}$& $ 1.43^{+ 0.19}_{- 0.18}$& $ 37.32^{+ 0.70}_{- 0.71}$& $ 1.25$	&$2.64^{+ 0.08}_{- 0.07}$& $ 5.27^{+ 0.49}_{- 0.39}$& $ 39.57^{+ 1.08}_{- 1.02}$& $ 1.30$\\
&215000&$2.88^{+ 0.06}_{- 0.06}$& $ 4.67^{+ 0.15}_{- 0.14}$& $ 17.62^{+ 0.19}_{- 0.19}$& $ 1.22$	&$2.71^{+ 0.08}_{- 0.07}$& $ 1.56^{+ 0.16}_{- 0.16}$& $ 37.07^{+ 0.51}_{- 0.50}$& $ 1.17$	&$2.73^{+ 0.05}_{- 0.05}$& $ 5.09^{+ 0.35}_{- 0.30}$& $ 39.92^{+ 0.69}_{- 0.69}$& $ 1.27$\\
&225000&$2.87^{+ 0.07}_{- 0.07}$& $ 4.68^{+ 0.16}_{- 0.15}$& $ 18.63^{+ 0.23}_{- 0.23}$& $ 1.14$	&$2.74^{+ 0.09}_{- 0.08}$& $ 1.51^{+ 0.17}_{- 0.17}$& $ 39.06^{+ 0.58}_{- 0.58}$& $ 1.15$	&$2.73^{+ 0.06}_{- 0.06}$& $ 5.07^{+ 0.39}_{- 0.33}$& $ 42.17^{+ 0.82}_{- 0.81}$& $ 1.18$\\
&235000&$2.88^{+ 0.05}_{- 0.05}$& $ 4.60^{+ 0.14}_{- 0.14}$& $ 19.06^{+ 0.18}_{- 0.18}$& $ 1.17$	&$2.74^{+ 0.06}_{- 0.06}$& $ 1.45^{+ 0.14}_{- 0.14}$& $ 39.85^{+ 0.49}_{- 0.48}$& $ 1.16$	&$2.74^{+ 0.05}_{- 0.05}$& $ 5.22^{+ 0.35}_{- 0.29}$& $ 42.94^{+ 0.64}_{- 0.64}$& $ 1.21$\\
&245000&$2.77^{+ 0.06}_{- 0.05}$& $ 4.47^{+ 0.14}_{- 0.13}$& $ 20.27^{+ 0.20}_{- 0.20}$& $ 1.08$	&$2.63^{+ 0.06}_{- 0.06}$& $ 1.65^{+ 0.11}_{- 0.10}$& $ 42.15^{+ 0.38}_{- 0.38}$& $ 1.73$	&$2.66^{+ 0.05}_{- 0.05}$& $ 5.17^{+ 0.35}_{- 0.30}$& $ 43.80^{+ 0.75}_{- 0.74}$& $ 1.12$\\
&255000&$2.86^{+ 0.06}_{- 0.06}$& $ 4.37^{+ 0.14}_{- 0.14}$& $ 20.04^{+ 0.21}_{- 0.21}$& $ 1.06$	&$2.76^{+ 0.07}_{- 0.07}$& $ 1.24^{+ 0.15}_{- 0.15}$& $ 41.23^{+ 0.56}_{- 0.55}$& $ 1.07$	&$2.74^{+ 0.05}_{- 0.05}$& $ 5.61^{+ 0.48}_{- 0.39}$& $ 44.41^{+ 0.73}_{- 0.73}$& $ 1.09$\\
&265000&$2.76^{+ 0.06}_{- 0.06}$& $ 4.26^{+ 0.14}_{- 0.13}$& $ 20.30^{+ 0.21}_{- 0.21}$& $ 0.98$	&$2.64^{+ 0.07}_{- 0.07}$& $ 1.26^{+ 0.15}_{- 0.15}$& $ 40.79^{+ 0.56}_{- 0.56}$& $ 0.99$	&$2.66^{+ 0.05}_{- 0.05}$& $ 5.54^{+ 0.45}_{- 0.37}$& $ 43.32^{+ 0.75}_{- 0.74}$& $ 1.01$\\
&275000&$2.86^{+ 0.10}_{- 0.10}$& $ 3.79^{+ 0.13}_{- 0.12}$& $ 19.01^{+ 0.32}_{- 0.32}$& $ 0.91$	&$2.77^{+ 0.12}_{- 0.11}$& $ 0.81^{+ 0.18}_{- 0.17}$& $ 37.96^{+ 0.75}_{- 0.76}$& $ 0.90$	&$2.81^{+ 0.10}_{- 0.09}$& $ 7.93^{+ 1.65}_{- 1.09}$& $ 40.52^{+ 0.93}_{- 0.91}$& $ 0.93$\\
&295000&$2.82^{+ 0.10}_{- 0.10}$& $ 3.90^{+ 0.14}_{- 0.13}$& $ 16.52^{+ 0.29}_{- 0.29}$& $ 0.99$	&$2.73^{+ 0.13}_{- 0.12}$& $ 0.91^{+ 0.19}_{- 0.18}$& $ 32.98^{+ 0.68}_{- 0.69}$& $ 0.99$	&$2.77^{+ 0.10}_{- 0.09}$& $ 7.12^{+ 1.32}_{- 0.91}$& $ 35.24^{+ 0.89}_{- 0.86}$& $ 1.02$\\
&305000&$3.06^{+ 0.06}_{- 0.06}$& $ 3.71^{+ 0.11}_{- 0.10}$& $ 16.00^{+ 0.18}_{- 0.18}$& $ 1.08$	&$3.02^{+ 0.06}_{- 0.06}$& $ 0.54^{+ 0.12}_{- 0.11}$& $ 32.94^{+ 0.42}_{- 0.42}$& $ 1.07$	&$2.98^{+ 0.06}_{- 0.06}$& $ 10.35^{+ 2.33}_{- 1.52}$& $ 35.25^{+ 0.63}_{- 0.62}$& $ 1.09$\\
&315000&$3.01^{+ 0.05}_{- 0.05}$& $ 3.79^{+ 0.11}_{- 0.10}$& $ 16.10^{+ 0.16}_{- 0.16}$& $ 0.92$	&$2.95^{+ 0.06}_{- 0.06}$& $ 0.65^{+ 0.12}_{- 0.11}$& $ 33.03^{+ 0.40}_{- 0.40}$& $ 0.92$	&$2.92^{+ 0.06}_{- 0.05}$& $ 8.91^{+ 1.48}_{- 1.05}$& $ 35.46^{+ 0.56}_{- 0.56}$& $ 0.94$\\
&325000&$2.97^{+ 0.07}_{- 0.07}$& $ 3.99^{+ 0.13}_{- 0.13}$& $ 16.34^{+ 0.21}_{- 0.21}$& $ 1.02$	&$2.90^{+ 0.08}_{- 0.08}$& $ 0.83^{+ 0.15}_{- 0.14}$& $ 33.56^{+ 0.50}_{- 0.50}$& $ 1.03$	&$2.87^{+ 0.07}_{- 0.07}$& $ 7.35^{+ 1.11}_{- 0.81}$& $ 36.29^{+ 0.65}_{- 0.66}$& $ 1.04$\\
&335000&$3.12^{+ 0.05}_{- 0.05}$& $ 4.19^{+ 0.13}_{- 0.13}$& $ 15.38^{+ 0.17}_{- 0.17}$& $ 0.85$	&$3.06^{+ 0.06}_{- 0.06}$& $ 0.87^{+ 0.13}_{- 0.13}$& $ 32.87^{+ 0.41}_{- 0.41}$& $ 0.85$	&$2.97^{+ 0.06}_{- 0.06}$& $ 7.19^{+ 0.90}_{- 0.69}$& $ 36.16^{+ 0.59}_{- 0.59}$& $ 0.86$\\
&345000&$3.17^{+ 0.06}_{- 0.05}$& $ 4.25^{+ 0.15}_{- 0.14}$& $ 14.92^{+ 0.18}_{- 0.18}$& $ 0.94$	&$3.10^{+ 0.06}_{- 0.06}$& $ 0.88^{+ 0.15}_{- 0.14}$& $ 32.32^{+ 0.44}_{- 0.44}$& $ 0.93$	&$3.02^{+ 0.06}_{- 0.06}$& $ 7.25^{+ 1.04}_{- 0.77}$& $ 35.71^{+ 0.67}_{- 0.65}$& $ 0.96$\\
&355000&$3.24^{+ 0.09}_{- 0.09}$& $ 4.23^{+ 0.18}_{- 0.17}$& $ 13.88^{+ 0.28}_{- 0.28}$& $ 0.93$	&$3.20^{+ 0.11}_{- 0.10}$& $ 0.77^{+ 0.19}_{- 0.18}$& $ 30.22^{+ 0.57}_{- 0.56}$& $ 0.92$	&$3.10^{+ 0.11}_{- 0.10}$& $ 7.83^{+ 1.81}_{- 1.17}$& $ 33.65^{+ 0.86}_{- 0.84}$& $ 0.94$\\
&365000&$3.15^{+ 0.11}_{- 0.11}$& $ 3.99^{+ 0.16}_{- 0.15}$& $ 13.92^{+ 0.32}_{- 0.31}$& $ 1.15$	&$3.09^{+ 0.13}_{- 0.12}$& $ 0.71^{+ 0.20}_{- 0.19}$& $ 29.62^{+ 0.63}_{- 0.62}$& $ 1.13$	&$3.06^{+ 0.12}_{- 0.12}$& $ 8.94^{+ 2.90}_{- 1.64}$& $ 32.22^{+ 0.89}_{- 0.86}$& $ 1.16$\\
&375000&$3.07^{+ 0.11}_{- 0.11}$& $ 4.20^{+ 0.16}_{- 0.15}$& $ 14.20^{+ 0.35}_{- 0.34}$& $ 1.01$	&$2.91^{+ 0.15}_{- 0.15}$& $ 1.05^{+ 0.23}_{- 0.22}$& $ 30.16^{+ 0.70}_{- 0.70}$& $ 1.01$	&$2.94^{+ 0.13}_{- 0.12}$& $ 6.98^{+ 1.48}_{- 0.97}$& $ 32.93^{+ 0.86}_{- 0.86}$& $ 1.06$\\
\hline
     
     \end{tabular}
     }
     \label{tab:t5}      
\end{table*}

\begin{figure}
    \centering
    \includegraphics[width=0.5\textwidth]{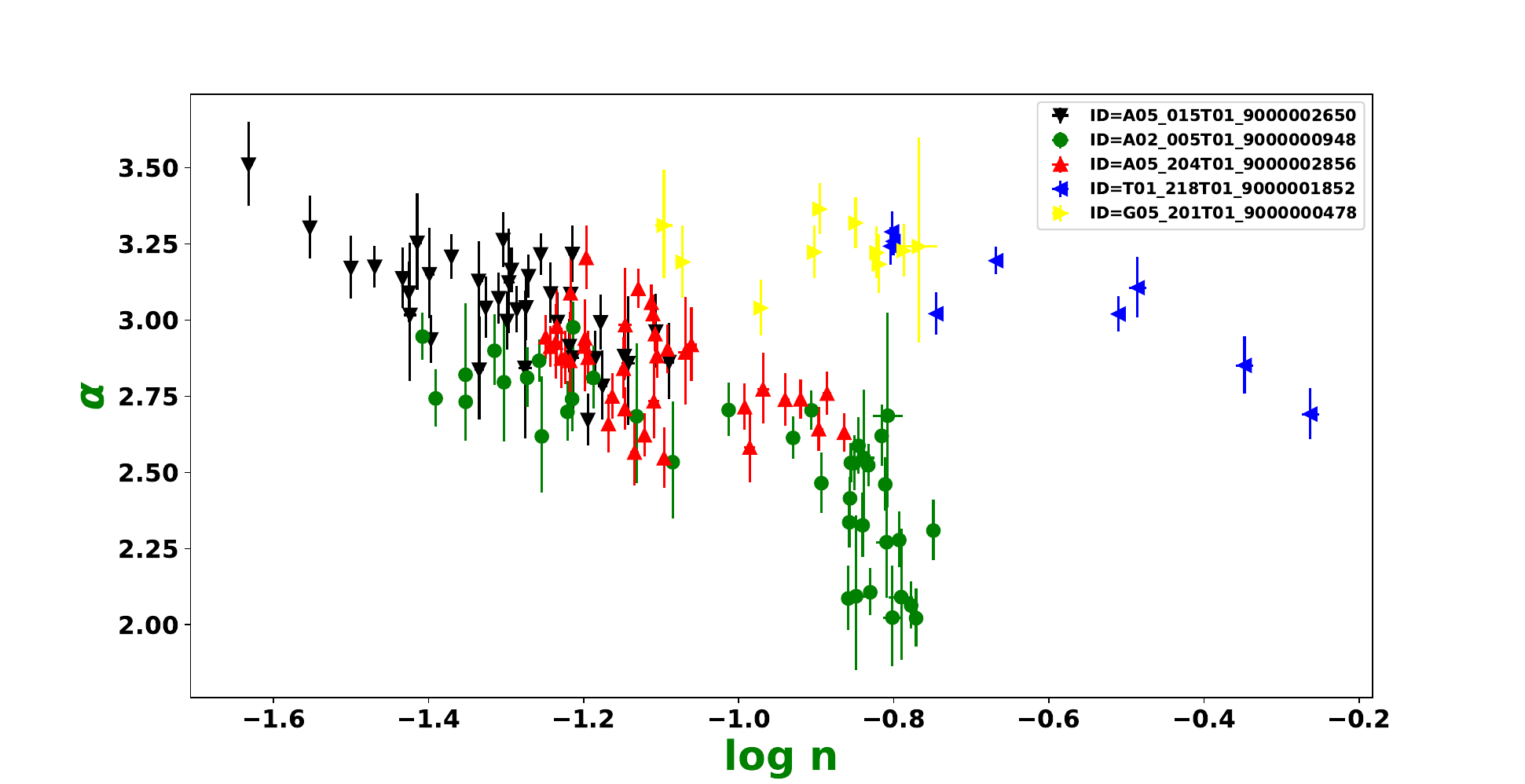}
    \includegraphics[width=0.5\textwidth]{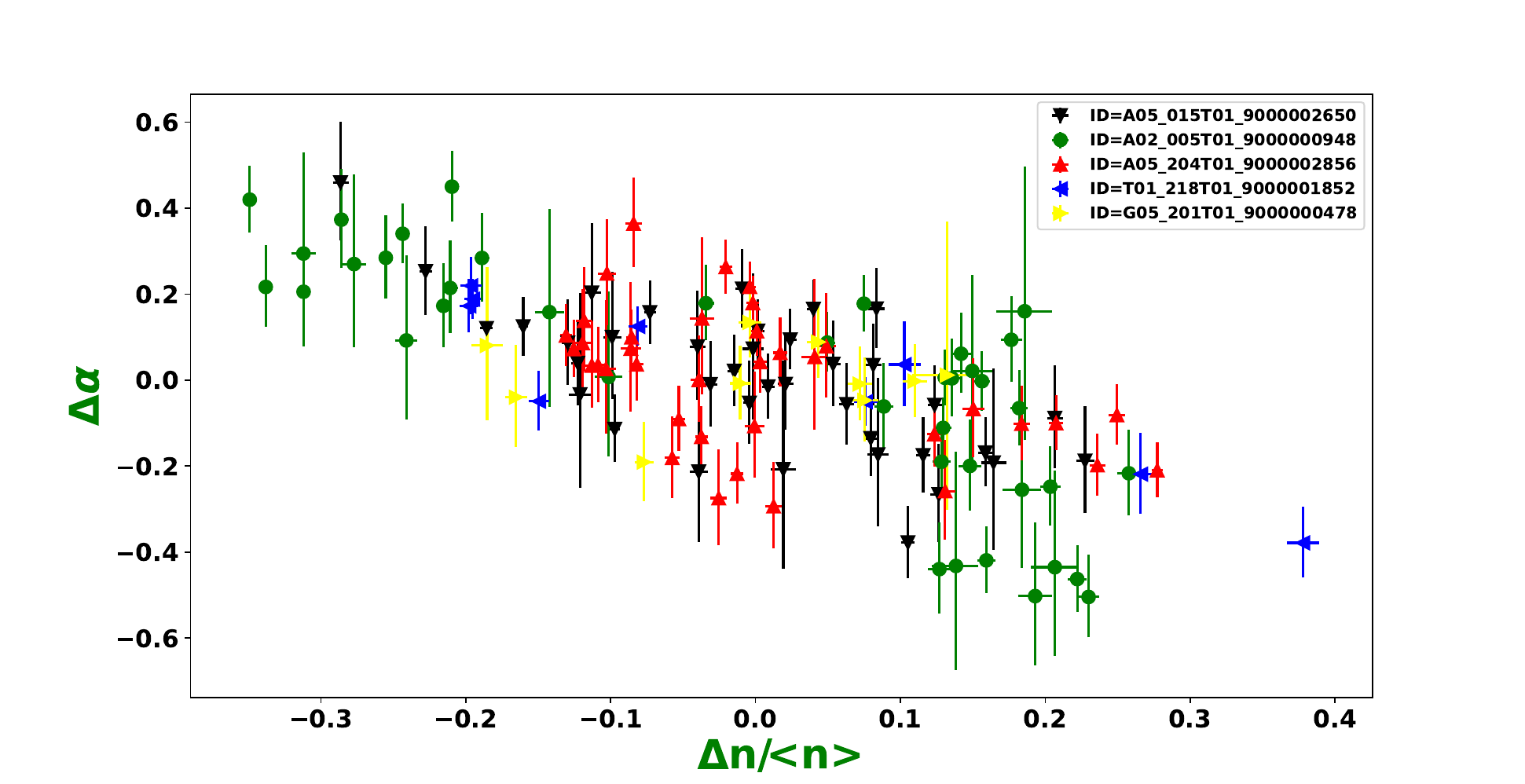}
    \includegraphics[width=0.5\textwidth]{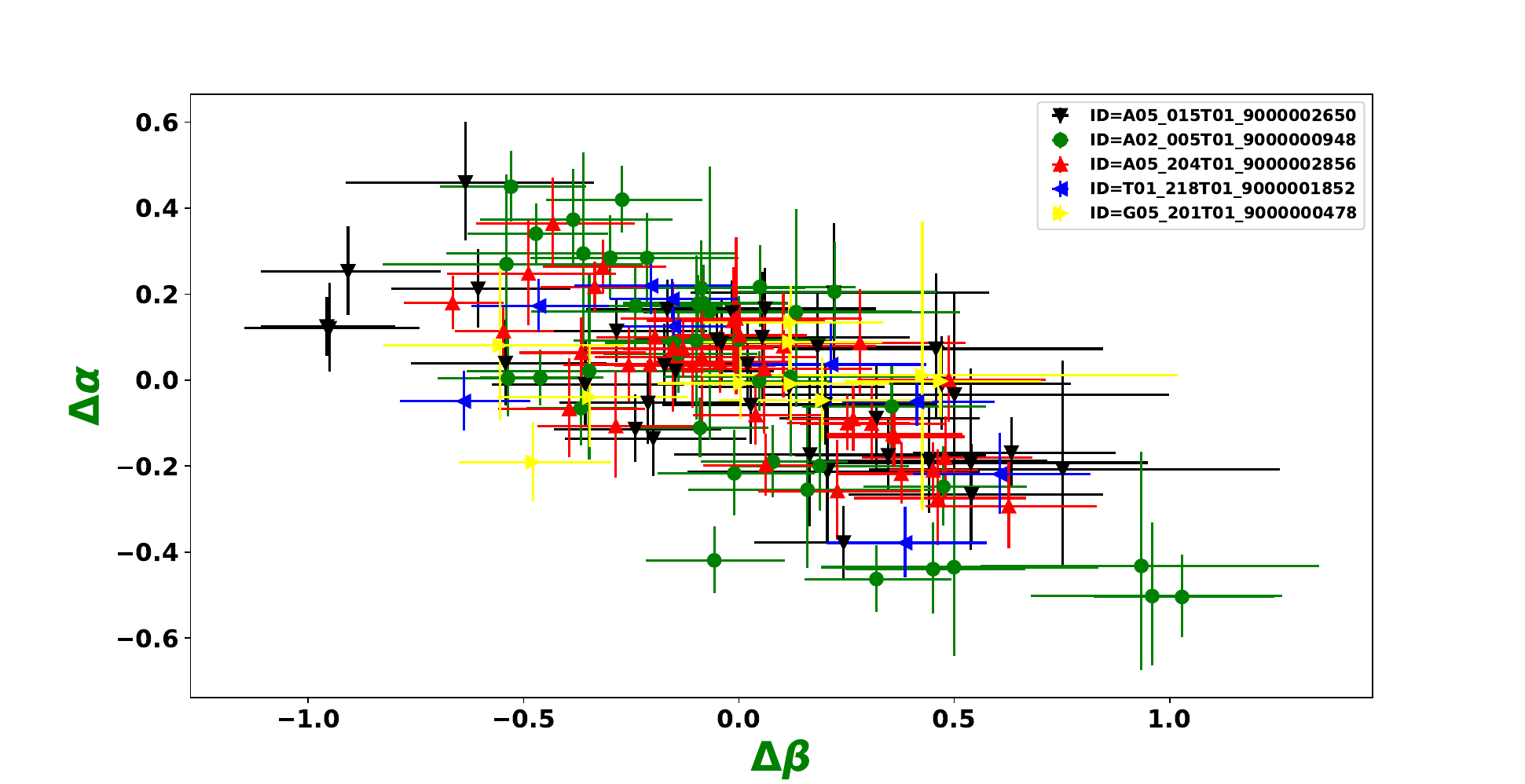}
    \caption{ The correlations between the spectral parameters obtained by fitting the synchrotron convolved LP model to the joint SXT and LAXPC spectrum in each time segment. Different colors in the plot depicts individual observations. %These observations were obtained using the model ($constant\times TBabs(Synconv\otimes log-parabola)$, fitted to the time-resolved spectral segments. 
 The Upper pannel represents variation of $\alpha$ with $\log(n)$ at $\xi=\xi_{ref}$. The  middle pannel represents  correlation between $\Delta\alpha$ and $\Delta$$n$/$<n>$. The Bottom pannel represents  correlation between  $\Delta\alpha$ and $\Delta\beta$.}
    \label{fig:com-lp}
\end{figure}

\begin{figure}
    \centering
    \includegraphics[width=0.80\textwidth]{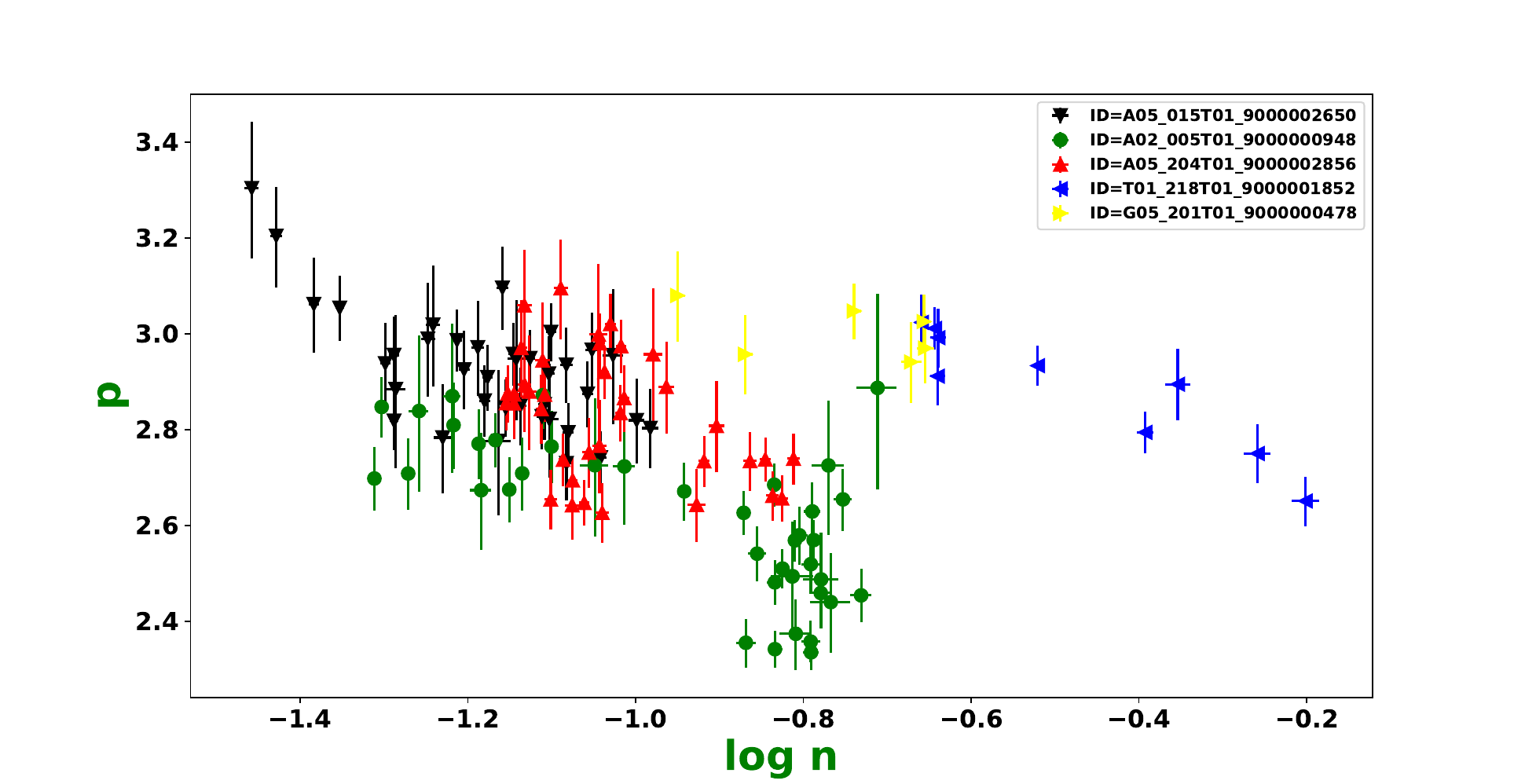}
    \includegraphics[width=0.80\textwidth]{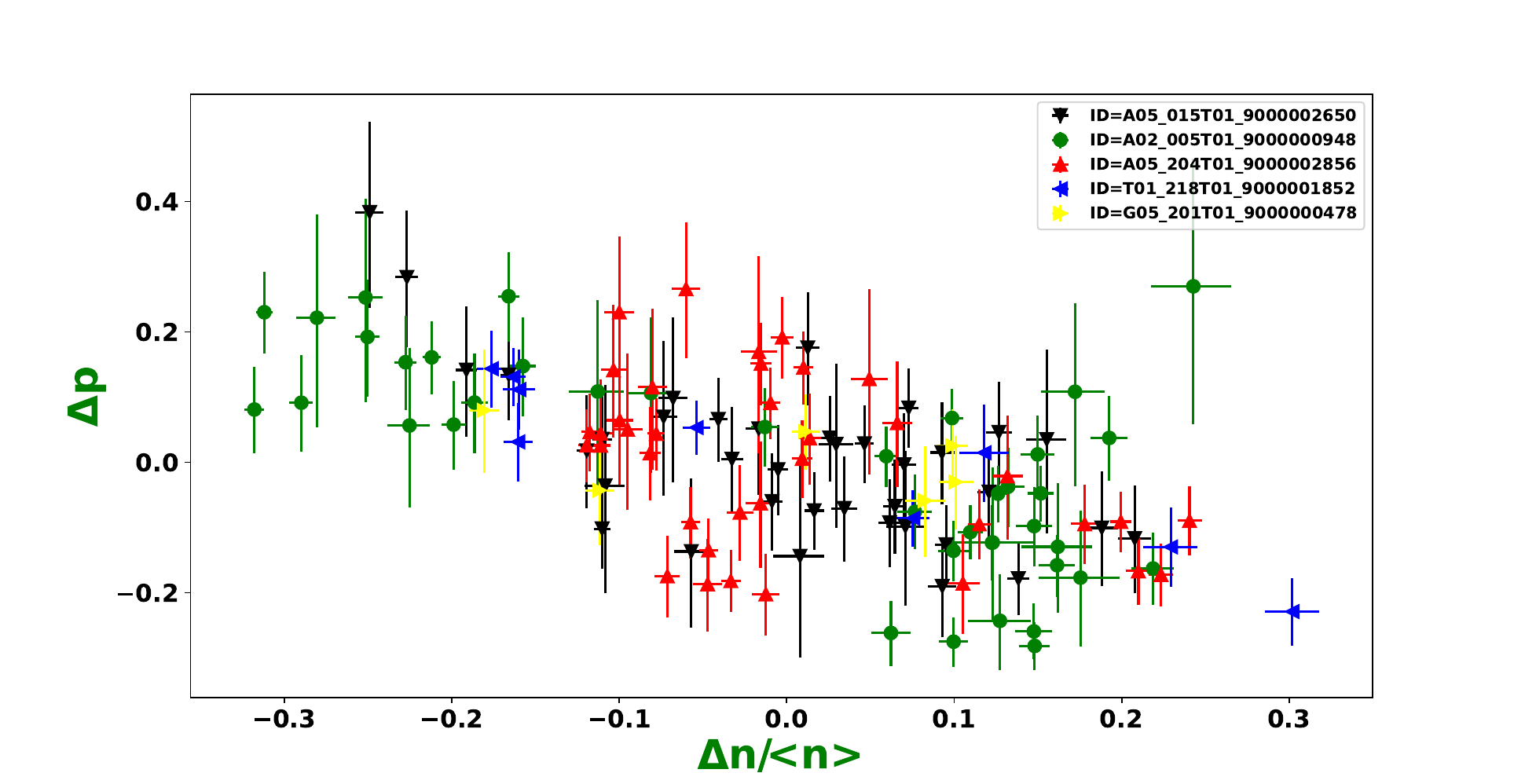}
    \includegraphics[width=0.80\textwidth]{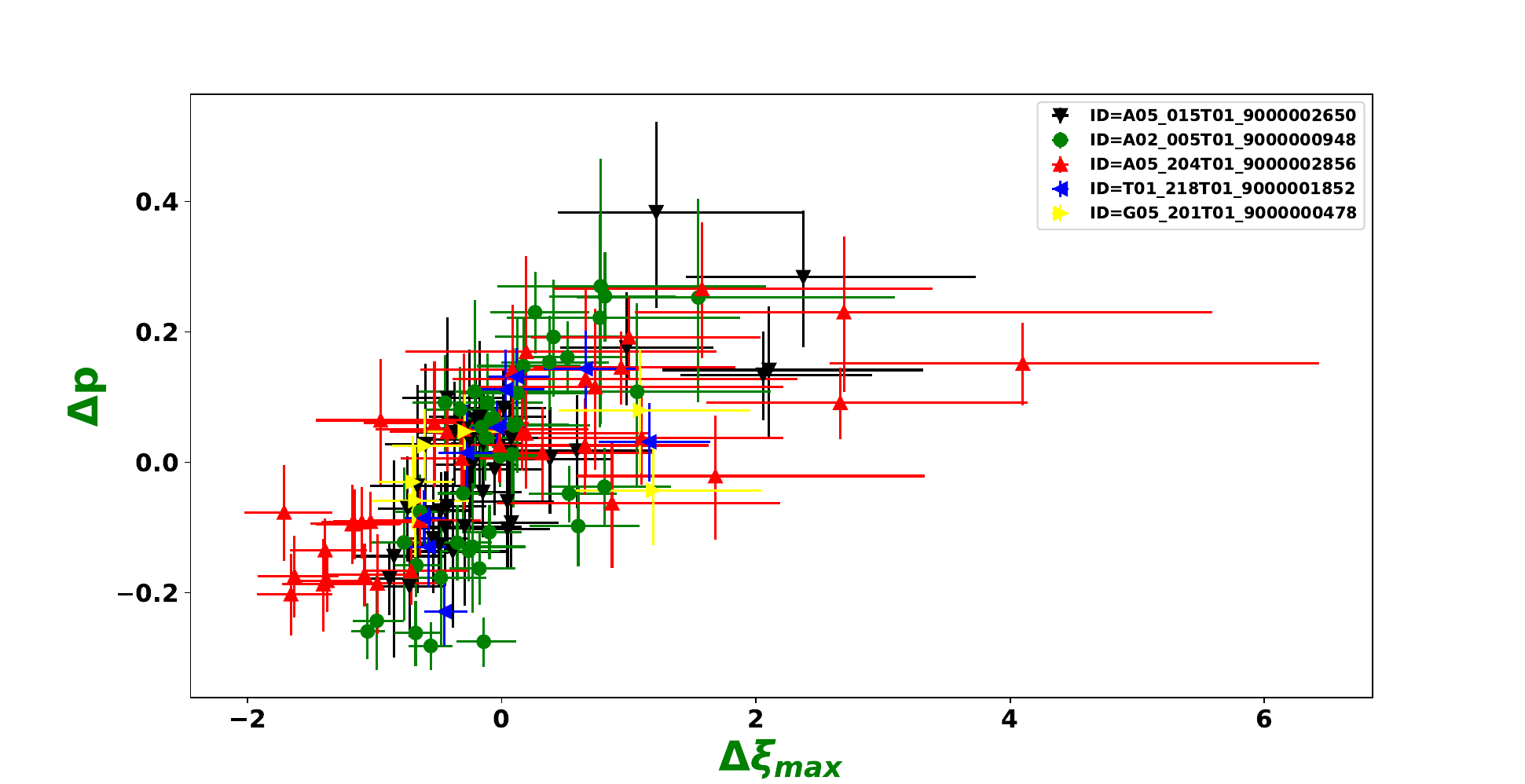}
    \caption{ The correlation  between the spectral parameters obtained by fitting the synchrotron convolved $\xi-max$ model to the joint SXT and LAXPC spectrum in each time segment. Different colors in the plot depicts individual observations. The upper pannel represents relationship between  $p$ and $\log(n)$  at $\xi=\xi_{ref}$. The middle panel represents  correlation between $\Delta p$ and $\Delta$$n$/$<n>$. The Bottom panel represents correlation between  $\Delta p$ and $\Delta \xi_{max}$.}
    \label{fig:combined-gamma}
\end{figure}

%Observation Ids $=$ 
%$ A02_005T01_9000000948$ (ritaban), %$A05_015T01_9000002650 $ (Almar), %$A05_204T01_9000002856$ (ritaban), %$G05_201T01_9000000478$ (Varsha), %$T01_218T01_9000001852$ (subir).

\begin{figure*}
  \centering
 
 \includegraphics[scale=0.7,angle=0]{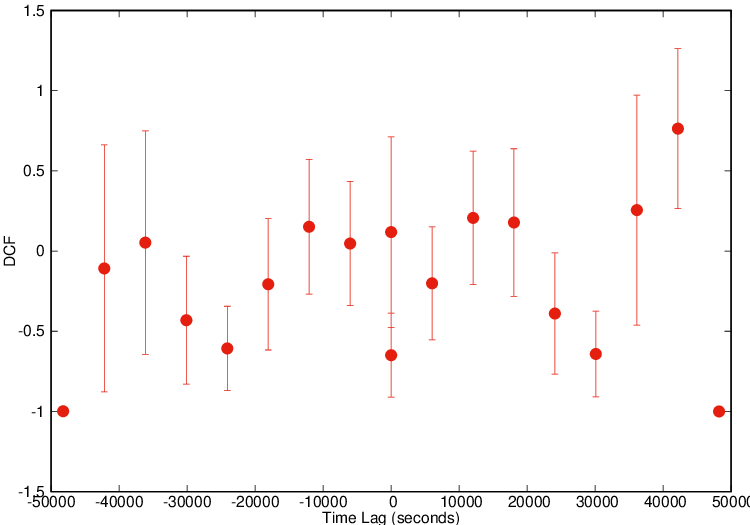}
  \includegraphics[scale=0.7,angle=0]{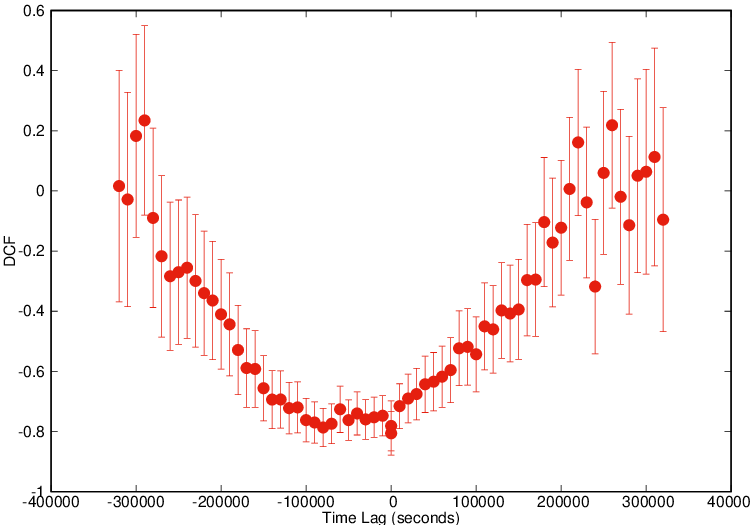}
  \includegraphics[scale=0.7,angle=0]{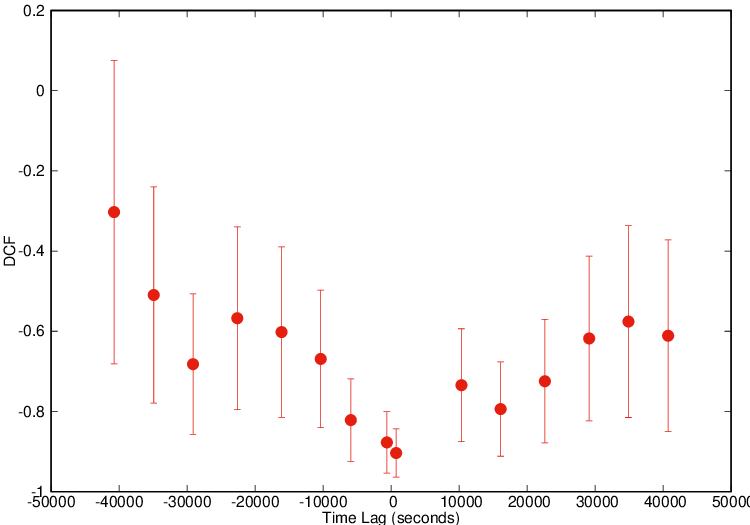} 
  \includegraphics[scale=0.7,angle=0]{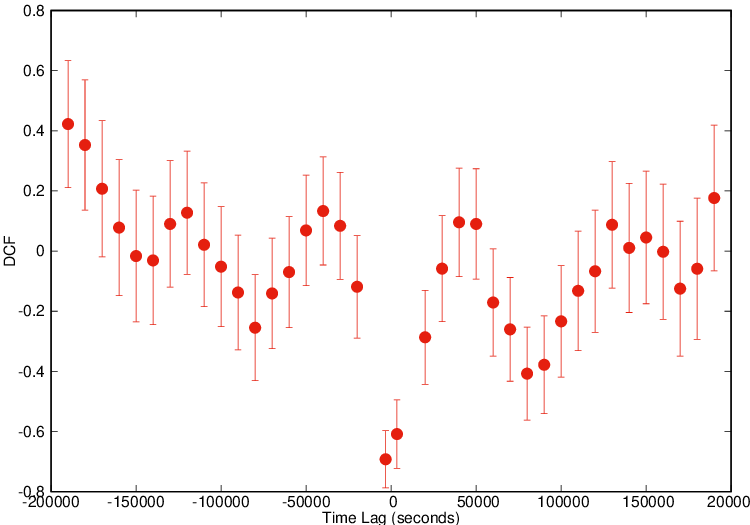} 
  \includegraphics[scale=0.7,angle=0]{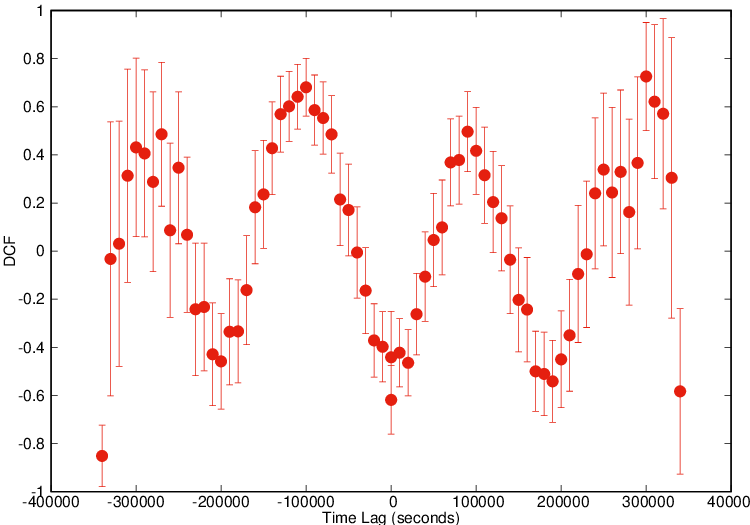} 
  
 \caption{The discrete correlation function  obtained between the variation in index and normalize particle density variation in case of BPL model. In the upper panel, DCF plots for observation S1 (left) and S2 (right) are presented. The middle panel displays DCF plots for observations S3 (left) and S4 (right). The bottom panel features the DCF plot for observation S5. The time lag is expressed in seconds.}

\label{fig:dcf}
\end{figure*}

\end{document}